\newcommand{\myfig}[5]{
\begin{figure}[{#5}]
\includegraphics[keepaspectratio,width=#4,angle=0]{#1}%
\caption{#2}\label{#3}%
\end{figure}}

\newcommand{\figref}[1]{Fig.\ \ref{#1}}
\newcommand{\secref}[1]{Sec.\ \ref{#1}}
\newcommand{\eqnref}[1]{Eq.\ (\ref{#1})}

\documentclass[aps,pre,twocolumn,showpacs,amsmath,amssymb]{revtex4}
\usepackage{graphicx,dcolumn,bm,verbatim,color,hyperref,subfigure}

\begin{document}

\title{A Replica Inference Approach to Unsupervised Multi-Scale Image Segmentation}

\begin{abstract}

We apply a replica inference based Potts model method to
unsupervised image segmentation on multiple scales. This approach
was inspired by the statistical mechanics problem of ``community
detection'' and its phase diagram. Specifically, the problem is cast
as identifying tightly bound clusters (``communities'' or
``solutes'') against a background or ``solvent''. Within our
multiresolution approach, we compute information theory based
correlations among multiple solutions (``replicas'') of the same
graph over a range of resolutions. Significant multiresolution
structures are identified by replica correlations as manifest in
information theory overlaps. With the aid of these correlations as
well as thermodynamic measures, the phase diagram of the
corresponding Potts model is analyzed both at zero and finite
temperatures. Optimal parameters corresponding to a sensible
unsupervised segmentation correspond to the ``easy phase'' of the
Potts model.  Our algorithm is fast and shown to be at least as
accurate as the best algorithms to date and to be especially suited
to the detection of camouflaged images.

\end{abstract}

\author{Dandan Hu}
\affiliation{Department of Physics, Washington University in St.
Louis, Campus Box 1105, 1 Brookings Drive, St. Louis, MO 63130, USA}
\author{Peter Ronhovde}
\affiliation{Department of Physics, Washington University in St.
Louis, Campus Box 1105, 1 Brookings Drive, St. Louis, MO 63130, USA}
\author{Zohar Nussinov$^*$}
\affiliation{Department of Physics, Washington University in St.
Louis, Campus Box 1105, 1 Brookings Drive, St. Louis, MO 63130, USA}

 \maketitle{} \vskip 0.1in

\section{Introduction}\label{sec:introduction}

``Image segmentation'' refers to the process of partitioning a
digital image into multiple segments based on certain visual
characteristics \cite{book1,book2,book3}. Image segmentation is
typically used to locate objects and boundaries in images. The
result of image segmentation is a set of segments that collectively
cover the entire image or a set of extracted contours of the image.
This problem is challenging (see, e.g., Fig.
(\ref{fig:hardzebradog})) and important in many fields. Examples of
its omnipresent use include, amongst many others, medical imaging \cite{medical}
(e.g., locating tumors and anatomical structure), face recognition
\cite{face}, fingerprint recognition \cite{fingerprint}, and machine
vision \cite{mechine}. Numerous algorithms and methods have been
developed for image segmentation. These include thresholding
\cite{thresholding}, clustering \cite{clustering}, compression
\cite{compression} and histogram based  \cite{histogram} approaches,
edge detection \cite{edge}, region growing \cite{region}, split and
merge \cite{split}, gradient flows and partial differential equation
based  approaches \cite{kim,levelset}, graph partitioning methods
and normalized cuts  \cite{graphpartition,normalizedcuts}, Markov
random fields and mean field theories
\cite{markov1,markov2,markov3,markov4}, watershed transformation
\cite{watershed}, random walks \cite{randomwalks},  isoperimetric
methods \cite{isoperimetric}, neural networks \cite{neural}, and a
variety of other approaches, e.g., \cite{model,multi,semi}.

\myfig{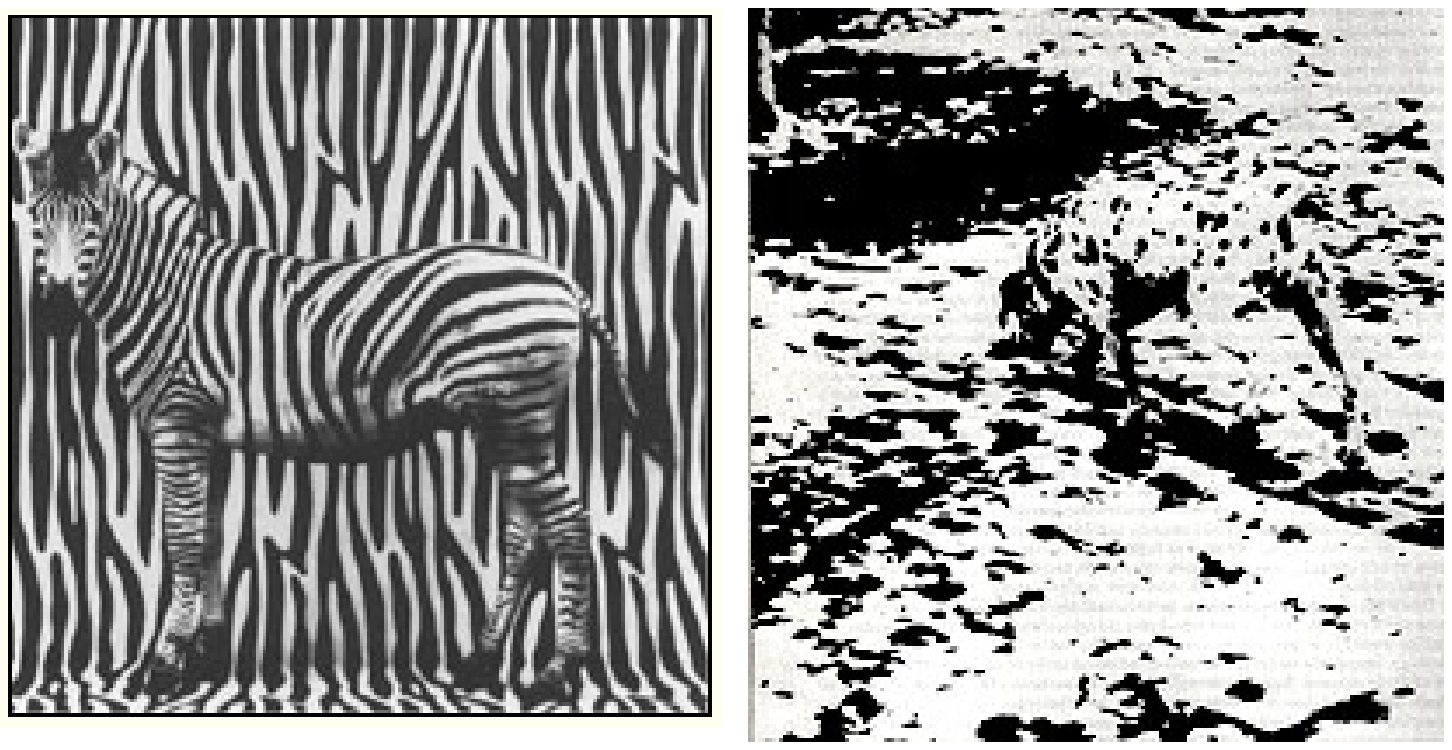}{Examples of currently challenging problems in
image segmentation. Left:  The left image is that of  zebra
(courtesy of Ref.\cite{berkeley-zebra}) with the a similar``stripe''
background. Right: The image on the right is that of a dalmatian dog
\cite{dalmatian}. Most people do not initially recognize the dog
before given clues as to its presence. Once the dog is seen it is
nearly impossible to perceive the image in a meaningless way.
\cite{dalmatian} }{fig:hardzebradog}{1\linewidth}{}

In this work, we will apply a ``community detection'' algorithm to
image segmentation. This method belongs to the graph partitioning
category. Community detection \cite{newman,fortunato,newman2,gmnn} seeks
to identify groups of nodes densely connected within their own group
(``community'') and more weakly connected to other groups. A
solution enables the partition of a large physically interacting system
into optimally decoupled communities. The image is then divided into
different regions (``communities'') based on a certain criterion,
and each resulting region corresponds to an object in the
original image.

It is notable that by virtue of its graph theoretical nature,
community detection is suited for the study of arbitrary dimensional
data.  However, unlike general high dimensional graphs, images are
two (or three) dimensional. Thus, real images are far simpler than
higher dimensional data sets as, e.g., evinced by the
four color theorem stating that four colors suffice to mark different neighboring
regions in a segmentation of any two dimensional image.
Thus, geometrical (and topological) constraints can be used to
further improve the efficiency of the bare graph theoretical method.
In \cite{glass1,glass2}, in the context of analyzing structures
of complex physical systems such as glasses, we used geometry dependent physical potentials to
set the graph weights in various two and three dimensional
systems. In the case of image segmentation, in the absence of
underlying physics, we will invoke geometrical cut-off scales.

In this work, we will discuss ``unsupervised'' image segmentation.
By this term, we allude to a general multi-purpose segmentation
method based on a general physical intuition. The current method
does not take into account initial ``training'' of the algorithm-
i.e., provide the system with known examples in which specific
patterns are told to correspond to specific objects. We leave the
study of supervised image segmentation and more sophisticated
extensions of our inference procedure to a future work. One possible
avenue which can be explored is the use of inference beyond that
relating to different ``replicas'' in the simple form discussed in
this manuscript that is built on prior knowledge (and prior
probabilities in a Bayesian type analysis) of expected patterns in
the studied images.

We will, specifically, apply the multiresolution community detection method, first
introduced in \cite{peter1}, to investigate the overall structure at different resolutions
in the test images. Similar to \cite{peter1}, we will employ
information based measures (e.g., the normalized mutual information
and the variation of information) to determine the significant
structures at which the ``replicas'' (independent solutions of the
same community detection algorithm) are strongly correlated. With
the aid of these information theory correlations, we illustrate how we may
discern structures at different pertinent resolutions (or spatial scales).
An image may be segmented at different
levels of detail and scales by setting the resolution parameters
to these pertinent values.  We demonstrate in a detailed study
of various test cases, how our method works in practice and
 the resulting high accuracy of our image
segmentation method.

\section{Outline}

The outline of our work is as follows.  In Section \ref{sec:potts},
we introduce the Potts model representation of image segmentation
and Potts model Hamiltonians that we will use. These Hamiltonians
were earlier derived for graph theory applications. In Section
\ref{translation}, we discuss how we represent images as graphs. In
Section \ref{def}, we briefly define the key concepts of trials and
replicas which are of great importance in our approach.  In
\secref{sec:algorithm}, we present our community detection
algorithm. In \secref{sec:multiresolution}, we discuss the
multiresolution method and the information based measures. In
Section \ref{corr_g}, we illustrate how replica correlations may be
used to set graph  weights. For the benefit of the reader, we
compile the list of parameters in Section \ref{par}. We discuss the
computational complexity of our method in Section
\ref{sec:complexity}. In \secref{sec:results}, we provide {\em in silico}
 ``experimental results'' of our image
segmentation method when applied to many different examples. These
examples include, amongst others, the
Berkeley image segmentation and the Microsoft Research Benchmarks.
We conclude in \secref{sec:conclusion} with a summary of our
results. Specific aspects are further detailed in
the appendices.

\section{Potts Models} \label{sec:potts}

In what follows, we will briefly elaborate on our particular Potts model representations of
images and the corresponding Hamiltonians (energy or cost
functions).


\subsection{Representation}

As is well appreciated, different objects in an image or more
general communities in complex graph theoretical problems are
ultimately denoted by a ``Potts type'' \cite{potts} variable
$\sigma_{i}$. That is, if node $i$ lies in a community number $w$
then $\sigma_{i} =w$. If there are $q$ communities in the graph then
$\sigma_{i}$ can assume values $1 \le \sigma_{i} \le q$. A state
$\{\sigma_{s}\}_{s=1}^{N}$ corresponds to a particular partition (or
segmentation) of the system into $q$ communities (or objects). In
the context of image segmentation, Potts model representations can,
e.g., also be found in
\cite{pottsmodel1,pottsmodel2,pottsmodel3,pottsmodel4}.

\subsection{Potts model Hamiltonian for unweighted graphs}

In \cite{peter2}, a particular Potts model Hamiltonian was
introduced for community detection. The ground states of this
Hamiltonian (or lowest energy states) correspond to optimal
partition of the nodes into communities. This Hamiltonian {\em does
not involve a comparison relative to random graphs (``null
models'')}  \cite{fortunato} and as such was free of the
``resolution limit'' problems \cite{fortunato, rlimit1, rlimit2}
wherein the number of found communities or objects scaled with the
system size in a way that was independent of the actual system
studied. In what follows below, there are $N$ elementary nodes in a
graph (or pixels in an image), we consider general {\em unweighted}
graphs in which any pair of nodes may be either linked with a
uniform weight or not linked at all. Specifically, a link between
sites $i$ and $j$ is associated with edge weights $A_{ij}$ and
$J_{ij}$. In these unweighted graphs, $A_{ij}$ is an element of the
adjacency matrix. That is, $A_{ij}=1$ if nodes $i$ and $j$ are
connected by an edge and $A_{ij} = 0$ otherwise. The weights $J_{ij} =
(1-A_{ij})$.

The goal of the general (or ``absolute'') Potts model Hamiltonian
\cite{peter2} was
to energetically favor any pair of linked nodes to be in the same
community, to penalize for a pair of unlinked nodes to
be in the same community and conversely for nodes in different
communities (penalize for having two linked nodes be in
different communities and favor disjoint nodes being
in different communities). Putting all of these bare energetic
considerations together (sans any
comparisons to random graphs),
the resulting Potts model Hamiltonian
(or energy function)
for a system of $N$ nodes simplifies to
\cite{peter2,explain_solute}
\begin{eqnarray}
 {\cal{H}} (\{\sigma_{s}\}_{s=1}^{N}) =
-\frac{1}{2}\sum_{i\neq j}(A_{ij}-\gamma
J_{ij})\delta(\sigma_i,\sigma_j).
 \label{eq:ourpotts}
\end{eqnarray}

In Eq.(\ref{eq:ourpotts}), we emphasize the dependence of the
Hamiltonian on the $N$ different variables $\{\sigma_{s}\}$ at each
lattice site $s$ (each of which can assume $q$ values).
In what follows, the dependence of the Hamiltonian on
$\{\sigma_{s}\}_{s=1}^{N}$ will always be understood.

The Kronecker delta $\delta(\sigma_i,\sigma_j)=1$ if
$\sigma_i=\sigma_j$ and $\delta(\sigma_i,\sigma_j)=0$ if
$\sigma_i\neq \sigma_j$. In this Hamiltonian, by virtue of the
$\delta_{\sigma_{i} \sigma_{j}}$ term, each spin $\sigma_i$
interacts only with other spins in its own community. As such,
the resulting model is {\em local}-- a feature that
enables high accuracy along with
rapid convergence \cite{peter2}.

As noted above, minimizing this Hamiltonian corresponds to
identifying strongly connected clusters of nodes. The parameter
$\gamma$ is the so called ``resolution parameter'' which adjusts the
relative weights of the linked and unlinked edges, as in
\eqnref{eq:ourpotts}. This is easily seen by inspecting
Eq.(\ref{eq:ourpotts}). A high value of $\gamma$ leads to forbidding
energy penalties unless all intra-community nodes ``attract'' one
another and lie in the same community. By contrast, $\gamma=0$ does
not penalize the inclusion of any additional nodes in a given
community and the lowest energy solution generally corresponds to
the entire system.

\subsection{Potts Model Hamiltonian for Weighted Graphs} In
\emph{weighted} graphs, we assign edges between nodes with the
respective weights based on the interaction strength (e.g., the
(dis-) similarity of the intensity or color defines the edge weight
in image segmentation problem). Specifically, in image segmentation
problems, we determine (based on, e.g., color or intensity
differences) the weight $V_{ij}$ between each member of a node pair.
We then shift each such value by an amount set by a background
$\bar{V}$, i.e., $V_{ij}'=(V_{ij}- \bar{V})$. The subtraction
relative to the background of $\bar{V}$ allows for our community
detection algorithm to better partition the network of pixels. In
principle, this background can be set to be spatially non-uniform.
However, in this work we set $\bar{V}$ to be a constant. Thus, we
generalize the earlier model of \cite{peter1} in
\eqnref{eq:ourpotts} by the inclusion of a background $\bar{V}$ and
by allowing for continuous weights $V_{ij}$ instead of discrete
weights that are prevalent in graph theory. The resulting
Hamiltonian \cite{glass1, glass2} reads

\begin{eqnarray}
{\cal{H}} = \frac{1}{2}\sum_{a=1}^q
(V_{ij}-\bar{V})\Big[\Theta(\bar{V}-V_{ij})+\gamma\Theta(V_{ij}-\bar{V})\Big]
\delta(\sigma_{i}, \sigma_{j}).
 \label{eq:newpotts}
\end{eqnarray}

The form of this Hamiltonian and that of Eq. (\ref{eq:ourpotts}) was
inspired by positive and negative energy terms that favor the
formation of tightly bound clusters (or ``solutes'') that are more
weakly coupled to their surroundings \cite{explain_solute}. Similar
to the important effects of the solute found in physical systems
\cite{mossa}, the Hamiltonian of Eq.(\ref{eq:newpotts}) captures all
interactions in the system \cite{explain_solute}. Earlier
\cite{glass1, glass2}), we invoked the Hamiltonian of Eq.
(\ref{eq:newpotts}) to analyze static and dynamic structures in
glasses.

In Eq. (\ref{eq:newpotts}), the number of communities $q$ may be specified from the input or left
arbitrary and have the algorithm decide by steadily increasing the
number of communities $q$ for which we have low energy solutions.
The Heavyside functions $\Theta(x)$ ``turns on'' or ``off'' the edge
designation [$\Theta(x>0)=1$ and $\Theta(x<0)=0$] relative to the
aforementioned background $\bar{V}$. As before, minimizing the
Hamiltonian of Eq. (\ref{eq:newpotts})
corresponds to identifying strongly connected clusters
of nodes.

While in Eq. (\ref{eq:ourpotts} (or Eq. \ref{eq:newpotts}),)
the input concerns two-point ($p=2$) edge weights $V_{ij}$ (or $A_{ij}$) , it is,
of course, possible to extend these Hamiltonian to allow for more general motifs
(such as $p=3$ node triangles) and include $p \ge 3$ point weights $V_{ijk}$ (and extensions thereof).
These correspond to $p$ spin interactions. In the current study, however, we limit ourselves
to $p=2$ node weights.

\section{Casting Images as Networks} \label{translation}

We will now detail how we translate images into networks with general edge weights
that appear in Eqs.(\ref{eq:ourpotts}, \ref{eq:newpotts}). We will represent
pixels as the nodes in a graph. Edge weights define the (dis-)
similarity between the neighborhood pixels.

Images may be broadly divided into two types: (a) those with the
uniform and (b) those with varying intensity. ``Uniform intensity''
means that the entire object or each component is colored by one
intensity or color. By the designation of ``varying intensity'', we allude to objects or
components that exhibit alternating intensities or colors, e.g.,
the stripes and spots seen in Fig. (\ref{fig:hardzebradog}).

Regarding the above two types of images, two different methods may
be employed to define the edge weights: (i) The intensity/color
difference between nodes is defined as the edge weight in images
with uniform intensity. (ii) The overlap between discrete Fourier
transforms of blocks is defined as the edge weight in images with
varying intensity. The second method is designed to distinguish the
target and the background by their specific frequencies. We will
detail both methods below in \secref{sec:uniform} (where we discuss
images with uniform intensities) and \secref{sec:nonuniform} (spatially varying intensities).

\subsection{Edge definition for images with uniform
intensity}\label{sec:uniform}

For images of uniform intensity, we will define edges based on the
color (dis-) similarity. For the unweighted Potts model of
\eqnref{eq:ourpotts}, we will assign an edge between two pixels ($i$
and $j$ ) if the ``color'' difference ($D_{ij}$) between them is
less than some threshold ($\bar{V}$). That is,
\begin{eqnarray}
A_{ij} = \Theta(\bar{V} - D_{ij}).
\end{eqnarray}

For weighted Potts model in \eqnref{eq:newpotts}, we will, as we
will elaborate on momentarily, set the weights $V_{ij}$ to be the
``color'' difference ($D_{ij}$) between nodes $i$ and $j$, i.e.,
\begin{eqnarray}
V_{ij}=D_{ij}.
\end{eqnarray}
 As seen from the energy functions of
Eqs. (\ref{eq:ourpotts}, \ref{eq:newpotts}), a large dis-similarity
$V_{ij}$ favors nodes $i$ and $j$ being in different clusters.

A grey scale image is an image that in which the value of each pixel
carries only intensity information. Images of this sort are composed
exclusively of shades of gray, varying from black at the weakest
intensity ($I=0$) to white at the strongest ($I=255$). For a
grey-scale image, the ``color'' difference is the absolute value of
the intensity difference, i.e.,
\begin{eqnarray}
D_{ij}=|I_i-I_j|.
\end{eqnarray}
 A ``color image'' is an image that includes color
information for each pixel. Each pixel contains three
color components: red, green and blue (or RGB). The value of the
intensity of each of these three components may attain any
of $2^{8}$ values (any integer in the interval $[0,255]$). For a
color-scale image, we define the ``color'' difference as the average
of the differences between the color components red, green and blue.
That is, with $R_{i}, G_{i},$ and $B_{i}$ respectively denoting the
strengths of the red, green, and blue color components at site $i$,
we set
\begin{eqnarray}
D_{ij}=\frac{1}{3}(|R_i-R_j|+|G_i-G_j|+|B_i-B_j|). \label{drgb}
\end{eqnarray}
We do not store edges between every pair of nodes. Rather, edges
connect nodes whose distance is less than a tunable value $\Lambda$.

\subsection{Edge definition for images with varying
intensities}\label{sec:nonuniform}

Typically, images with varying intensities contain different
patterns. To separate these patterns, we construct a
``block-structure'' containing the quintessential pattern
information. We next introduce a method to divide blocks and then
elaborate on two different ways to connect edges between blocks.

General contending pertinent scales may be determined by, e.g.,
examining the peaks of the Fourier transform of an entire image
(whose location yields the inverse wave-length and whose width
reveals the corresponding spatial extent of these modulations).
While such simple transforms may aid optimization in determining
candidate parameter scales,
our algorithm goes far beyond such simple global measures.\\

\subsubsection{Overlapping blocks} \label{sec:blocks}

\myfig{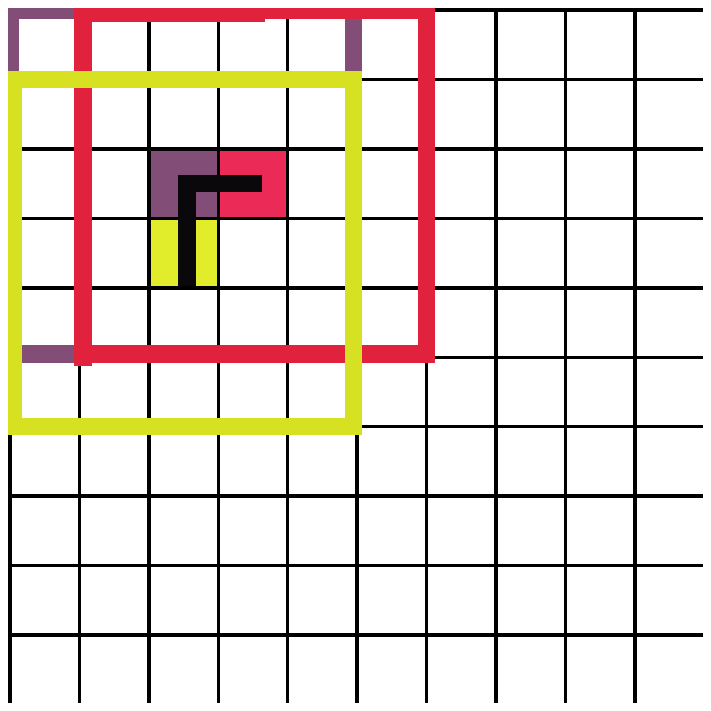}{[Color Online.] An example of overlapping blocks.
The block size is $L_x \times L_y = 5\times5$. The nearest neighbor
of the block enclosed by ``purple'' in x-direction is the one
enclosed by ``red'', and its nearest neighbor in y-direction is the
one enclosed by ``yellow''. They are connected due to the nearest
neighbor condition.}{fig:block}{0.5\linewidth}{t}

We will divide an entire image of size $N=N_x\times N_y$ into $N$
overlapping blocks. These blocks are centered about each (of  the $N$)
pixels and are of size $L_x\times L_y$. The dimensions of the
individual blocks are, generally, far smaller than that of the
entire system, $L_{x,y}\ll N_{x,y}$. General scales can be gleaned
from a Fourier transform of the entire image.

To construct the connection matrix between the blocks, we connect
edges between each pair of blocks and set the distance between the nearest block pair to be $1$. 
This choice has the benefit of overlapping the nearest neighbor
blocks, which share more commons. \figref{fig:block} gives
a schematic plot of  the ``overlapping block'' structure.\\

\subsubsection{Average intensity difference between blocks}
\label{sec:average}

Following the construction of the overlap blocks structure, we next compute
the average intensity of each block and connect the edges between
blocks based on the difference. In this case, each block can be
treated as a ``super-node'' which contains the pattern information
of the studied image.

To further incorporate geometrical scales, we multiply
the edge weights by $\exp(-|{\bf{r}}_m-{\bf{r}}_n|/\ell)$ (where $\ell$
is a tunable length scale and the vectors ${\bf{r}}_{m}$ and ${\bf{r}}_{n}$
denote the spatial locations of points $m$ and $n$). We remind the reader
that there are $N$ basic blocks and thus $N$ possible values of $m$
(and $N$ possible values of $n$). We will set in \eqnref{eq:newpotts}, the weights $V_{mn}$
between block $m$ and $n$ to be

\begin{eqnarray}
V_{mn}=\frac{D_{mn} \exp(-|{\bf{r}}_m-{\bf{r}}_n|/\ell)}{L_x  L_y},
\label{eqn:vmn}
\end{eqnarray}

where

\begin{eqnarray}
D_{mn} =(1- \delta(m,n))
|\sum_{i=0}^{i=L_x-1}\sum_{j=0}^{j=L_y-1}(I^m(i,j)-I^n(i,j))|.
\label{eqn:dmn}
\end{eqnarray}

As seen in
\eqnref{eqn:dmn}, $D_{mn}$ is the sum of the absolute values of the
intensity differences between blocks $m$ and $n$ with each of these
blocks being of size $L_x\times L_y$. In \eqnref{eqn:vmn},
$|{\bf{r}}_m-{\bf{r}}_n|$ is the physical distance between block $m$
and $n$ (i.e., the distance between the central nodes of each
block).

The geometrical factor of
($\exp(-|{\bf{r}}_m-{\bf{r}}_n|/\ell)$) in \eqnref{eqn:vmn} with a
tunable length scale $\ell$ can be set to prefer (and, as we will illustrate also
to detect) certain scales in the image. This enables the algorithm
to detect clusters of varying sizes that contain rich textures. \\

\subsubsection{Fourier amplitude derived weights}\label{sec:fourier}

As it is applied to image segmentation, the utility of Fourier transformations is well appreciated.
We next discuss how to invoke these in our Potts model Hamiltonian. To highlight
their well known and obvious use, we note that, e.g.,
the stripes of the zebra in \figref{fig:hardzebradog} contain wave-vectors which are different from
those of the more uniformly modulated background. Thus, a spatially local Fourier
transform of this image may distinguish the zebra from the
background. We will now invoke Fourier transforms in a general way in order
to determine the edge weights in our network representation
of the image.

With the preliminaries of setting up the block structure in tow, we
now apply a discrete Fourier transform inside each block.
Rather explicitly,  excluding the spatial origin, the local discrete $2-D$ Fourier
transform of a general quantity $f_{m}$ within block $m$
with internal Cartesian coordinates $(a,b)$ is
\begin{eqnarray}
F_m(k,l)= \sum_{a=0}^{L_x-1}\sum_{b=0}^{L_y-1}f_{m}(a,b) e^{-2\pi i
(\frac{k a}{L_x}+\frac{l b}{L_y})} - f_{m}(0,0). \label{eq:fourier}
\end{eqnarray}
The wave-vector components $k=0,1,...,L_y-1$ and $l=0,1,...,L_x-1$.  In
applications, we set, for grey-scale images, $f_{m}(a,b)$ to be the
intensity $I$ at site $(a,b)$ in block $m$ (a whose location relative to the origin of the entire image we
denote by ${\bf r}_{ab;m}$). That is,
$f_{m}(a,b)= I({\bf r}_{ab;m})$. In color images,
we set $f$ to be the average over the intensity of the red, green and
blue components: $f(a,b)=\frac{1}{3}(R(a,b)+G(a,b)+B(a,b))$.
We fix the couplings $J_{mn}$
between blocks $m$ and $n$ to be
\begin{eqnarray}
J_{mn}=\sum_{k=0}^{L_x-1}\sum_{l=0}^{L_y-1} |F_m^{*}(k,l) F_n(k,l)|.
\label{eq:coupling}
\end{eqnarray}
We connect blocks whose spatial separation is less than the
aforementioned tunable cutoff distance $\Lambda$ by links having
edge weights $V_{mn}$. In practice, we fixed $\Lambda$. With Eq.
(\ref{eq:coupling}) in hand, we set
\begin{eqnarray}
\label{VJ}
V_{mn} =  (\delta(m,n)-1) J_{mn}
\exp(-|{\bf{r}}_m-{\bf{r}}_n|/\ell)
\end{eqnarray}
in \eqnref{eq:newpotts}. In this case, the background $\bar{V}$ would be negative.\\

When inverting the sign of the left hand side of Eq.(\ref{VJ})
(shown in Appendix C), our algorithm will be also suited for the
detection of changing objects against a more uniform background.

We now briefly comment on the relation between
the Fourier space overlaps and weights in Eqs.(\ref{eq:coupling},\ref{VJ}) and the real space overlaps and weights in
Eqs.(\ref{eqn:vmn},\ref{eqn:dmn}).  It is notable that
in Eq.(\ref{eq:coupling}), we sum over the {\em modulus}
of the products of the Fourier amplitudes. By Parseval's theorem,
sans the modulus in the summation in Eq.(\ref{eq:coupling}), $J_{mn}$ would be identical
to the overlap in real space between $f_{m}$ and $f_{n}$. Such real space overlaps directly relate
to the real-space overlaps in Eq.(\ref{eqn:dmn}) [following a replacement
of the absolute value in Eq.(\ref{eqn:dmn}) by its square and an overall innocuous multiplicative
scale factor]. Thus,
without the modulus in Eq.(\ref{eq:coupling}), the Fourier
space calculation outlined above affords no
benefit over its real space counter-part. Physically,
the removal of the phase factors when performing the summation
in Eq.(\ref{eq:coupling}) avoids knowledge of the relative location
of the origins between different blocks. This allows different regions of
a periodic pattern to be strongly correlated and clumped together.
By contrast, for a periodic wave
of a particular wave-vector, the real space overlaps between blocks $m$ and $n$
may vanish when the origins
of blocks $m$ and $n$ are displaced by a real space distance
that is equal to half of the wave-length of the periodic wave
along the modulation direction. Thus, the real space weights as derived
from Eqs.(\ref{eqn:vmn},\ref{eqn:dmn}) may vanish when the corresponding
Fourier space derived weights (Eqs.(\ref{eq:coupling},\ref{VJ})) are sizable.

It is possible to improve on the simple
Fourier space derived weights by a general wavelet analysis.

\section{Definitions: Trials and Replicas} \label{def}

In the following sections, we will discuss our specific algorithms for
(i) community detection and (ii) multi-scale community detection.
Before giving the specifics of our algorithms, we wish to introduce
two concepts on which our algorithms are based. Both pertain to the
use of multiple identical copies of the same system (image) which
differ from one another by a permutation of the site indices. Thus,
whenever the time evolution may depend on sequentially ordered
searches for energy lowering moves (as it will in our greedy
algorithm),  these copies may generally reach different final
candidate solutions. By the use of an {\em ensemble} of such
identical copies, we can attain accurate result as well as determine
information theory correlations between candidate solutions and
infer from these a detailed picture of the system.

In the definitions of ``trials'' and ``replicas'' given below, we build on the existence of a given
algorithm (any algorithm) that may minimize a given energy or cost
function. In our particular case, we minimize the Hamiltonian of
Eqs. (\ref{eq:ourpotts}, \ref{eq:newpotts}).
\bigskip

$\bullet$ {\underline{{\em Trials}.} We use trials alone in our bare
community detection algorithm \cite{peter1,peter2}. We run the
algorithm on the same problem $t$ independent times. This may
generally lead to different contending states that minimize Eqs.
(\ref{eq:ourpotts}, \ref{eq:newpotts}). Out of these $t$ trials, we
will pick the lowest energy state and use that state as the
solution.
\bigskip

$\bullet$ {\underline{{\em Replicas}.}  We use both trials and
replicas in our multi-scale community detection algorithm
\cite{peter2}. Each sequence of the above described $t$ trials is
termed a {\em replica}.  When using ``replicas'' in the current
context, we run the aforementioned $t$ trials (and pick the lowest
solution) $r$ independent times. By examining information theory
correlations between the $r$ {\em replicas} we infer which features
of the contending solutions are well agreed on (and thus are likely
to be correct) and on which features there is a large variance
between the disparate contending solutions that may generally mark
important physical boundaries. We will compute the information
theory correlations within the ensemble of $r$ replicas.
Specifically, {\em information theory extrema} as a function of the scale
parameters, generally correspond to more pertinent solutions that
are locally stable to a continuous change of scale.  It is in this
way that we will detect the important physical scales in the system.
\bigskip

These definitions might seem fairly abstract for the moment. We will
flesh these out and re-iterate their definition anew when detailing
our specific algorithms to which we turn next.

\section{The community detection algorithm}\label{sec:algorithm}

Our community detection algorithm for minimizing Eqs. (\ref{eq:ourpotts}, \ref{eq:newpotts})
follows four steps \cite{peter2}.

\bigskip

{\bf (1)} We partition the nodes based on a ``symmetric'' or ``fixed
q'' initialization ($q$ is the number of community). \newline

 $\bullet$ ``Symmetric'' initialization
alludes to an initialization wherein each node forms its own community (and thus, initially,
there are $q=N$ communities).

  $\bullet$ ``Fixed q'' initialization corresponds to a random initial distribution of all
nodes into q communities.

\bigskip

For the
application of image segmentation, ``symmetric'' initialization is
used for the ``\textit{unsupervised}'' case. In this case, the
algorithm does not know what to look for, thus the ``symmetric''
initialization provides the advantage of no bias towards a
particular community.  The algorithm will decide the number of community $q$ by
merging nodes for which we have lower energy solution.

``Fixed q''
initialization may be used in a ``\textit{supervised}'' image
segmentation. The community membership of individual node will be
changed to lower the solution energy. One has to decide how much
information is needed by observing the original image and enter the
number of communities $q$ as an input. Different levels of
information correspond to different number of communities $q$. For
instance, if only one target needs to be identified, $q=2$ is
enough. The $q$ communities include the target and background.

In the following sections, we
will use the ``unsupervised'' image segmentation and let the
algorithm decide the community number $q$.
\bigskip

{\bf(2)} Next, we sequentially ``pick up'' each node and place it in
the community that best lowers the energy of Eqs.
(\ref{eq:ourpotts}, \ref{eq:newpotts}) based on the current state of
the system.\\

{\bf(3)} We repeat this process for all nodes and continue iterating
until no energy lowering moves are found after one full cycle through all nodes.\\

{\bf(4)} We repeat these processes ``t'' times (trials) and select
the lowest energy result as the best solution.\\

\section{Multi-scale networks}\label{sec:multiresolution}

After determining for the adjacency matrix in \secref{sec:uniform}
and \secref{sec:nonuniform}, we now turn to the-so called
``resolution parameter'' ($\gamma$) in
\eqnref{eq:ourpotts}/\eqnref{eq:newpotts}. In \cite{peter1}, we
introduced the multiresolution algorithm to select the best
resolution. Our multi-scale community detection was inspired by the
use of overlap between replicas in spin-glass physics. In the
current context, we employ information-theory measures,  to examine
contending partitions for each system scale.  Decreasing $\gamma$,
the minima of Eqs. (\ref{eq:ourpotts}, \ref{eq:newpotts}) lead to
solutions progressively lower intra-community edge densities,
effectively ``zooming out'' toward larger structures. We determine
all natural graph scales by identifying the values of $\gamma$ for
which the earlier mentioned ``replicas''  exhibit extrema in the
average of information theory overlaps such as the normalized mutual
information ($I_N$) and the variation of information ($V$) when
expressed as functions of $\gamma$, $\ell$. The extrema and plateau
of the average information theory overlaps as a function of
$\gamma$, $\ell$ over all replica pairs indicate the natural network
scales \cite{peter1}. The \emph{replicas} can be chosen to be
identical copies of the same system for the detection of static
structures, e.g., the image segmentation.

We will briefly introduce the information theory measures in the
following section.

\subsection{Information theory measures}\label{sec:vinmi}

The normalized mutual information $I_N$ and the variation of
information $V$ are the accuracy parameters which are employed to
calculate the similarity (or overlap) between replicas.

We begin with a list of definitions of the information theory overlaps as they
pertain to community detection.
\bigskip

$\bullet$ {\em Shannon Entropy:}  If there are $q$
communities in a partition $A$, then the Shannon entropy is

\begin{eqnarray}
H_A=-\sum_{a=1}^q\frac{n_a}{N}\log_2\frac{n_a}{N}, \label{eq:HA}
\end{eqnarray}

where $\frac{n_a}{N}$ is the probability for a randomly selected
node to be in a community $a$, $n_a$ is the number of nodes in
community $a$ and $N$ is the total number of nodes.

\bigskip

$\bullet$ {\em Mutual Information:}

The mutual
information $I(A,B)$ between partitions found by two replicas $A$
and $B$ is

\begin{eqnarray}
I(A,B)=\sum_{a=1}^{q_A}\sum_{b=1}^{q_B}\frac{n_{ab}}{N}\log_2\frac{n_{ab}N}{n_an_b},
\label{eq:I}
\end{eqnarray}

where $n_{ab}$ is the number of nodes of community $a$ of partition
$A$ that are shared with community $b$ of partition $B$, $q_A$ (or $q_B$)
is the number of communities in partition $A$ (or $B$) and $n_a$ (or $n_b$) is
defined the same as before, i.e., the number of nodes in community
$a$ (or community $b$).

\bigskip

$\bullet$ {\em Variation of information:}

The variation of information  $(0\leq V(A,B)\leq\log_2N)$ between two partitions $A$
and $B$ is given by

\begin{eqnarray}
V(A,B)=H_A+H_B-2I(A,B).
\label{eq:V}
\end{eqnarray}

\bigskip

$\bullet$ {\em Normalized Mutual Information:}

The normalized mutual information $0\leq I_N(A,B)\leq1$ is

\begin{eqnarray}
I_N(A,B)=\frac{2I(A,B)}{H_A+H_B}. \label{eq:IN}
\end{eqnarray}

\bigskip

Now, here is a key idea employed in \cite{peter1} which will be
of great use in our image segmentation analysis:
{\em when taken over an entire ensemble of replicas}, the average $I_N$ or $V$
indicates how strongly a given structure dominates the energy
landscape. High values of $I_N$ (or low values of $V$) corresponds
to more dominate and thus more significant structure. From a local
point of view, at resolutions where the system has well-defined
structure, a set of independent replicas should be highly correlated
because the individual nodes have strongly preferred community
memberships. Conversely, for resolutions ``in-between'' two strongly
defined configurations, one might expect that independent replicas
will be less correlated due to ``mixing'' between competing
divisions of the graph.

\subsection{The application of the multiresolution algorithm for a hierarchal network example}
\label{mra_ex}

We will shortly illustrate how the multiresolution algorithm \cite{peter1} works in practice by presenting
an example of the multiresolution algorithm as it is applied to a hierarchal test
system of $N=1024$ nodes.

To begin the multiresolution algorithm, we need to specify the
\emph{number of replicas} $r$ at each test resolution, the
\emph{number of trials per replica} $t$, and the starting and ending
resolution $[\gamma_0,\gamma_f]$. Usually, the number of replicas is
$8\leq r\leq 12$, the number of trials is $2\leq t\leq 20$. As
detailed in Section \ref{def}, we select the lowest energy solution
among the $t$ trials for each replica. The initial states within
each of the replicas and trials are generated by reordering the node
labels in the ``symmetric'' initialized state of one node per
community. These permutations $P$ simply reorder the node numbers
$(1,2,3, ..., i, ..., N) \to (P1, P2, ..., PN)$ (with $Pi$ the image
of $i$ under a permutation) and thus lead to a different initial
state.
\bigskip

{\bf (1)} The algorithm starts from the initialization of the system
described in item (1) of Section \ref{sec:algorithm}.
\bigskip

{\bf(2)}  We then minimize Eq. (\ref{eq:ourpotts})
independently for all replicas at a resolution
$\gamma = \gamma_{i} \in [\gamma_{0}, \gamma_{f}]$  as described
in Section \ref{sec:algorithm}. Initially $i=0$ (i.e., $\gamma = \gamma_{0}$).
\bigskip

{\bf(3)} The algorithm then calculates the average inter-replica
information measures like $I_N$ and $V$ at that value of $\gamma$.
\bigskip

{\bf (4)} The algorithm then proceeds to the next
resolution point $\gamma_{i+1} \in(\gamma_0,\gamma_f]$
(with $\gamma_{i+1} > \gamma_{i}$).
\bigskip

{\bf (5)} We then return to step number {\bf(3)}.
\bigskip

{\bf (6)} After examining the case of $\gamma = \gamma_{f}$, the algorithm outputs the inter-replica information theory
overlaps for entire the range of the resolutions studied (i.e., $\gamma$ on the interval $[\gamma_{0}, \gamma_{f}]$).
\bigskip

{\bf (7)} We examine those values of $\gamma$ corresponding to extrema in the average inter-replica information theory overlaps. Physically, for these values the resulting image segmentation is locally insensitive to the change of scale (i.e., the change in $\gamma$) and generally highlights prominent features of the image.

With $A$ and $B$ denoting graph partitions in two different replicas
and $Q(A,B)$ their overlap, these average inter-replica overlaps for a general quantity $Q$ \cite{peter1} are
explicitly
\begin{eqnarray}
\langle Q \rangle = \frac{1}{r(r-1)} \sum_{A \neq B} Q(A,B).
\end{eqnarray}
Similarly, for a single replica quantity (such as the Shannon entropy $H$ for partitions $A$ in different replicas), the average is, trivially, $\langle Q \rangle = \sum_{A} H(A)/r$.
(Averages for higher order inter-replica cumulants may be similarly written down
with a replica symmetric probability distribution function \cite{peter1}.)
\bigskip

\myfig{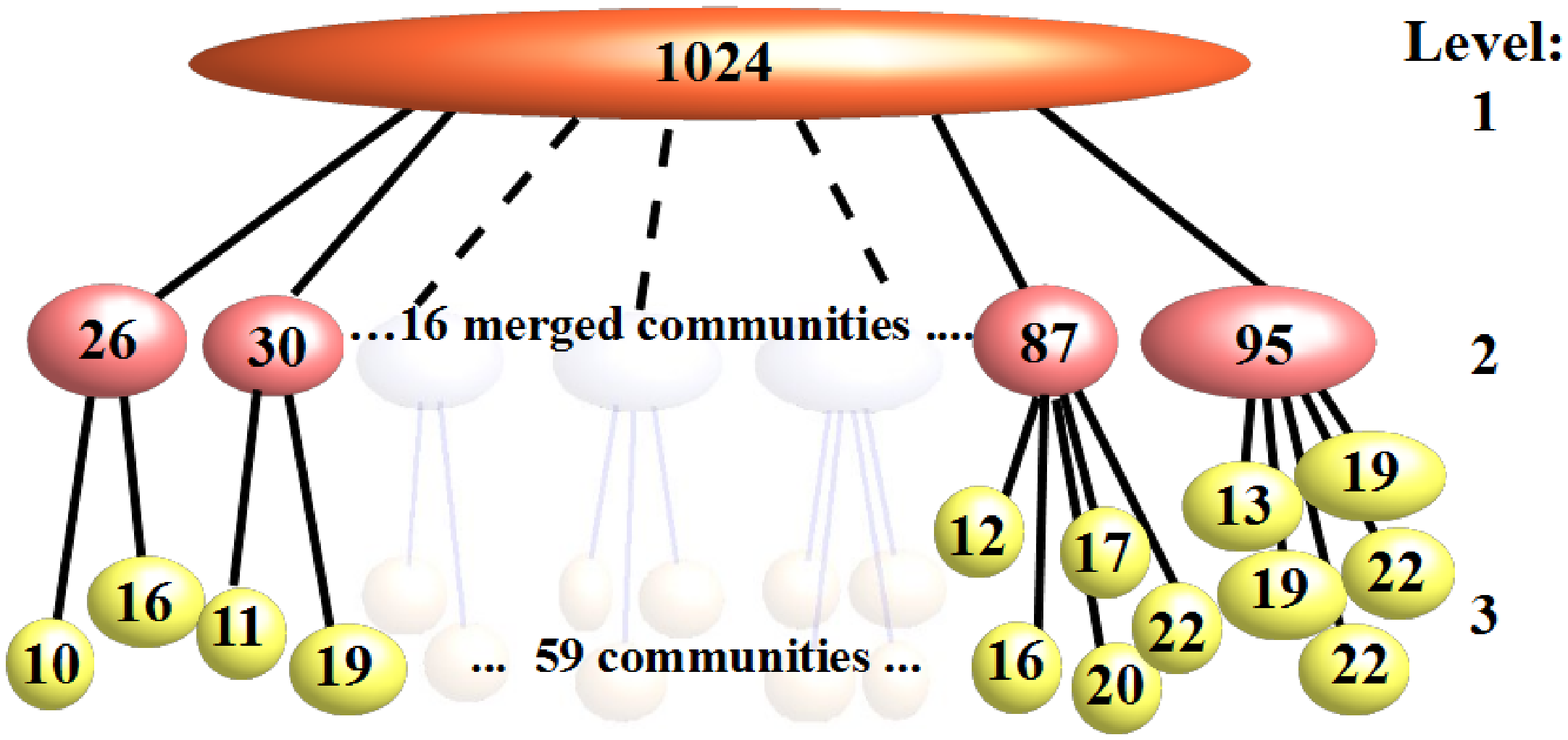}{Heterogeneous hierarchical system corresponding to
the plots in \figref{fig:diagramnmivi}. In this figure, the $1024$
node system is divided into a three-level hierarchy. Level $3$ has
$59$ communities with sizes from $10$ to $24$ nodes. Level $2$ has
$16$ communities with sizes from $26$ to $95$ nodes. Level $1$ is
the completely merged system of $1024$ nodes. The average edge
density is $p=0.054$. This system has $28185$ edges.
}{fig:diagram}{0.9\linewidth}{b}

\begin{figure}[]
\begin{center}
\subfigure[Plot of information measures $I_N$, $I$ and the community
number $q$ vs the Potts model weight $\gamma$ in
\eqnref{eq:ourpotts} for the three-level heterogeneous hierarchy
depicted in \figref{fig:diagram}.]{\includegraphics[width=
3in]{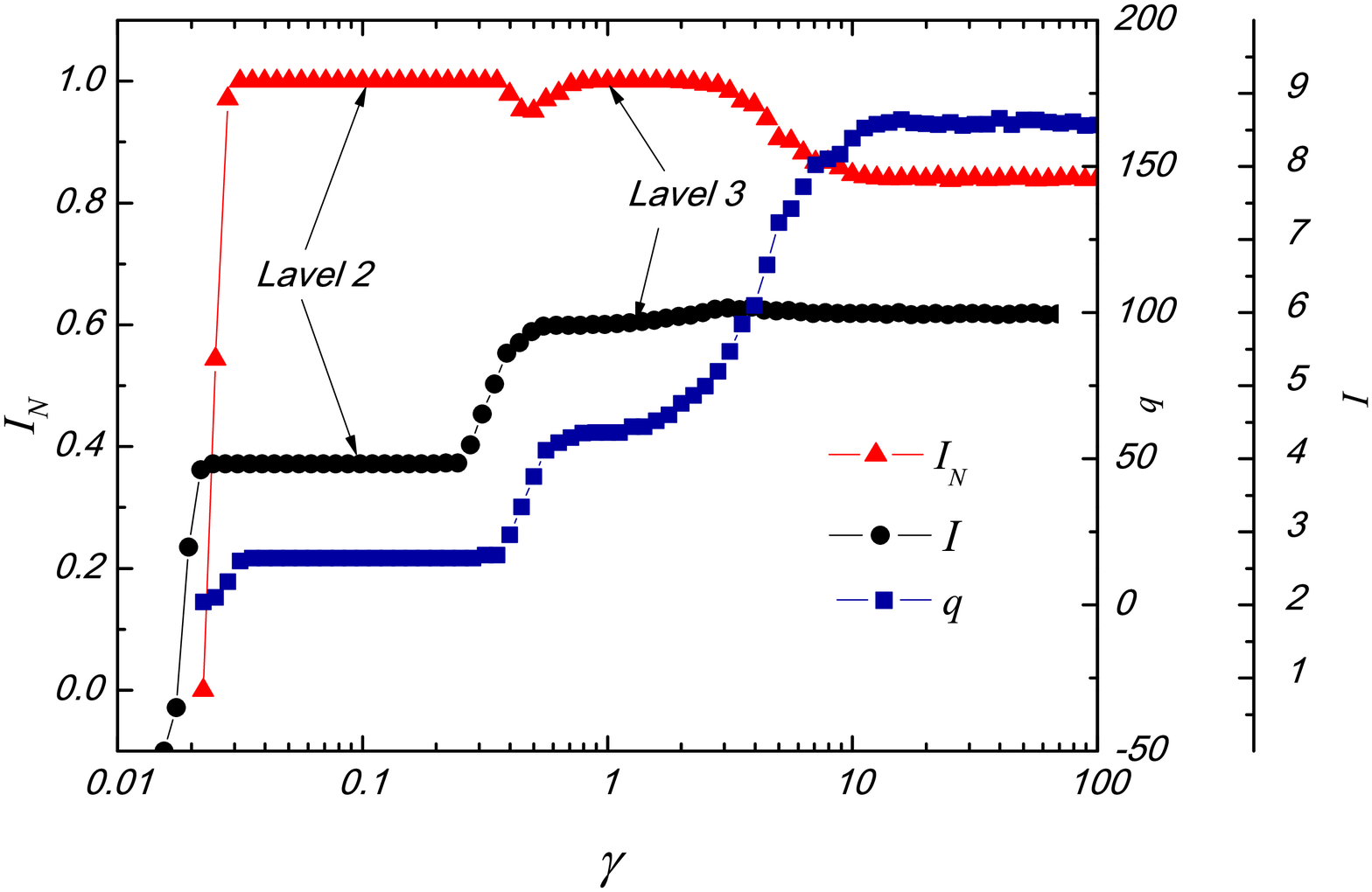}} \subfigure[Plot of information measures $V$,
$H$ and the community number $q$ vs Potts model weight $\gamma$ in
\eqnref{eq:ourpotts} for the three-level heterogeneous hierarchy
depicted in \figref{fig:diagram}. .]{\includegraphics[width=
3in]{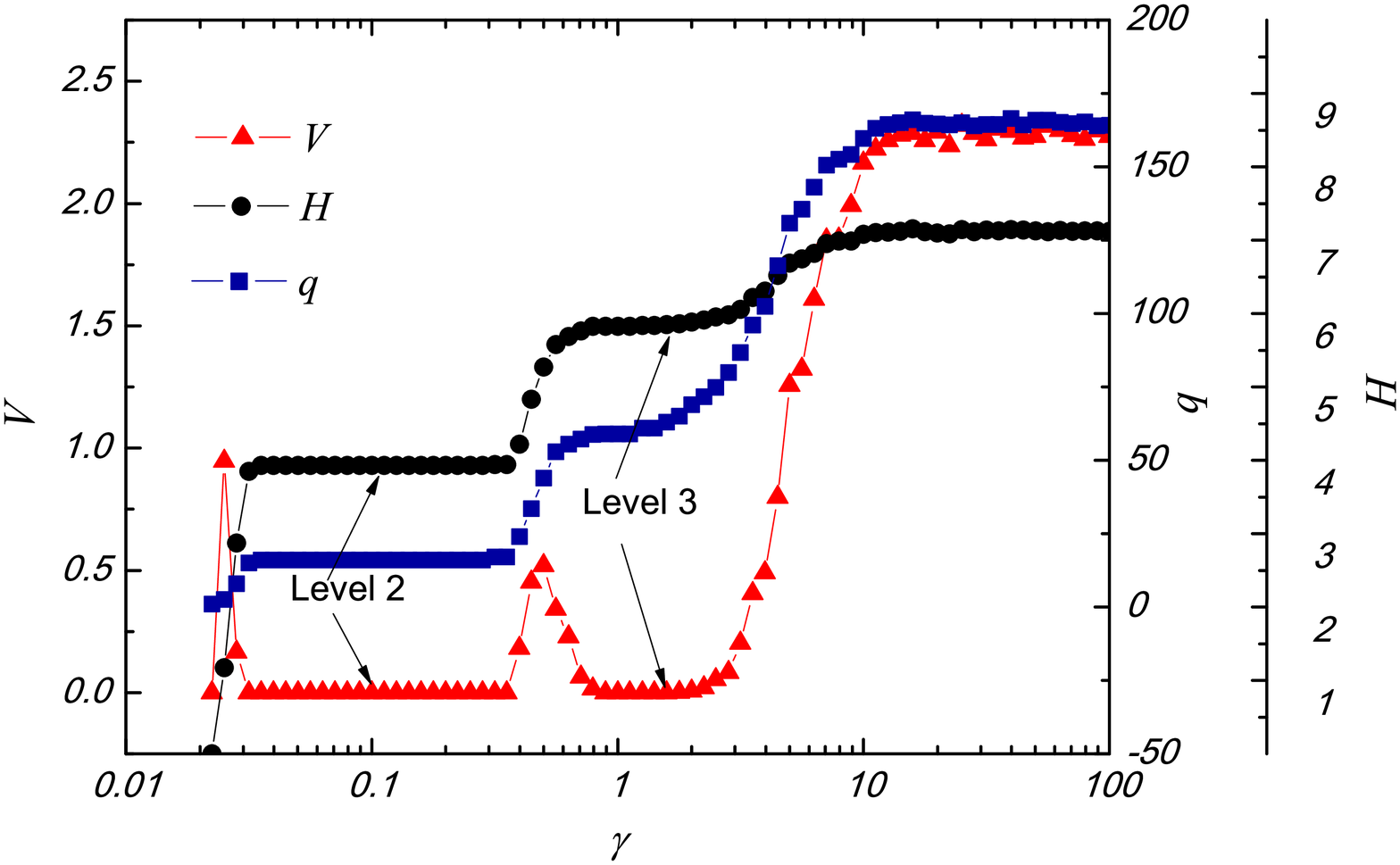}} \caption{Plot of information measures $I_N$,
$V$, $H$ and $I$ vs the Potts model weight $\gamma$ in
\figref{fig:diagram}. In panel(a), the peak (plateau) $I_N$ denoted
by the arrows correspond to levels 2 and 3 of the hierarchy depicted
in \figref{fig:diagram}. Similarly in panel (b), the minimal $V$
values, indicated by arrows, accurately correspond to levels 2 and 3
of the hierarchy. The number of communities $q$ is $16$ and $59$ in
disparate plateau regions (denoted by the arrows) in both panels.
These communities assignments (and, obviously, also their numbers)
are exactly the same as those of the communities in levels 2 and 3
of the original hierarchical graph of \figref{fig:diagram}. In
panels (a) and (b), both the mutual information $I$ and the Shannon
entropy $H$ display a plateau behavior corresponding to the correct
solutions.}\label{fig:diagramnmivi}
\end{center}
\end{figure}

\figref{fig:diagramnmivi} shows the result of multiresolution
algorithm applied to the three-level hierarchy system in
\figref{fig:diagram}. The system investigated is that of
a standard simple graph with unweighted links
(i.e., in Eq.(\ref{eq:ourpotts}), $A_{ij}= 1$ if nodes $i$ and $j$ share an edge
and is zero otherwise).  In \figref{fig:diagram}, ``level $3$'' communities exhibit a
density of links $p_3=0.9$ (i.e., a fraction $p_{3}$ of the intra-community node
pairs are connected by a link ($A_{ij} =1$)). The individual communities in
level $3$ have sizes that
range between $5$ to $24$ nodes. The less dense communities in level $2$
harbor a density of links $p_2=0.3$;  the nodes in this case, are divided into five groups with
sizes that vary from $26$ to $95$. Highest up in the hierarchy is the trivial level $1$ ``partition''- that of
a completely merged system of $1024$ nodes. Thus, as a function of $\gamma$, this easily solvable
system exhibits ``transitions'' between different stable solutions corresponding to different
regions (or basins) of $\gamma$. In Sections(\ref{sec:complexity}, \ref{sec:phasediagram}),
we will further discuss additional transitions between easy solvable regions and
regions of parameter space which are very ``hard'' or impossible (``unsolvable'').

 \figref{fig:diagramnmivi}(a) depicts the averages of $I_N$ (on the left
axis) and $I$ (right axis) over all replica pairs. (We further provide
in this figure the number of communities $q$.) A {\em ``cascade'' composed
of three plateaus} is evident in these information theory measures.
Similarly, \figref{fig:diagramnmivi}(b) shows the $V$ in left axis and $H$ in
right axis average over all replica pairs. The extrema denoted by
the arrows in both panel (a) and (b) are the correctly identified
levels $2$ and $3$ respectively of the hierarchy depicted in
\figref{fig:diagram}. The two plateaus with the peak values in panel(a)
correspond to a normalized mutual information of size $I_N=1$
(the highest theoretically possible) and similarly the corresponding minima
 in panel(b) have a variation of information $V=0$ (the smallest value possible)
for the same range of $\gamma$ values. These extreme values
of $I_{N}$ and $V$ indicate perfect correlations
among the replicas for both levels of the hierarchy. The
``plateaus'' in $H$, $I$ and $q$ are also important indicators of
system structure. These plateau (and more general extrema elsewhere)
illustrate when the system is insensitive to parameter changes and
approaches stable solutions. In Section \ref{sec:complexity} (and
in Eq.(\ref{QZ}) in particular), we will discuss this more generally
in the context of the phase diagram of the community detection
problem.

\section{Replica correlations as weights in a graph} \label{corr_g}

Within the multiresolution method, significant structures are
identified by strongly-correlated replicas (multiple copies of the
studied system). Thus, if a node pair is always in the same
community in all replicas, the two nodes must have strong preference
to be connected or have a large edge weight. Similarly, if a node
pair is not always in the same community in all replicas, they must
have preference not to be connected or have a small edge weight. We
re-assign edge weights based on the correlations between replicas.

Specifically, we first generate $r$ replicas by permuting the
``symmetric'' initialized state of one node per cluster of the
studied system, then apply our community detection algorithm to each
replica and record the community membership for each node. We then
calculate the probability of each node pair based on the statistics
of replicas. The probability is defined as follows

\begin{eqnarray}
p_{ij}=\frac{1}{r}\sum_{\alpha=1}^{r}\omega_{ij}^{\alpha},
\label{eqn:pij}
\end{eqnarray}
where
\begin{eqnarray}
\omega_{ij}^{\alpha}=\delta_{\sigma_i^\alpha,\sigma_j^\alpha}
+(1-\delta_{\sigma_i^\alpha,\sigma_j^\alpha})\frac{\exp(-|{\bf
r}_i^\alpha-{\bf r}_j^\alpha|/\ell)}{N_{A}^\alpha N_{B}^\alpha}.
\label{eqn:wij}
\end{eqnarray}
In Eq. (\ref{eqn:wij}), when node $i$ and $j$ belongs to the same
community in replica $\alpha$, i.e.,
$\delta_{\sigma_i^\alpha,\sigma_j^\alpha}=1$,
$\omega_{ij}^\alpha=1$. When node $i$ and $j$ are not in the same
community in replica $\alpha$, i.e., $\delta_{\sigma_i^\alpha,
\sigma_j^\alpha}=0$ (we use $A$ and $B$ to represent these two
different communities, where $i\in A$ and $j\in B$. $N_A^\alpha$ and
$N_B^\alpha$ denote the size of cluster $A$ and $B$ in replica
$\alpha$), $\omega_{ij}^\alpha=\frac{\exp(-|{\bf r}_i^\alpha-{\bf
r}_j^\alpha|/\ell)}{N_{A}^\alpha N_{B}^\alpha}$. As throughout,
$|{\bf r}_i^\alpha-{\bf r}_j^\alpha|$ is the distance between node
$i$ and $j$ in replica $\alpha$. In \eqnref{eqn:pij}, we sum the
probability in each replica to define the edge weight. The assigned
weights given by Eq. (\ref{eqn:pij}) are based on a frequency type
inference. Although we will not report on it in this work, it is possible
to perform Bayesian analysis with weights (``priors'') that are
derived from a variant of Eq. (\ref{eqn:wij}); this enables an
inference of the correlations from the sequence of results
concerning the correlations between nodes $i$ and $j$ in a sequence
of different replicas.

In unweighted graphs, we connect nodes if the edge weight between
the node pair is larger than some threshold value $\bar{p}$ in
\eqnref{eq:ourpotts}, i.e.,
\begin{eqnarray}
A_{ij}=\Theta(p_{ij} - \bar{p}).
\end{eqnarray}

In weighted graph, the analog of \eqnref{eq:newpotts}
is the Hamiltonian given by

\begin{eqnarray}
{\cal{H}}=  \frac{1}{2}\sum_{a=1}^q
(\bar{p}-p_{ij})\Big[\Theta(p_{ij}-\bar{p})+\gamma\Theta(\bar{p}-
p_{ij}))\Big] \delta(\sigma_{i}, \sigma_{j}).
 \label{eq:newpotts2}
\end{eqnarray}

That is, when there is a high probability $p_{ij}$, relative to a
background threshold $\bar{p}$, that nodes $i$ and $j$ are linked,
we assign a positive edge weight to the link $(ij)$ of size $(p_{ij}
- \bar{p})$. Similarly, if the probability of a link $(ij)$ is low,
we assign a negative weight of size $\gamma (p_{ij}-\bar{p})$.

Armed with Eq. (\ref{eq:newpotts2}), we then minimize in an
identical fashion to the minimization of Eq. (\ref{eq:newpotts})
that we discussed earlier.
Specifically, we follow the 4 steps outlined in Section \ref{sec:algorithm}
for non multi-scale images and the 7 steps of Section \ref{mra_ex} in
the analysis of general multi-scale systems.

\section{Summary of parameters} \label{par}

We now very briefly collect and list anew the parameters that define our Hamiltonians and
appear in our methods.

$\bullet$ The resolution parameter $\gamma$ in Eqs.(\ref{eq:ourpotts}, \ref{eq:newpotts},
\ref{eq:newpotts2}). This parameter sets the graph scale over which
we search for communities. This parameter is held fixed (typically with a value of
$\gamma = {\cal{O}}(1)$) in the community detection method and varies
within our multi-scale analysis. We determine the optimal value
of $\gamma$ by determining the local extrema of the average information
theory overlaps between replicas.

$\bullet$ The spatial scale $\ell$ in Eq.(\ref{eqn:vmn}). Similar to
the more general graph scale set by $\gamma$, we may determine
optimal $\ell$ by examining extrema in the average inter-replica
information theory correlations.  In practice, in all but the
hardest cases (i.e., the case of the dalmatian dog in
Fig.(\ref{fig:hardzebradog})), we ignored this scale and fixed
$\ell$ to be infinite. Fixing $\ell =1$ led to good results in the
analysis of the dalmatian dog.

$\bullet$ The spatial cutoff scale $\Lambda$ for defining link weights- see the brief discussions
after Eqs.(\ref{drgb}, \ref{eq:coupling}). Whenever the spatial
distance between two sites or blocks exceeded a threshold distance $\Lambda$
we set the link weight to be zero.  We did not tune this parameter in any
of the calculations. It was fixed to the value of $\Lambda = 30$.

$\bullet$  The scale of the block size $L_{x,y}$ introduced in
Section \ref{sec:blocks}. This parameter is far smaller than the
image size $N_x\times N_y$, yet large enough to cover the image
features. We usually set $L_x\times L_y$ to be $9\times 9$ for an
image size $N_x\times N_y$ of around $200 \times 200$.

$\bullet$ The background intensities $\bar{V}$ in
Eq.(\ref{eq:newpotts}) and $\bar{p}$ in Eq.(\ref{eq:newpotts2}).
Similar to the graph scale set by $\gamma$ and $\ell$, we may
determine the optimal $\bar{V}$ and $\bar{p}$ by observing the local
extrema of the average information correlations between replicas.

As we will elaborate on briefly, all optimal parameters can be found by determining the local extrema
of the information theory correlations that signify no change in
structure over variations of scale. In reality, we may fix some
parameters and vary others--usually, $\Lambda$ fixed as $30$,
$L_x\times L_y$ in the range of $7\times 7$ to $11\times 11$, and
$\gamma$, $\bar{V}$ and $\bar{p}$ been changed.

As an aside, in this brief paragraph, we briefly note for readers
inclined towards spin glass physics and optimization theory that, in
principle, in the large $N$ limit (images with a large number of
pixels) the effective optimal values for the likes of the parameters
listed above may be derived by solving the so-called ``cavity''
equations \cite{mezard2} that capture the maximal inference possible
(in their application without the aid of replicas that we introduced
here) \cite{infer,lenka}. In the current context, in applying these
equations anew to image segmentation, we arrive at the maximal
inference possible of objects in an image. While these equations are
tractable for simple cases, solving these equations is relatively
forbidding for general cases. In practice, we thus efficiently
directly examine our Potts model Hamiltonians of Eqs.
(\ref{eq:ourpotts}, \ref{eq:newpotts}, \ref{eq:newpotts2}) and, when
needed, directly infer optimal values of the parameters by examining
inter-replica correlations as described in the earlier sections.
This will be expanded on in the next section (specifically, in
Eq.(\ref{QZ})). [Detailed applications of this method are provided
in Section \ref{sec:phasediagram}.]

\section{Computational Complexity, the Phase Diagram,
and the determination of optimal parameters} \label{sec:complexity}

Our community detection algorithm is very rapid. For a system with
$L$ links, the typical convergence time scales as
${\cal{O}}(L^{1.3})$ \cite{peter2}. In an image with $N$ pixels,
with all of the constants $\Lambda, L_{x,y} = {\cal{O}}(1)$ (i.e.,
not scaling with the system size), the number of links $L \sim N$.

Our general multi-scale community detection algorithm (that with
varying $\gamma$) has a convergence time $\tau \sim L^{1.3} \ln N$
\cite{peter1}.  Thus, generally, for an image of size $N$, the convergence time
$\tau \sim N^{1.3} \ln N$. Rapid convergence occurs in all but
the ``hard phase'' of the community detection problem.

Specifically, we numerically investigated the phase diagram as a
function of noise and temperature (i.e., when different
configurations are weighted with a Boltzmann factor $\exp(-\beta
{\cal{H}})$ with $\beta = 1/T$ at a temperature $T$ for general
graphs with an arbitrary number of clusters in \cite{dandan}.)
Related analytic calculations were done for sparse graphs in
\cite{lenka}. In particular, in these and earlier works
\cite{peter2,peter1} it was found that there is a phase transition
between the detectable and undetectable clusters. The detectable
phase further splinters into an ``easy'' and a ``hard'' phase. These
three phases in the community detection problem constitute analogs
of  three related phases in the ``SAT-hard-unSAT'' in k-SAT problem
\cite{mezard1,mezard2}.  The found phase diagram \cite{dandan}
exhibits universal features. Increasing the temperature can aid the
detection of clusters \cite{dandan}. The universal features of the
phase diagram and the known cascade of transitions that appear on
introducing temperature enable better confidence in the results of
the community detection algorithm.  One of the central results of
Ref.\cite{dandan} is that the ``easy'' solvable phase(s) of the
community detection problem which leads to correct relevant
solutions (i.e., not noisy partitions of a structureless system)
universally appear in a ``flat'' \cite{peter1, dandan} phase(s) [see
also the flat information theoretic curves in Fig.
(\ref{fig:diagramnmivi}) and related discussion in
Section(\ref{mra_ex})] as ascertained by the {\em inter-replica
averages of  all thermodynamic and information theoretic quantities
$\{\langle Q \rangle\}$}. These may correspond to the internal
energy $(Q = {\cal{H}})$, average Shannon entropy ($Q  = H$),
average inter-replica normalized mutual information and variation of
information ($Q  = I_{N}, V$), the complexity $(Q=\Sigma)$
\cite{mezard2} or an associated ``susceptibility'' $(Q=\chi)$
\cite{peter1,dandan} that monitors the onset of large complexity.
[This susceptibility will be defined with the aid in the change in
the average normalized mutual information $I_{N}$ as a function of
the number of trials $t$. It is defined as $\chi(n) = [I_{N}(t=n) -
I_{N}(t=4)]$.] That is, with $z$ denoting a set of generalized
parameters (e.g., artificially added additional noise in networks
$(z=p_{out})$ \cite{dandan}, temperature $(z=T)$ \cite{dandan}, or
resolution parameter $(z=\gamma)$ \cite{peter1}), pertinent
partitions appear for those values of the parameters $z$ for which
\begin{eqnarray}
\frac{\partial \langle Q \rangle}{\partial z} = 0.
\label{QZ}
\end{eqnarray}
As alluded to above, a particular realization of Eq.(\ref{QZ}) appears
in the hierarchal system discussed in Section \ref{mra_ex} wherein $z=\gamma$ and $Q=I_{N},V$.
In that case, Eq.(\ref{QZ}) was satisfied in well defined plateaus.

When present, crisp solutions are furthermore generally
characterized by relatively high values of $I_{N}$, and these
correspond to the ``easy phase'' of the image segmentation problem.
In Sec. \ref{sec:phasediagram}, we will provide explicit analysis of
the phase diagram and optimal parameters as they pertain to several
example images.

All of the results (except the ones in \secref{sec:phasediagram})
presented below in the current manuscript were attained at zero
temperature and may be improved by the incorporation of thermal
annealing as the results of \cite{dandan} illustrate for general
systems.

\section{Results}\label{sec:results}

\subsection{Brain Image}

\subsubsection{Unweighted graphs}

\begin{figure}[t]
\begin{center}
\subfigure[The result of the ``multiresolution'' algorithm applied
to the unweighted brain image shown in (b): $I_N$, $V$ and $q$ in
terms of the resolution $\gamma$.]{\includegraphics[width=
3in]{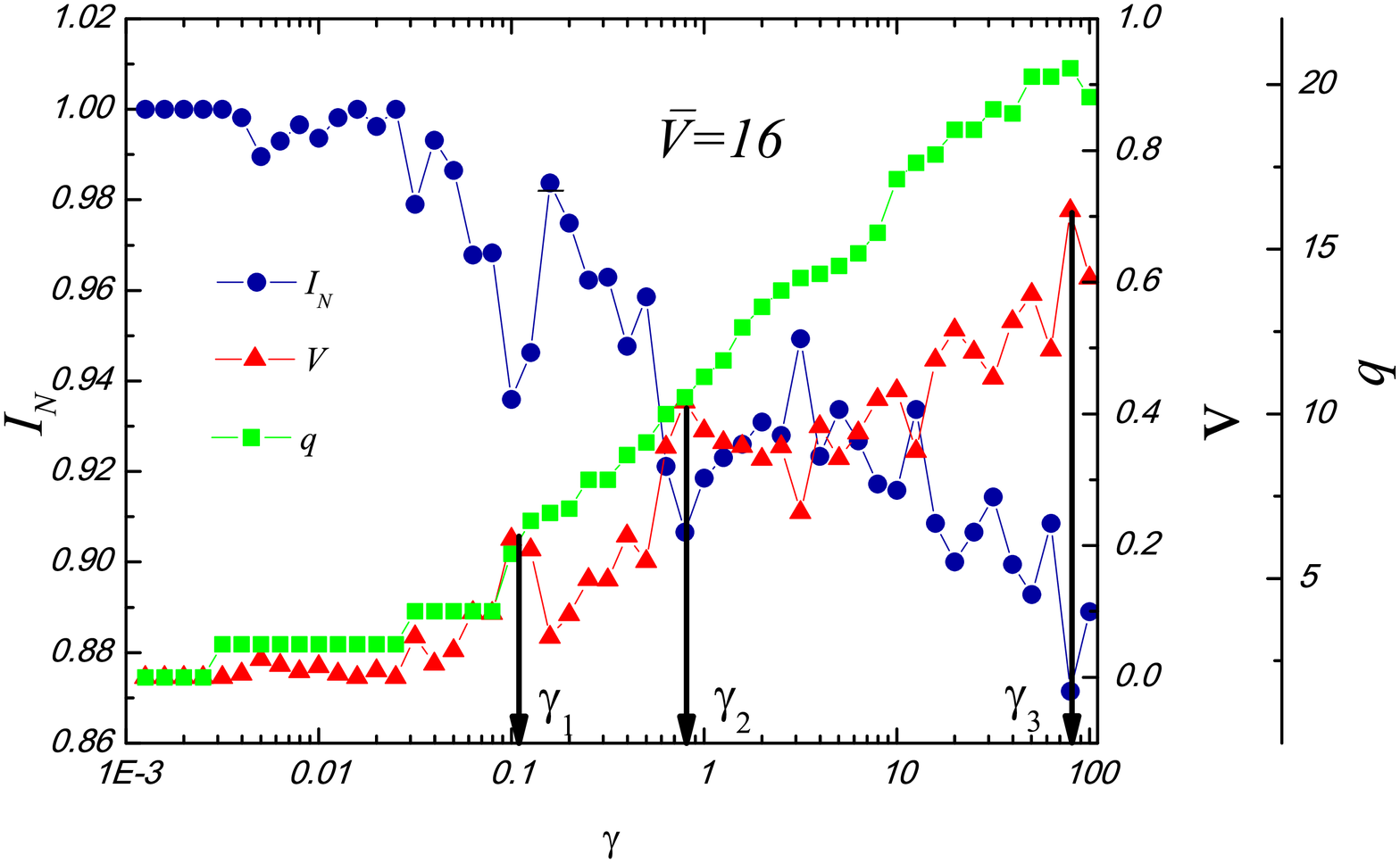}} \subfigure[The unweighted result of the brain
images with different $\gamma$, which correspond to $\gamma_1=0.1$,
$\gamma_2=0.8$, $\gamma_3=79.4$.]{\includegraphics[width=
3in]{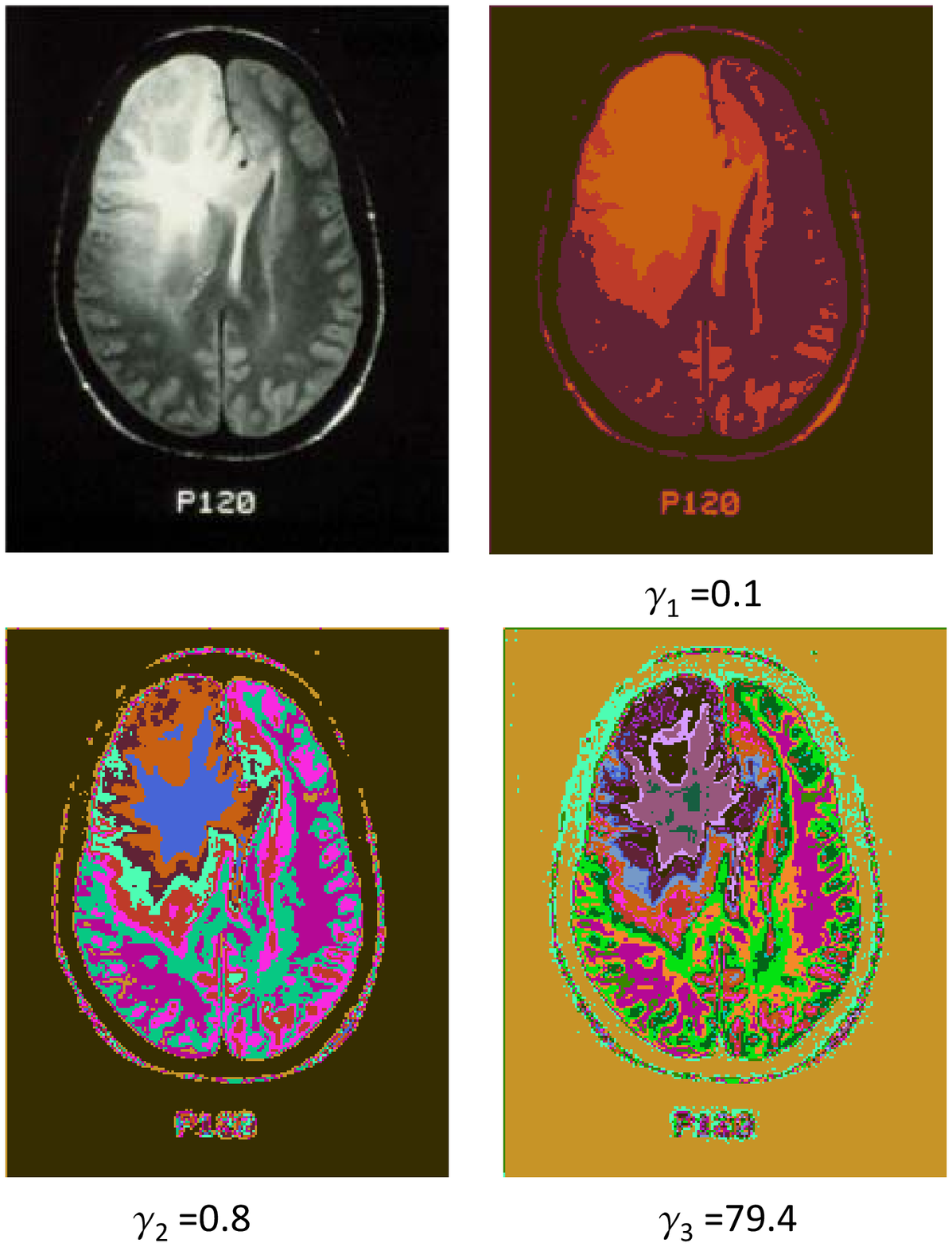}} \caption{[Color Online.] The plot of the
normalized mutual information $I_N$, variation of information $V$
and the number of communities $q$ as a function of $\gamma$ for the
brain image. This image is reproduced with permission from the Iowa
Neuroradiology Library. The axis for $\gamma$ is on a logarithmic
scale. There are three prominent peaks in the $V$ curve. We apply
our community detection algorithm to the grey-scale brain image at
these three values of $\gamma$s. The corresponding results are shown
in panel (b). Note that the results show three-level hierarchy as
$\gamma$ varies.}\label{fig:unweightMRA}
\end{center}
\end{figure}

\begin{figure}[t]
\begin{center}
\subfigure[The result of ``multiresolution'' algorithm for the
weighted brain image shown in (b):  $I_N$, $V$ and $q$ in terms of
the resolution $\gamma$.]{\includegraphics[width=
3in]{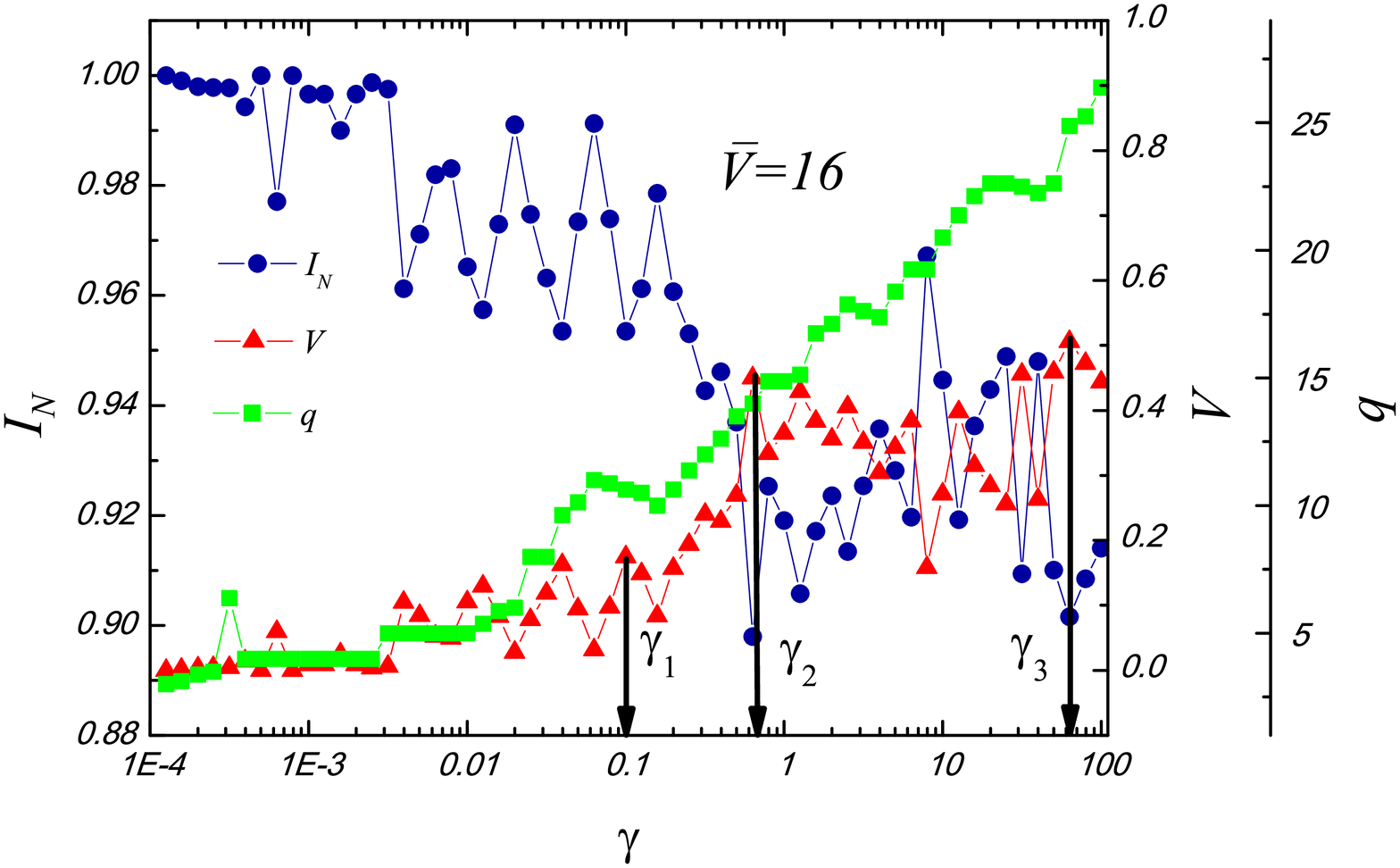}} \subfigure[The weighted result of the brain
images with different $\gamma$, which correspond to $\gamma_1=0.1$,
$\gamma_2=0.64$, $\gamma_3=64$.]{\includegraphics[width=
3in]{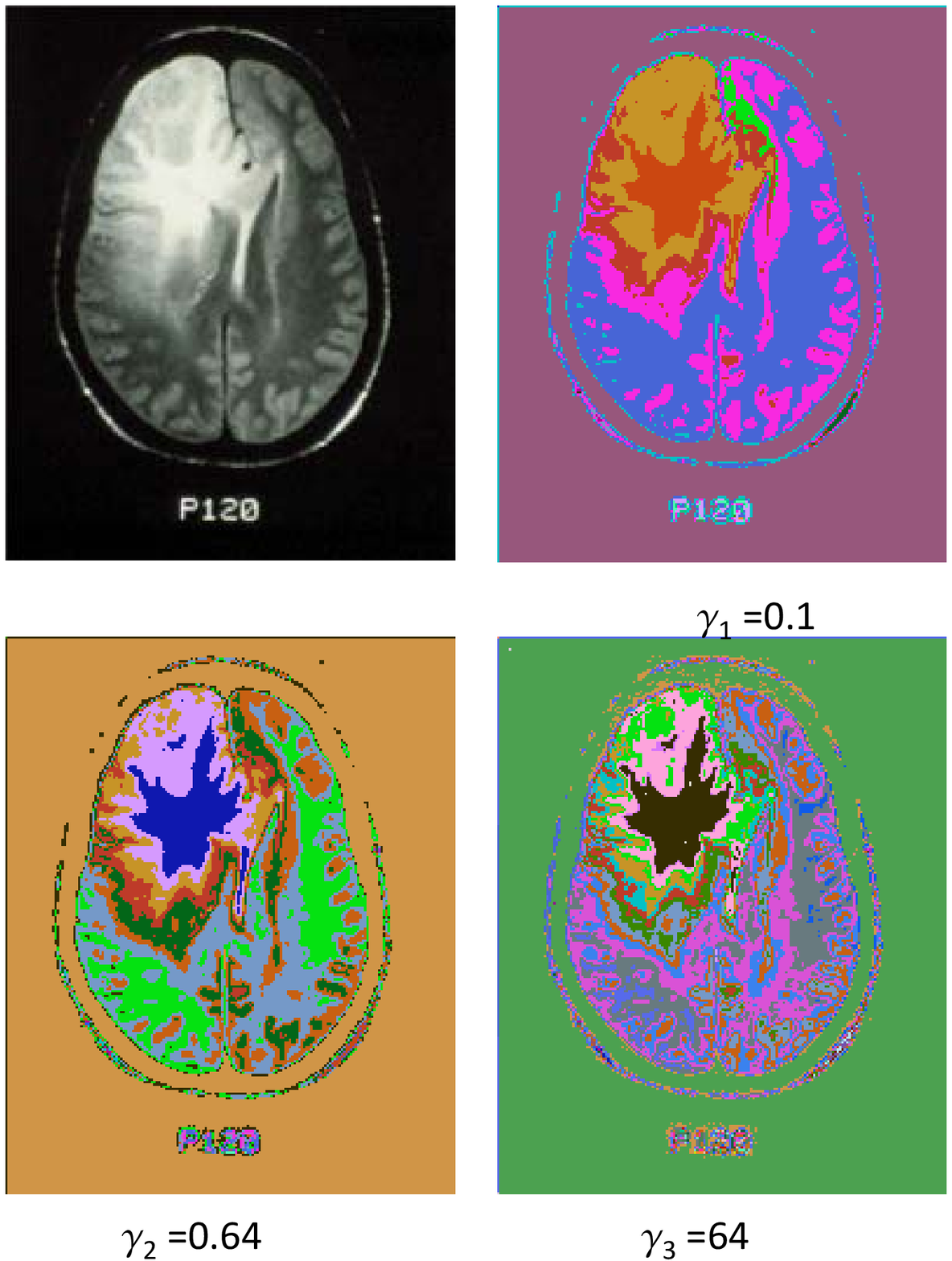}} \caption{[Color Online.] The result
of ``multiresolution'' for the weighted brain image shown in panel
(b). In panel (a), the ``multiresolution'' result here behaves
different from \figref{fig:unweightMRA} but keeps the same trend.
The structure is only stable in the resolution range of
$\gamma<0.01$, compared to the wider range of $\gamma<0.1$ in
\figref{fig:unweightMRA}. This illustrates that the weighted graph
is more sensitive to the change of resolution.
}\label{fig:weightMRA-gamma}
\end{center}
\end{figure}

We start the review of the results of our methods by analyzing an
\textit{unweighted} graph (\eqnref{eq:ourpotts}) for the grey-scale
brain image as shown in \figref{fig:unweightMRA}. We will assign
edges between pixels only if the intensity difference is less than
the threshold $\bar{V}=16$ as denoted in panel (a) of
\figref{fig:unweightMRA}. The algorithm uses \eqnref{eq:ourpotts} to
solve for a range of resolution parameters $ \gamma$ in the interval $[\gamma_0,\gamma_f]$. In the
particular case in \figref{fig:unweightMRA}, $\gamma_0=10^{-3}$ and
$\gamma_f=100$. There are two more input parameters that are needed
in our algorithm: the number of independent replicas $r$ that will
be solved at each tested resolution and the number of trials per
replica $t$. We use $r=10$ and $t=4$ in \figref{fig:unweightMRA}
respectively.

As noted earlier (see Section \ref{def}), for each \emph{replica}, we select the lowest energy
solution among the $t$ \emph{trials}. The $r$ replicas are generated
by reordering the ``symmetric'' initialized state of one node per
community. We then use the information based measures (i.e., $I_N$
or $V$) to determine the multiresolution structure.

The plots of $I_N$, $V$ and $q$ as a function of $\gamma$ in
\figref{fig:unweightMRA} exhibit non-trivial behaviors. Extrema in
$I_N$ and $V$ correspond to jumps in the number of communities $q$.
In the low $\gamma$ region, i.e., $\gamma<0.1$, the number of
communities is stable. However, when $\gamma>0.1$, the number of
communities $q$ sharply increase. This indicates that the structure
changes rapidly as the resolution $\gamma$ varies. There are three
prominent peaks in the $V$ (variation of information) curve. We show
the corresponding images at these resolutions, that is in panel(b)
in \figref{fig:unweightMRA}. These corresponding segmented images
show more and more sophisticated structures. The lower right image
at a resolution of $\gamma=79.4$ shows the information in detail.
Different colors in the image correspond to different clusters.
There are, at least, five contours surrounding the tumor, that
denote the degree by which the tissue was pushed by the tumor. The
lower left image at the resolution $\gamma=0.8$ is less detailed
than the one on the right. Nevertheless, it retains the details
surrounding the tumor. If we further decrease $\gamma$, the upper
right image at the resolution $\gamma=0.1$ will not keep the details
of the tumor boundary, only the rough location of the tumor. Thus,
neither too large nor too small resolutions are appropriate for
tumor detection in this image. The resolution around $\gamma=0.8$ is
the most suitable in this case. This is in accord with our general
found maxim in Section \ref{par} concerning a value of $\gamma  =
1$. We re-iterate that, in general, the optimal value
of $\gamma$ is found by Eq.(\ref{QZ}) (an example
of which is manifest in the information theory plateaus
discussed in Section \ref{mra_ex}).  In Section \ref{sec:phasediagram},
we will discuss, in depth, how the optimal values of $\gamma$ may
be determined in (weighted) example systems.

\subsubsection{Weighted graphs}

\begin{figure}[]
\begin{center}
\subfigure[The result of ``multiresolution'' algorithm for the
weighted brain image shown in (b):  $I_N$, $V$ and $q$ in terms of
the threshold $\bar{V}$.]{\includegraphics[width=
3in]{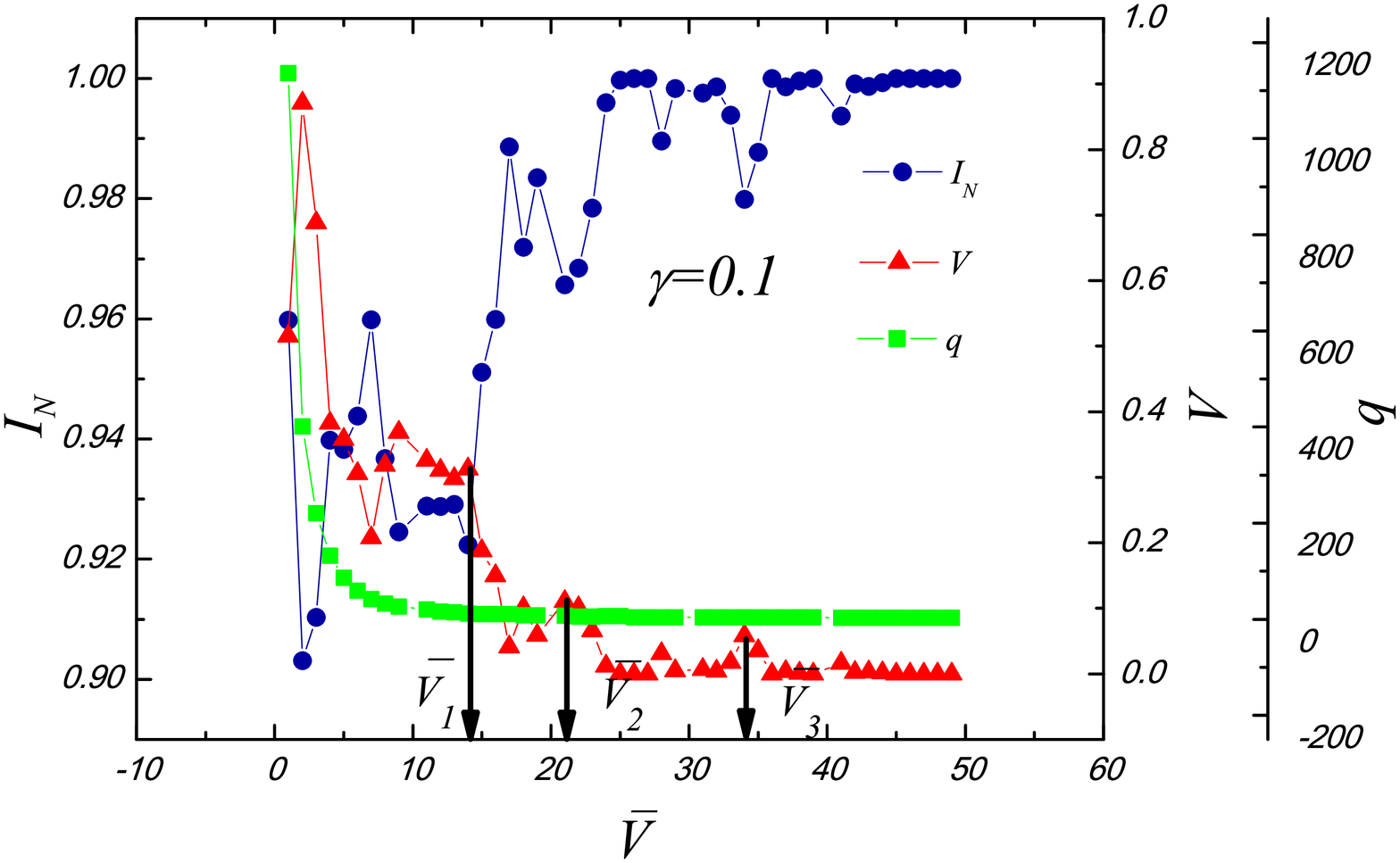}} \subfigure[The weighted result of the images
with different $\bar{V}$, which correspond to $\bar{V}_1=14$,
$\bar{V}_2=21$, $\bar{V}_3=34$.]{\includegraphics[width=
3in]{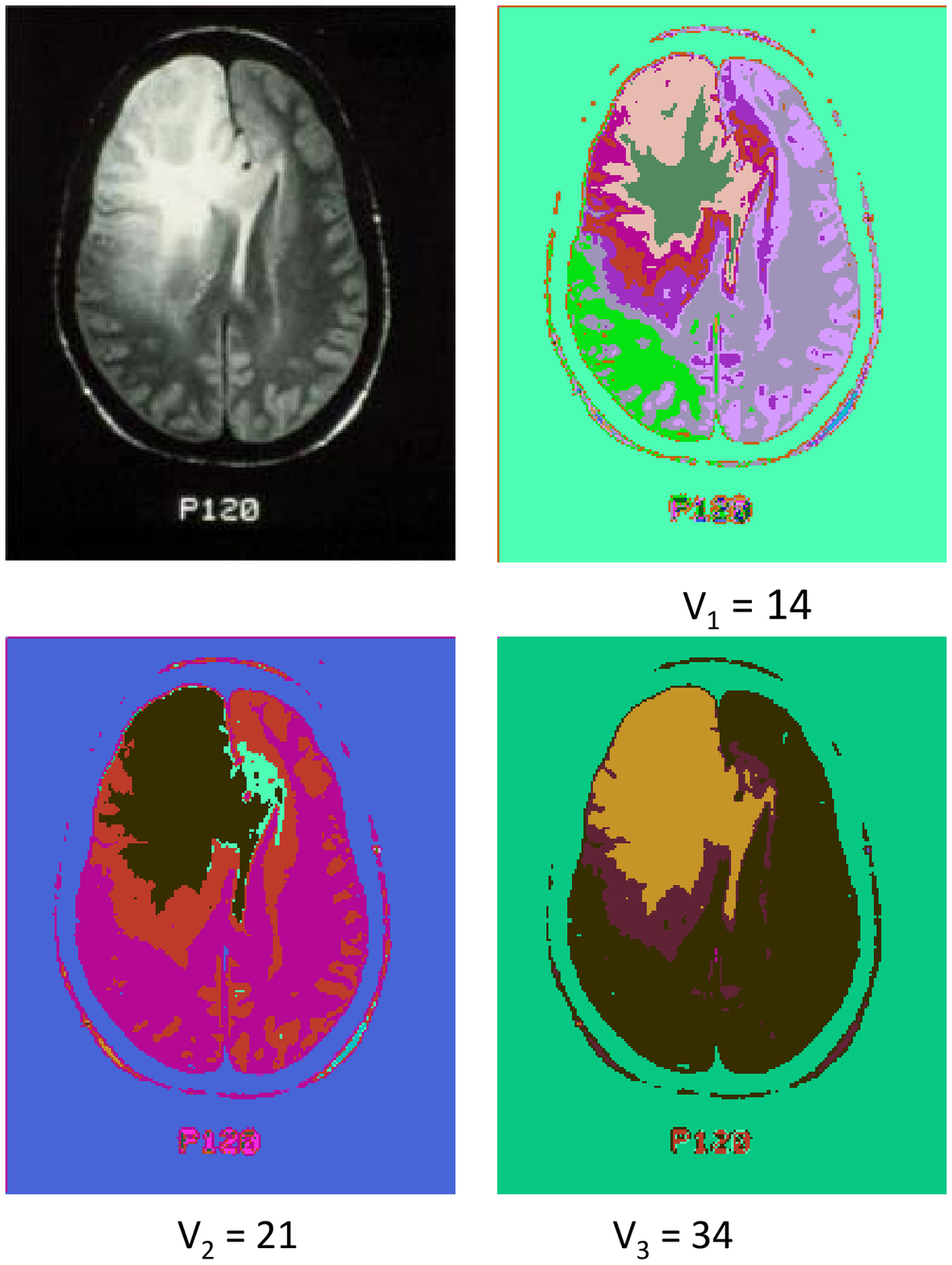}} \caption{[Color Online.] The
``multiresolution'' result also shows the hierarchy structure as the
threshold $\bar{V}$ varies, as in \figref{fig:weightMRA-gamma}.
Higher ``$\bar{V}$'' corresponds to the lower ''$\gamma$'', which
means pixels intend to merge in higher ``$\bar{V}$''. The structure
is stable in the range of $\bar{V}>25$, below which the structure is
sensitive. }\label{fig:weightMRA-th}
\end{center}
\end{figure}

In Figs. (\ref{fig:weightMRA-gamma}, \ref{fig:weightMRA-th}), we
provide the ``multiresolution'' results for the \textit{weighted}
graph (\eqnref{eq:newpotts}) of $\gamma$ and $\bar{V}$ for the brain
image. Both the resolution $\gamma$ and the threshold $\bar{V}$
control the hierarchy structures: the peaks in the normalized mutual
information $I_N$ and variation of information $V$ always correspond
to the jumps in the number of communities $q$. The jumps in $q$
correlate with the changes in hierarchal structures on different scales. We can combine
both parameters to obtain the desirable results in the test images.
See, e.g., the $3D$ plot of $I_N(\bar{V},\gamma)$.

The results of our method with weighted edges are more sensitive to the changes of
parameters (as seen from a comparison of \figref{fig:weightMRA-gamma} with
\figref{fig:unweightMRA}). According to \eqnref{eq:newpotts},  edges $(ij)$
with small (or large) difference $|D_{ij}|$ will decrease (or
increase) the energy by $|\bar{V}-D_{ij}|$ (or
$\gamma |\bar{V}-D_{ij}|$). However, if the
unweighted graphs and the Potts model with discrete weights
(\eqnref{eq:ourpotts}) are applied, the edges with small or large
``color'' difference will decrease or increase the energy by the
amount of $1$ or $\gamma$. Thus, considerable information
(e.g., the ``color'' of each pixel) is omitted when
using an unweighted graph approach.

\subsection{A painting by Salvador Dali}

\begin{figure}[t]
\begin{center}
\subfigure[The variation of information $V$ as a function of
resolution $\gamma$ for the image shown in Panel(b)
.]{\includegraphics[width= 3in]{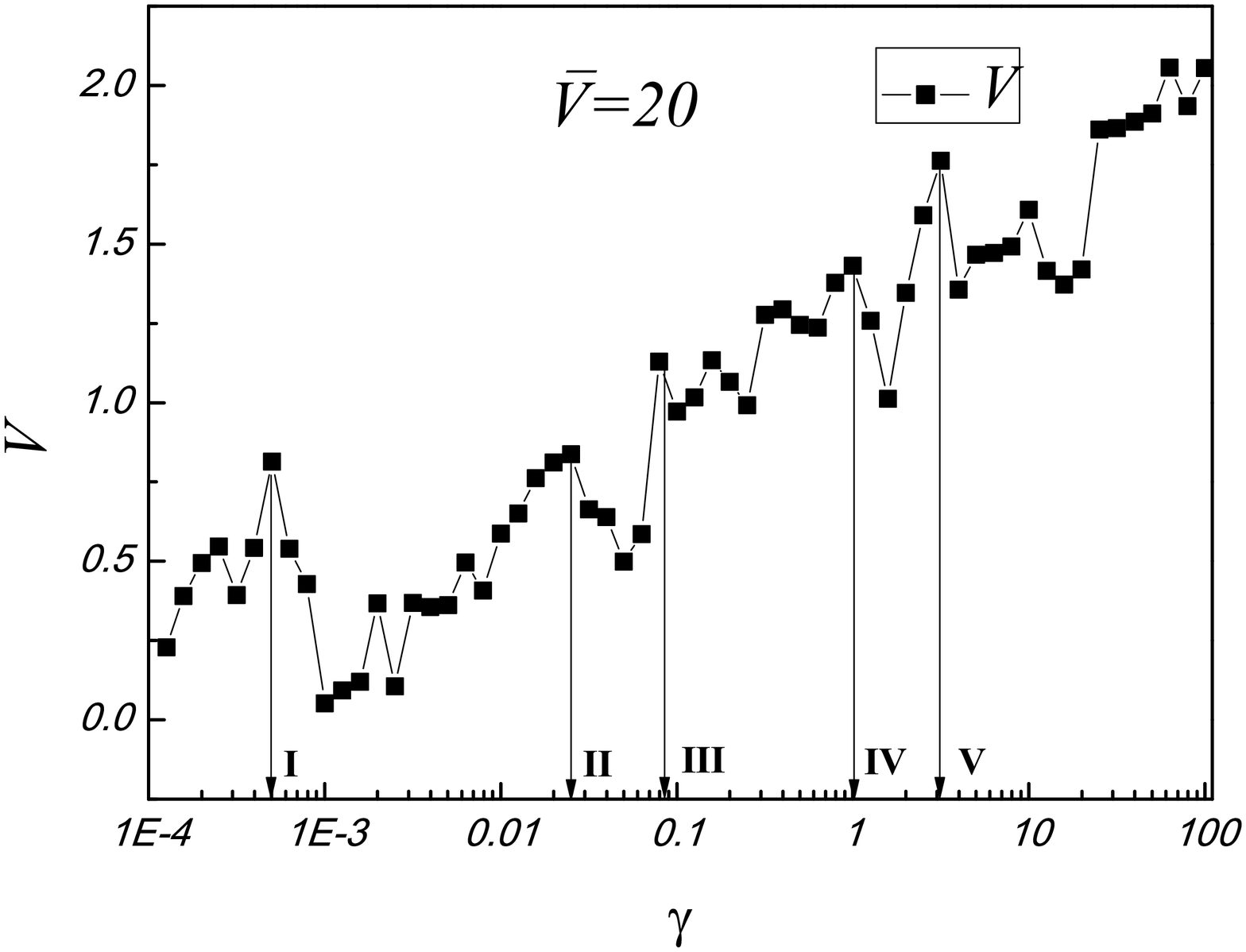}} \subfigure[The
original image and the corresponding segmentation for the specific
resolution marked (I) in panel(a).]{\includegraphics[width=
3.5in]{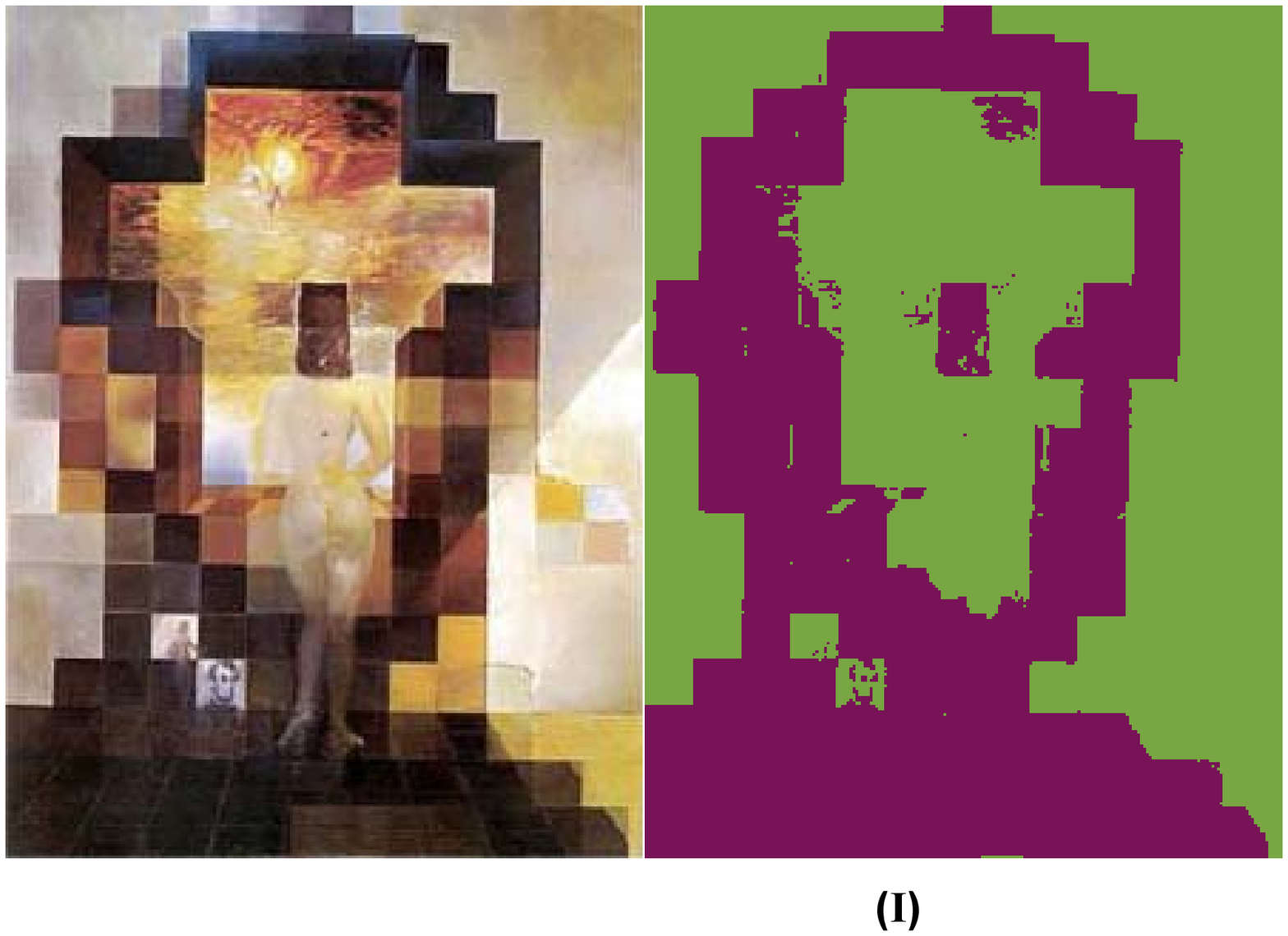}}
\end{center}
\end{figure}
\begin{figure}[t]
\begin{center}
\subfigure[The corresponding images in the specific resolution
marked (II),(III),(IV) and (V) in panel(a).]{\includegraphics[width=
2.3in]{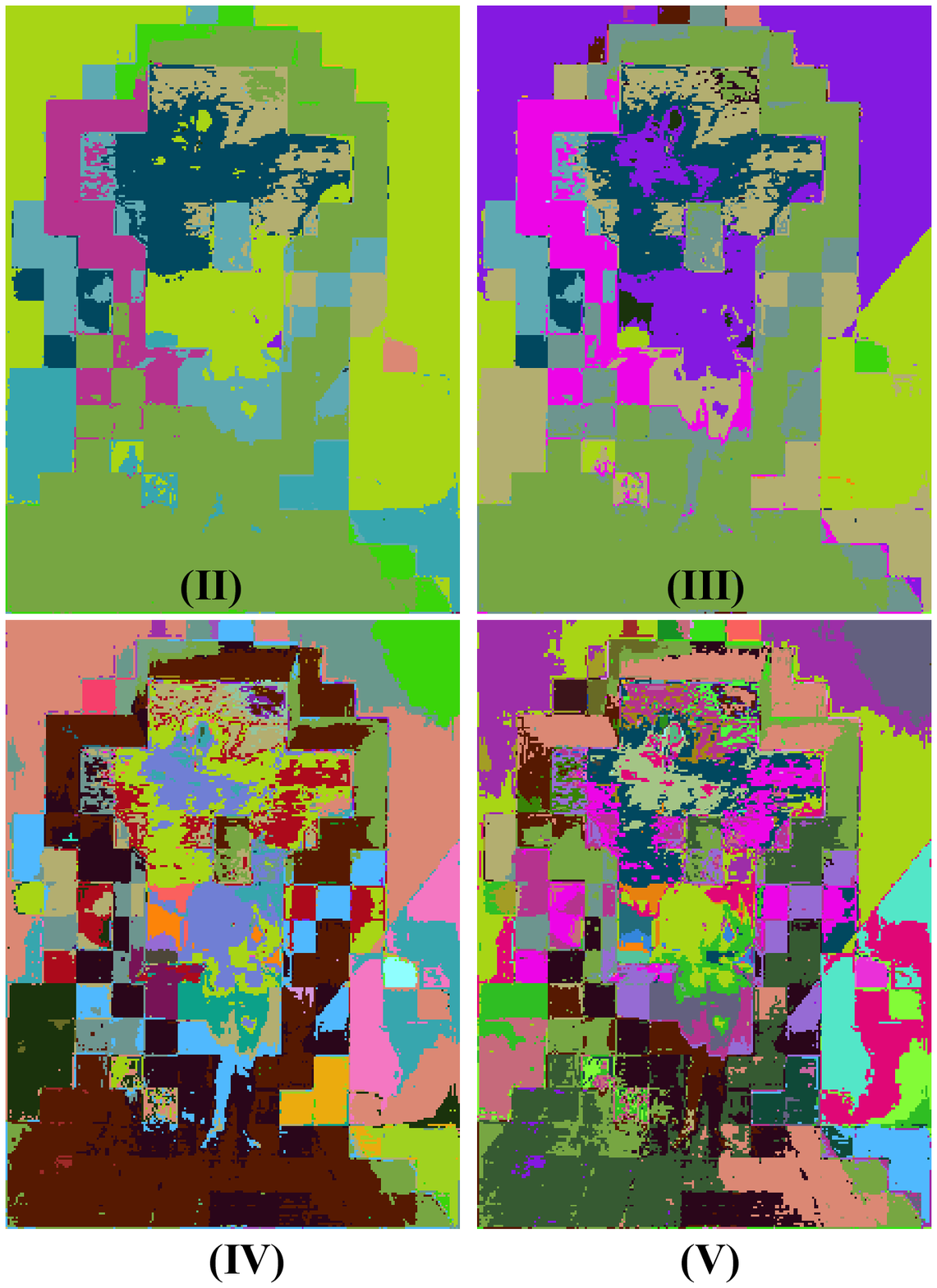}} \caption{[Color Online.] The specific
image is from \cite{lincoln}. At close distance, this is ``Gala
contemplating the Mediterranean sea'' while at larger distance is
``a portrait of Abraham Lincoln''. Panel(a) shows the variation of
information as a function of resolution. We pick the resolution at
each ``peak'' position and apply our algorithm at these particular
resolutions. Panels (b) and (c) show the resulted images at the
corresponding resolutions marked in panel (a). Note that at low
resolution, the resulting segmentation clearly depicts ``the
portrait of Abraham Lincoln'' as shown in panel (b) on the right. In
particular, notwithstanding noise, as $\gamma$ increases, the
segmentation results show more details and we could detect the lady
in the middle in (II)-(V) of Panel(c).  }\label{fig:lincoln}
\end{center}
\end{figure}

We next apply our multiresolution community detection algorithm to
the images that are by construction truly multi-scale. The results
at different resolutions are shown in \figref{fig:lincoln}. The
original image is that of Salvador Dali's famous painting ``Gala
contemplating the Mediterranean sea which at twenty meters becomes a
portrait of Abraham Lincoln''. Our algorithm perfectly detected
the portrait of Lincoln at low resolution as shown
in \figref{fig:lincoln} in the segmentation result
appearing in panel (I) of (b). As the resolution parameter
$\gamma$ increases, the algorithm is able to detect more details.
However, due to the non-uniform
color and the similarity of the surrounding colors to those of the targets, the results are
very noisy. At the threshold of $\bar{V}=20$, the algorithm has
difficulty in merging pixels to reproduce the lady in the image. For
example, in image (II) in Panel (c), the lady's legs are merged into
the background. In image (III), only one leg is detected. In images
(IV) and (V), both legs can be detected but belong to different
clusters.

\subsection{Benchmarks}
In order to assess the success of our method and ascertain general features,
we applied it to standard benchmarks. In particular, we
examined two known benchmarks:  (i) The Berkeley image segmentation benchmark
and (ii) that of Microsoft Research.

\subsubsection{Berkeley Image Segmentation
Benchmark}\label{sec:berkeley}

We were able to accurately
detect the targets in test images, as in
Figs.(\ref{fig:precision}, \ref{fig:uniform}). The original images
in \figref{fig:precision} were downloaded from the Berkeley image
segmentation benchmark BSDS300 \cite{berkeley-paper}, and those of
 \figref{fig:uniform} are downloaded from the
Microsoft Research \cite{microsoft}. We will now compare our results
with the results by other algorithms. The Berkeley image
segmentation benchmark provides the platform to compare the boundary
detection algorithms by an ``F-measure''. This quantity is defined as
\begin{eqnarray}
\mbox{F-measure} =  \frac{\mbox{2 }\times \mbox{Precision
}\times\mbox{ Recall}}{\mbox{Precision+Recall}}.
\end{eqnarray}
``Recall'' is computed as the fraction of correct instances among
all instances that actually belong to the relevant subset, while
``Precision'' is the fraction of correct instances among those that
the algorithm believes to belong to the relevant subset. Thus, we
have to draw the boundaries in our results and compute the
F-measure. We use the tool ``EdgeDetect''  of Mathematica software
to draw the boundaries within our region detection results, as shown
in the right column in \figref{fig:precision}. The comparison of the
``F-measure'' of our algorithm (``F-Absolute Potts Model'') and the
best algorithm in the benchmark (``F-Global Probability of
boundary'')\cite{gPb1,gPb2} is shown in Table. \ref{table}. On the
whole, our results are better than the algorithm of the Berkeley
group.

\myfig{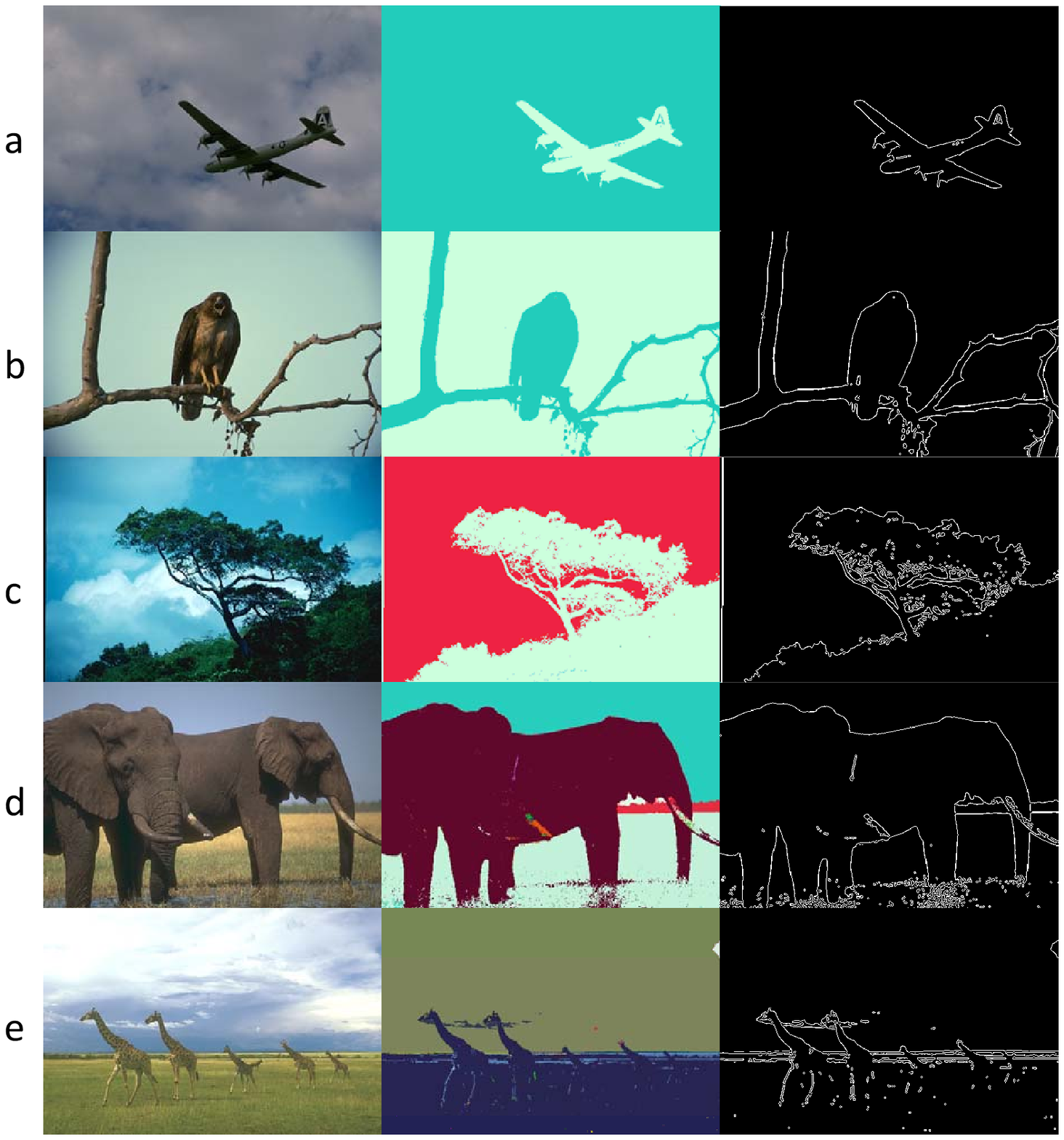}{[Color Online.] Image segmentation
results of our algorithm when tested with examples from the Berkeley BSDS300 benchmark.
Shown, in the left column, are the original images. The central column contains the
results of our method. The right column provides the boundaries of the
images in the middle by running ``EdgeDetect'' of Mathematica on the
results of our run in the central column. The parameters of
community detection algorithm used for these images are: in (a),
$\gamma=0.001$, $\bar{V}=15$. In panel (b), $\gamma=0.0001$,
$\bar{V}=20$. In (c), $\gamma=0.001$, $\bar{V}=20$. In (d),
$\gamma=0.01$, $\bar{V}=15$. In (e), $\gamma=0.01$, $\bar{V}=15$. We
performed the boundary detection on the results of our community detection algorithm
(i.e., the central column) and employed the ``F-measure'' accuracy parameter
in order to compare the results of our algorithm with earlier results reported for the Berkeley
image segmentation benchmark (shown in Table. \ref{table}).
}{fig:precision}{1\linewidth}{t}

\begin{table}
\begin{tabular}{|c|c|c|}
  \hline
   & F-Absolute Potts Model & F-Global Probability of boundary \\
   \hline
  a & 0.79 & 0.78 \\ \hline
  b & 0.94 & 0.91 \\ \hline
  c & 0.82 & 0.74 \\ \hline
  d & 0.79 & 0.83 \\ \hline
  e & 0.75 & 0.60 \\ \hline

  \hline
\end{tabular}
\caption{The comparison of ``F measure'' for \figref{fig:precision}
by our community detection algorithm (``F-Absolute Potts Model'')
with the algorithm ``Global Probability of boundary'' (``gPb'')
\cite{gPb1,gPb2} which has the highest score in the Berkeley image
segmentation benchmark (``F-Global Probability of boundary''). The
higher F-value corresponds to the better detection. Note that our
algorithm is performing better than the ``gPb algorithm'' in almost
all images except the fourth one. Our fourth (d) image gets lower
score is mostly because there are dots in the lower grass place.
These small dots will lead to small high accuracy features. These
features are unexpected in the ground truth and thus lower the
F-value. } \label{table}
\end{table}

\subsubsection{Microsoft Research Benchmarks}

In \figref{fig:uniform}, we compare our results (in the rightmost column) with the
ground truths provided by Microsoft Research (the central column).
By adjusting the $\gamma$ and $\bar{V}$
values, we can merge the background pixels and highlight the target.
In the segmentation of the image of the flower in the first row,
$\gamma=0.001$ and $\bar{V}=20$. For both the picnic table in
the middle row and that of the two sheep in the bottom
row, we set $\gamma=0.01$ and $\bar{V}=15$.

\myfig{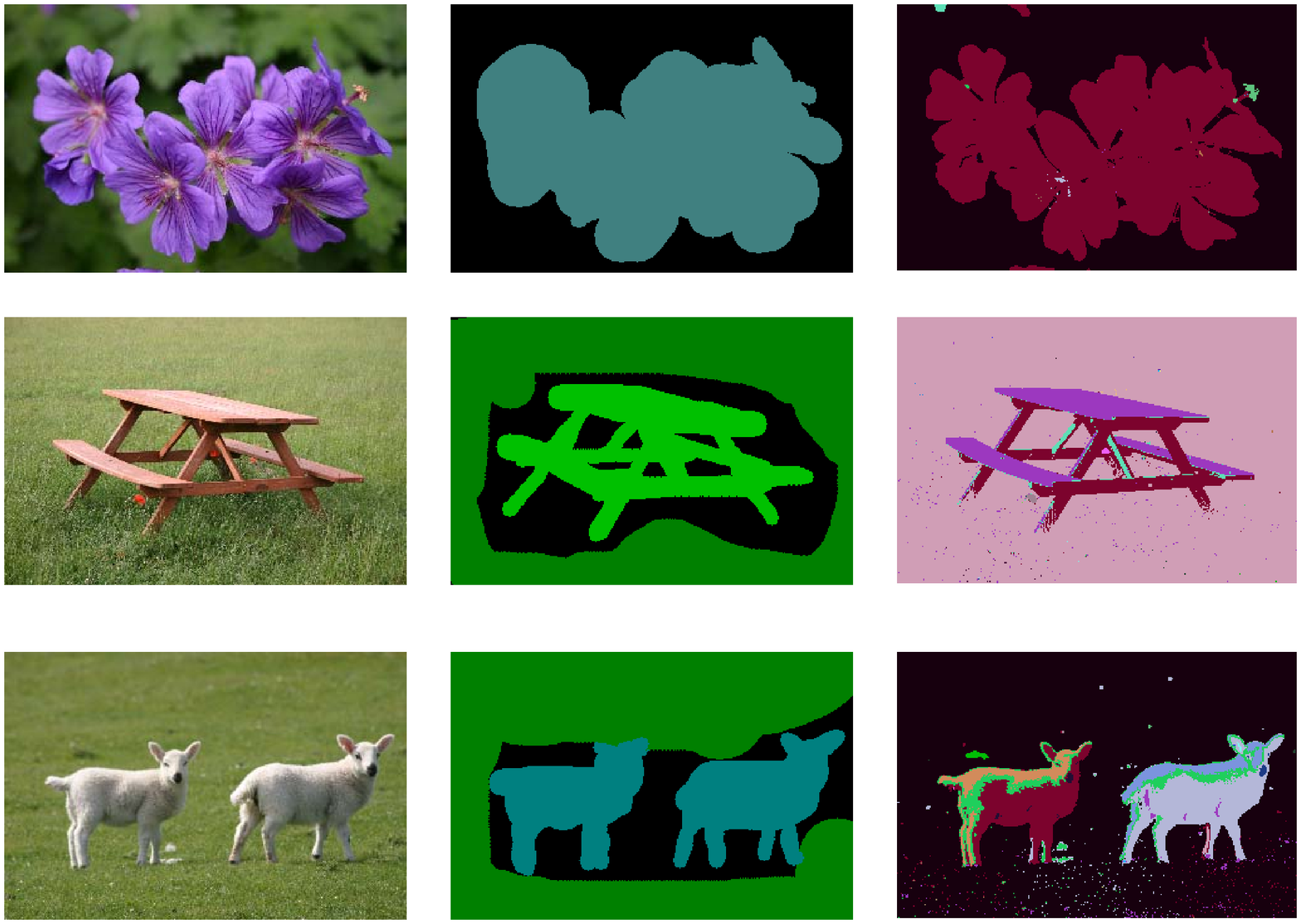}{[Color Online.] The results of some image
segmentations by our Potts model (\eqnref{eq:newpotts}) and
community detection algorithm (\cite{peter1}). The images are
downloaded from the website of Microsoft Research
(\cite{microsoft}). The left column are the original images. The
central column are the ground truths defined by the website of
Microsoft Research (\cite{microsoft}), which are the desirable image
segmentation results. The right column are the segmentation results
by our algorithm. The parameters used for each image are:
(1)$\gamma=0.001$, $\bar{V}=20$ for the flower image. (2)
$\gamma=0.01$, $\bar{V}=15$ for the image of the picnic table. (3)
Similarly, $\gamma=0.01$, $\bar{V}=15$ for the image of the two
sheep. Note that our algorithm works very well for this kind of
images in which the color is nearly uniform within each object.
}{fig:uniform}{1\linewidth}{}

\subsection{Detection of quasi-periodic structure in quasicrystals}
\label{sec:quasicrystal}

Quasicrystals \cite{quasicrystal} are ordered but not periodic (hence the name ``quasi'').
In \figref{fig:quasicrystal}, the image in row (a) is such a quasi-crystal formed by
``Penrose tiling''. We applied the Fourier
transform method to reveal the corresponding underlying structures.
In row (a), the image marked by (I) is the original image
(downloaded from \cite{quasi1}), the one with notation (II) is the
result of our algorithm, and (III) is the image of (II) with the
connections of the nearest neighbor nodes. The images marked by (II)
and (III) show the first Penrose tiling (tiling P1). Penrose's first
tiling employs a five-pointed pentagram, 3/5 pentagram shape and a thin
rhombus. Similarly, the result images of panels (II) and (III) in row (b)
reveal the underlying structure of the superlattice with $AB_4$
stoichiometry and the structural motif of the ($3^2.4.3.4$)
``Archimedean tiling'' of the original image (I) (from
\cite{quasi2}). The Archimedean tiling displayed in image (III) of
row (b) of  \figref{fig:quasicrystal} employs squares and triangles.
It is straightforward to analyze the quasi-periodic structure by
applying our image segmentation algorithm as shown in
\figref{fig:quasicrystal}. By iterating the scheme outlined herein,
structure on larger and larger scales was revealed.

\myfig{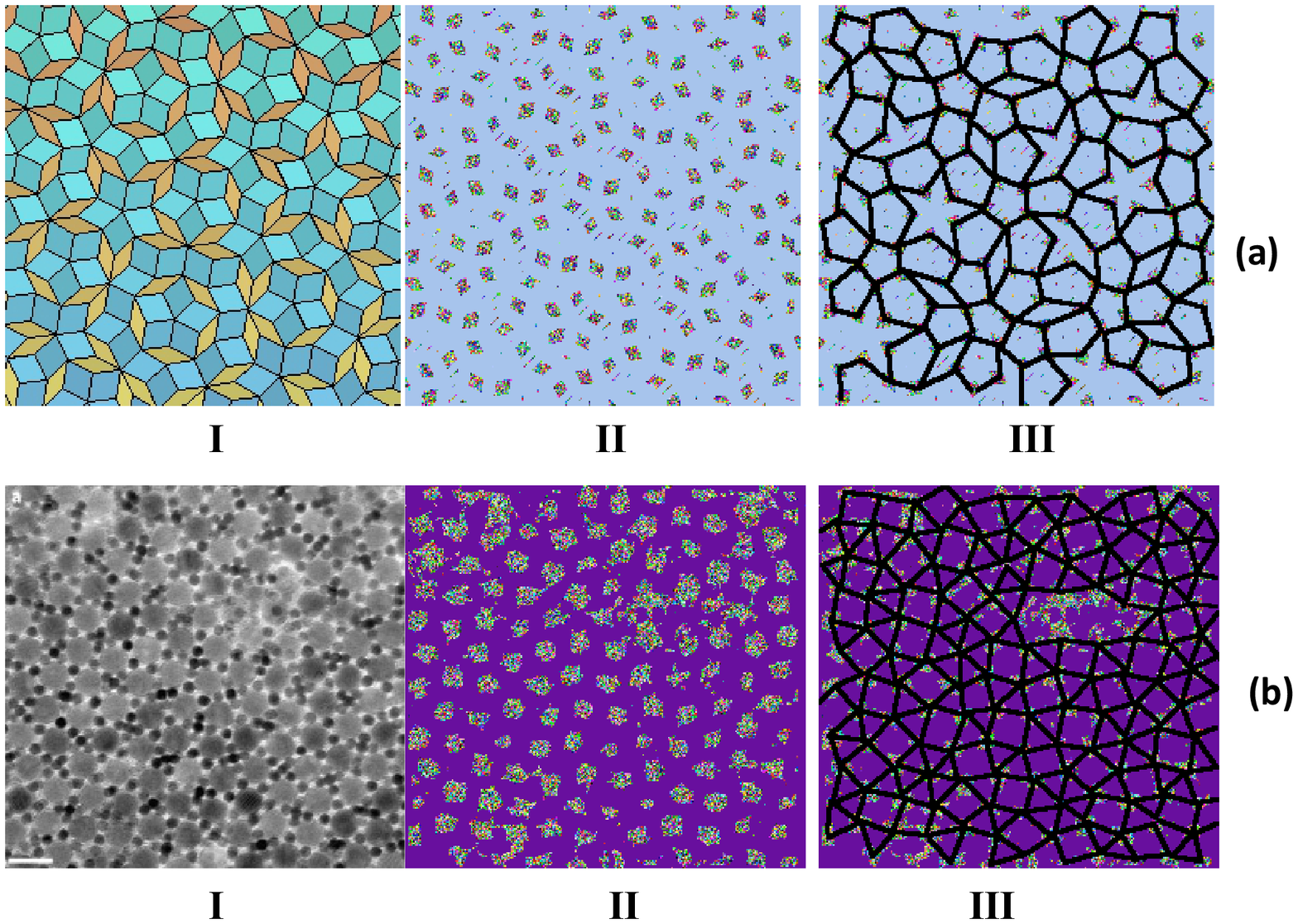}{[Color Online.] Quasicrystal images are
displayed in panels (I). The corresponding image segmentation
results by our algorithm are shown in (II). In (III), we connect the
basic object by line, resulting in large basic blocks. This process
can be repeated recursively leading to larger and larger scale
structures. Note that we are able to reveal the underlying
quasi-periodic structures in both row (a) (the original image in (I)
is from \cite{quasi1}) and (b) (the original image in (I) is from
\cite{quasi2}). We show the first Penrose tiling (tiling ``P1'') in
(a), and the structural motif of the ($3^2.4.3.4$) Archimedean
tiling in (b). }{fig:quasicrystal}{1\linewidth}{}

\subsection{Images with spatially varying intensities}

If the target is similar to the background (as in, e.g., animal
camouflage), then the simplest initialization of edges with linear
weights will, generally, not suffice. For example, in
\figref{fig:fourier} the zebra appears with black and white stripes.
It is hard to directly detect the stripes of the zebra because of
the large ``color'' difference between the black and white stripes
of the zebra. \figref{fig:zebra-hard} has the similar stripe-shaped
background which is very difficult to distinguish from the zebra
itself by using the weights of Eqs.(\ref{eqn:vmn}, \ref{eqn:dmn})
for the edges. Towards this end, we will next employ the Fourier transform method of
\secref{sec:fourier}.

\myfig{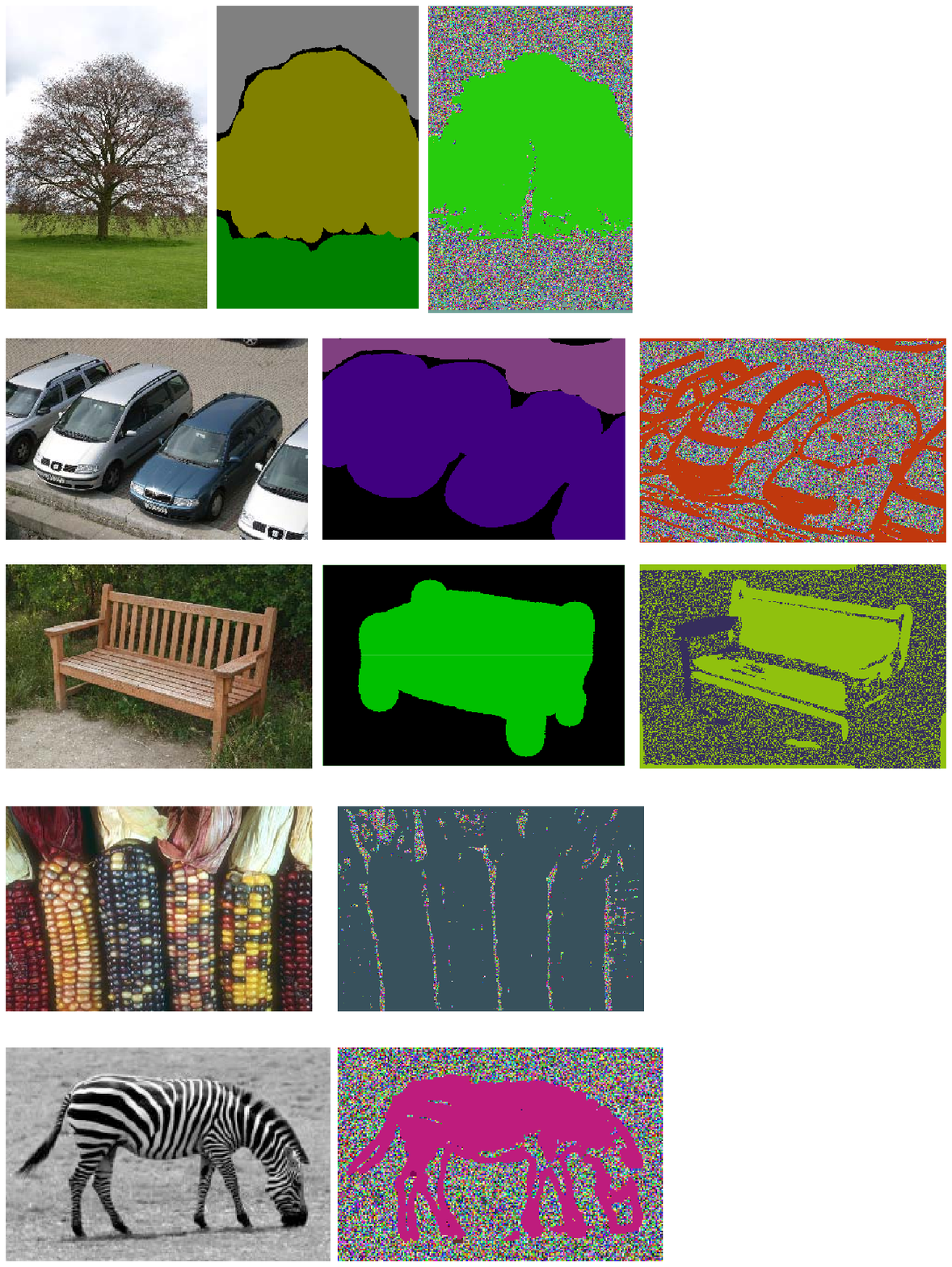}{[Color Online.] The image segmentation results by
the community detection algorithm with Fourier weights as described
in Section \ref{sec:fourier}. Some of the images are downloaded from
the Microsoft Research (\cite{microsoft}) and some of them are
download from the Berkeley image segmentation benchmark
(\cite{berkeley-paper}). The left column contains the original
images.  The central column (apart from the last two rows) provides
the ``ground truths''.  The right images on the right are our
results. The parameters used in each image are: (1) $\gamma=0.01$,
$\bar{V}=-300$ for the tree image. (2) $\gamma=0.1$ and
$\bar{V}=-300$ for the car image. (3) $\gamma=0.01$ and
$\bar{V}=-400$ for the bench image. (4) $\gamma=0.01$,
$\bar{V}=-100$ for the image of corn. (5) $\gamma=0.1$,
$\bar{V}=-900$ for the zebra image. Even though the color is not
uniform inside the targets, we can nevertheless easily detect the
targets by this method.}{fig:fourier}{1\linewidth}{}

As seen in \figref{fig:fourier}, the original images are not
uniform. Rather, these images are composed of different basic
components such as stripes or spots, etc. With the aid of Fourier
transform within each block, as discussed in Section
\ref{sec:fourier}, we are able to detect the target. For some of the
images such as the second one in \figref{fig:fourier}, when the
target is composed of more than one uniform color or style, our
community detection algorithm is able to detect the boundaries, but
the regions inside the boundary are hard to merge. This is because
the block size is smaller than that needed to cover both the target
and the background. That is, block size of $L_x\times L_y=5\times 5$
is much smaller than the image size of $N_x\times N_y=320\times213$
in the car image in the second row, so most of the blocks are within
one color of the target (car) or the background (ground). However,
the dominant Fourier wave-vector of the region within one color
component of the car is similar to that of the ground. Therefore,
the algorithm always treats them as the same cluster, rather than
merging the region inside the car with the boundary.

In other instances (e.g., all the other
rows except the second in \figref{fig:fourier}),
the targets are markedly different from the backgrounds.
Following the scheme discussed in Section \ref{sec:complexity}
(that will be fleshed out in Section \ref{sec:phasediagram}),
we may always optimize parameters such as
the resolution, threshold, or the block size to obtain better
segmentation.

\subsection{Detection of camouflaged objects}

\myfig{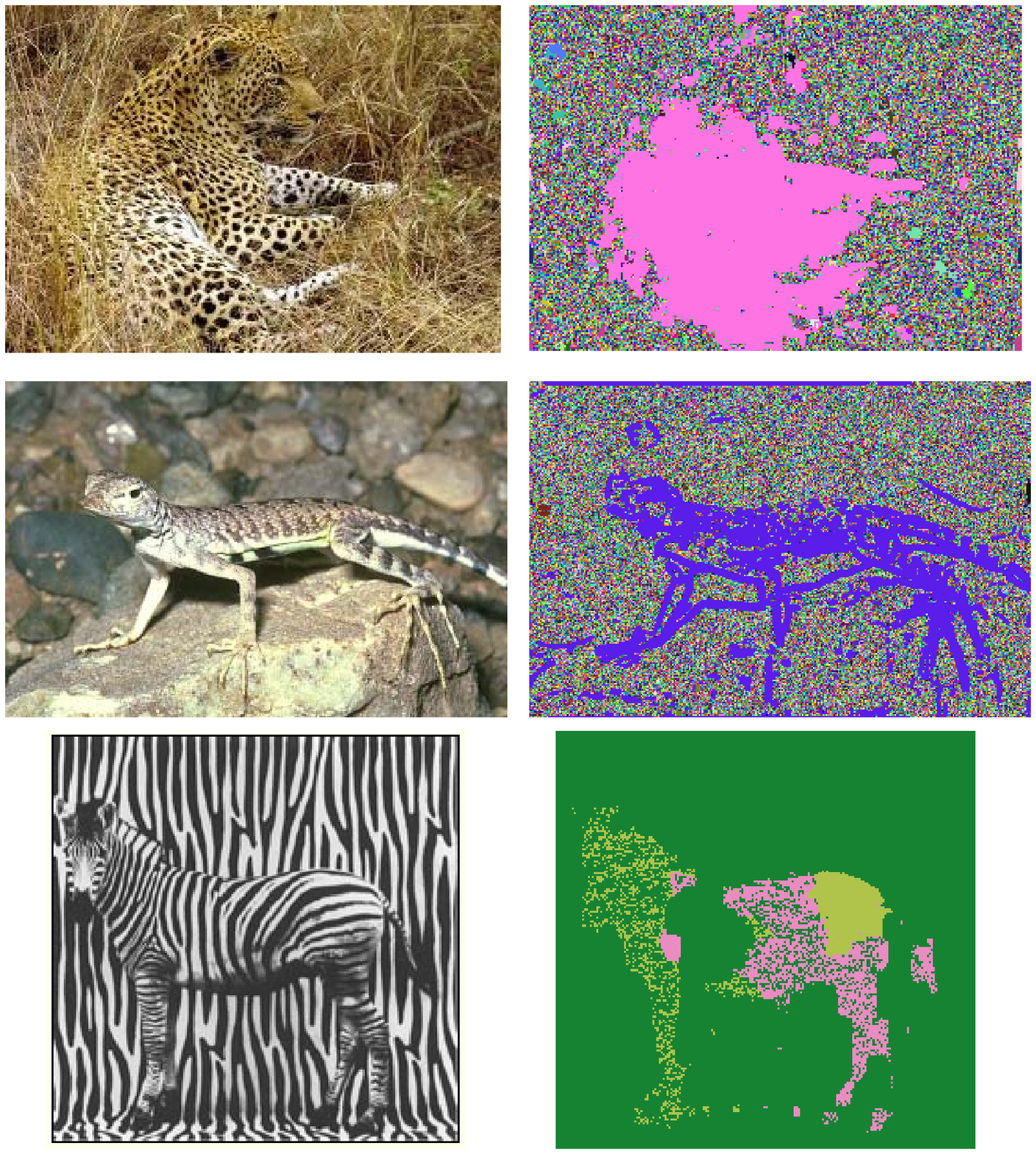}{[Color Online.] The results of the image
segmentation for a ``camouflaged image''.  The image
of the leopard is from (\cite{wikipedia}), the lizard is provided in
the Berkeley image segmentation benchmark(\cite{berkeley-paper}),
and the last image is from the website of the EECS department of
Berkeley (\cite{berkeley-zebra}). The parameters for the shown segmentations
are: (1) $\gamma=1$, $\bar{V}=-700$ for the image of the leopard, (2)
$\gamma=0.1$, $\bar{V}=-500$ in the image of the lizard, and (3) $\gamma=1$,
$\bar{V}=-1100$ for the zebra image.
}{fig:camouflage}{1\linewidth}{t}

\begin{figure}[t]
\begin{center}
\subfigure[The variation of information $V$ as a function of the
negative threshold $-\bar{V}$ for the zebra image in panel (b).
]{\includegraphics[width= 3in]{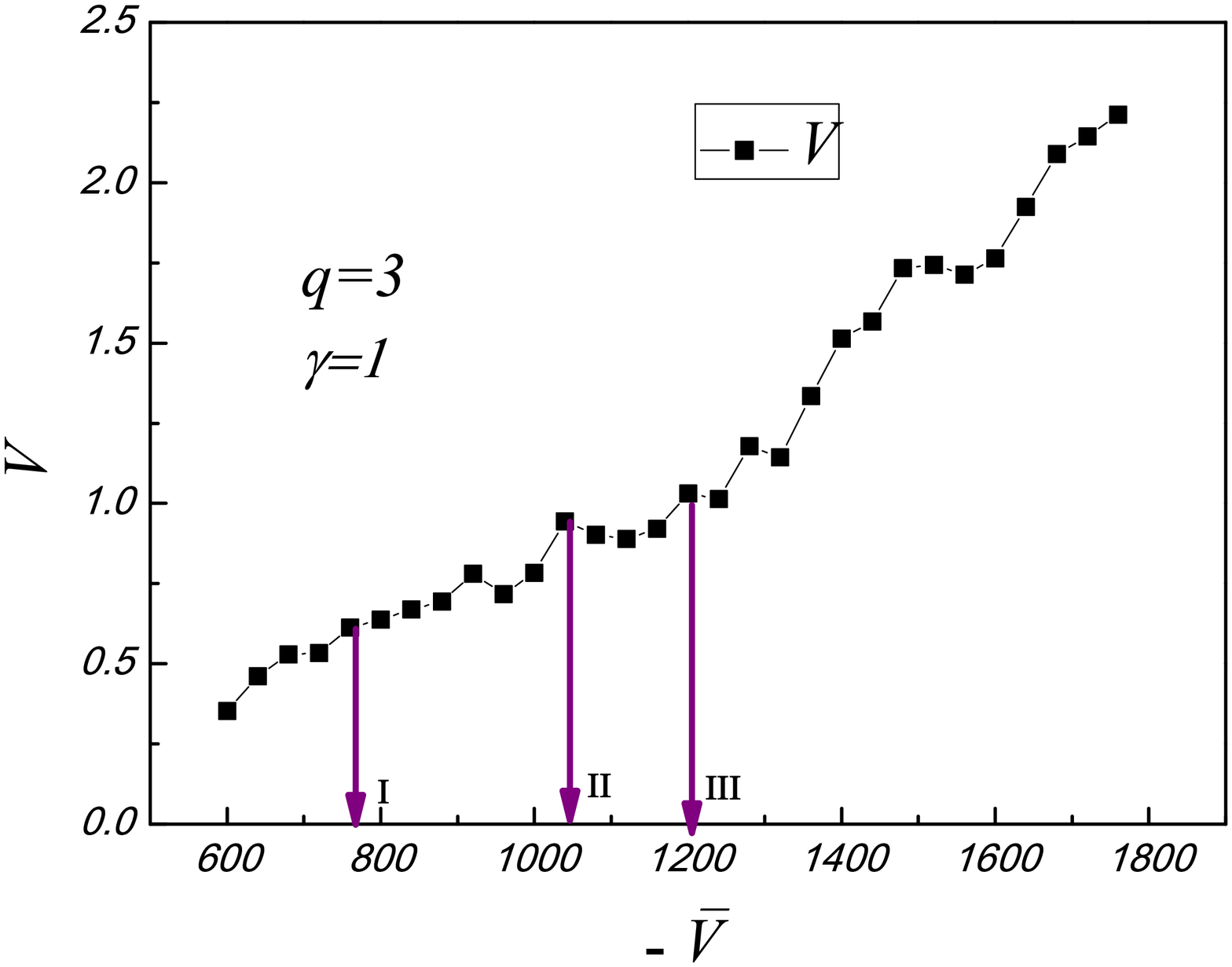}} \subfigure[The weighted
results of the zebra images at the corresponding thresholds:
$\bar{V}_1=-760$ (I), $\bar{V}_2=-1040$ (II), and $\bar{V}_3=-1200$
(III). ]{\includegraphics[width= 3in]{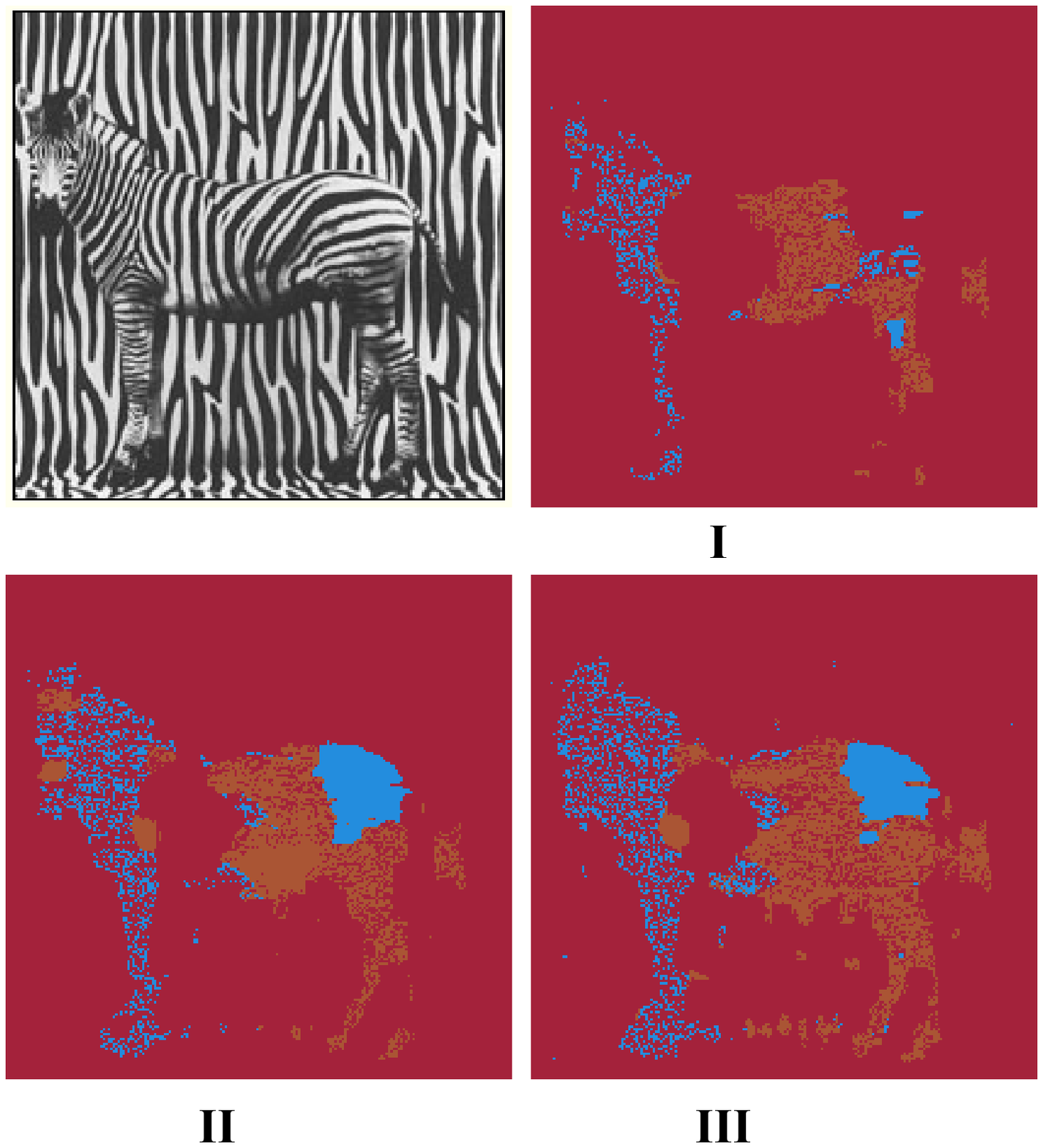}}
\caption{[Color Online.] The ``multiresolution'' result of zebra with
fixed community number $q=3$ and resolution $\gamma=1$. In panel(a),
we plot the variation of information $V$ as a function of negative
threshold $-\bar{V}$. The peaks in $V$ correspond to the changes of
structures. We choose three peaks and run the algorithm at these
three particular thresholds, and the result images are shown in
panel (b). As $|\bar{V}|$ increases, less regions in the zebra merge
to the background, and the boundary becomes more clear. If we
increase the threshold further, the result is more noisy as the last
image of $\bar{V}=-1200$ (III) shows. }\label{fig:zebra-hard}
\end{center}
\end{figure}

In the images of \figref{fig:fourier}, the target objects are very
different from their background.
However, there are images wherein (camouflaged) objects are similar
to their background. In what follows, we will report on the results
of our community detection algorithm when these challenging  images
were analyzed. In all of the cases below in Section
\ref{sec:leopard}, the edge weights were initialized by the Fourier
amplitudes discussed earlier (Section \ref{sec:fourier}). In the
case of the dalmatian dog image in Section \ref{sec:dalmatian}, we
further employed the method of average intensity difference between
blocks discussed in Section \ref{sec:average}. In all cases but this
last one of the dalmatian dog, we fixed the length scale parameter
$\ell$ of Section \ref{sec:fourier} to be infinite.

\subsubsection{Images of a leopard, a lizard, and
a zebra}\label{sec:leopard}

\begin{figure}[]
\begin{center}
\subfigure[The variation of information $V$ as a function of length
$\ell$ at $\gamma=0.1$]{\includegraphics[width= 2.5in]{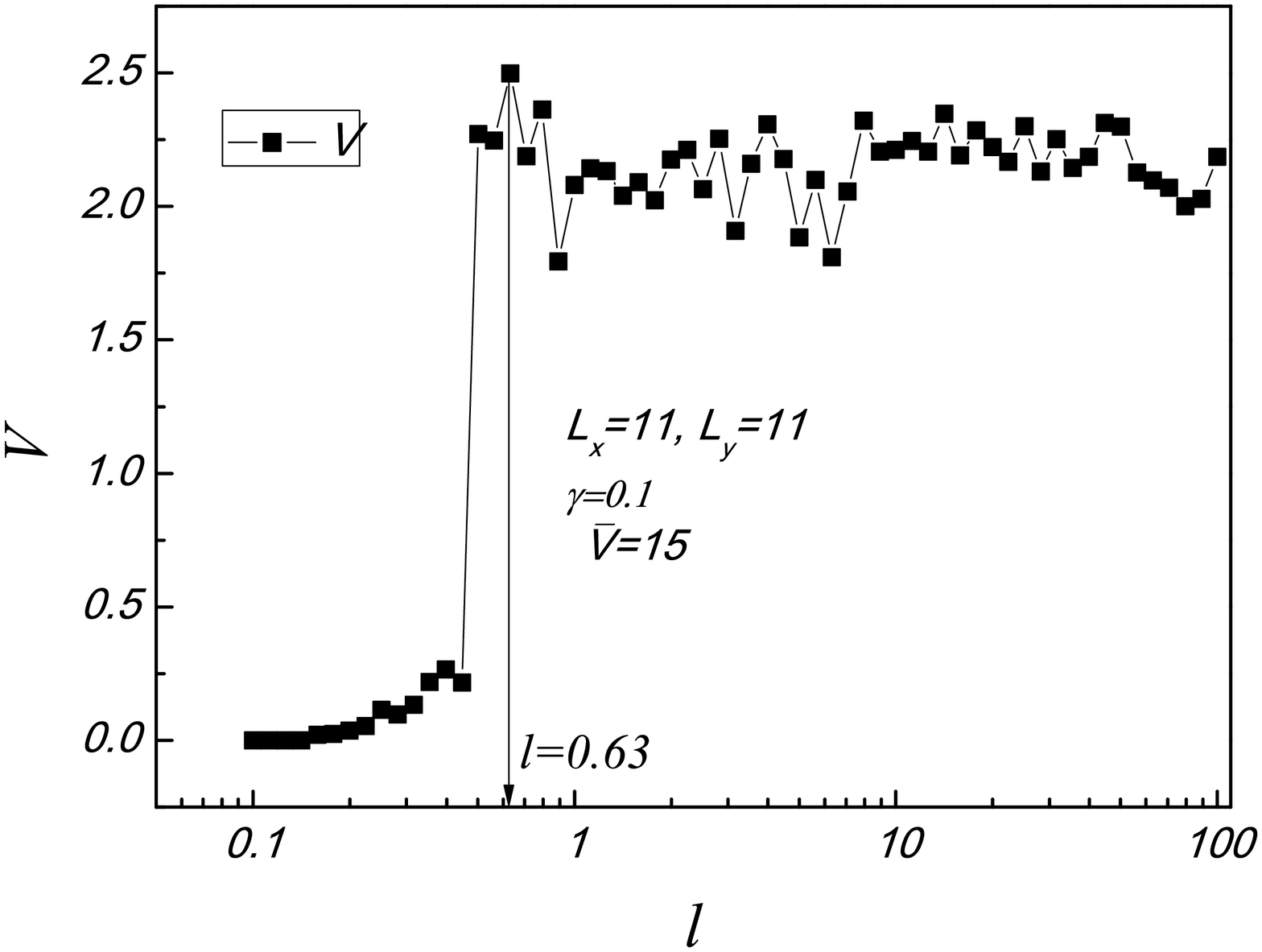}}
\subfigure[The normalized mutual information $I_N$ as a function of
length $\ell$ at $\gamma=0.1$]{\includegraphics[width=
2.5in]{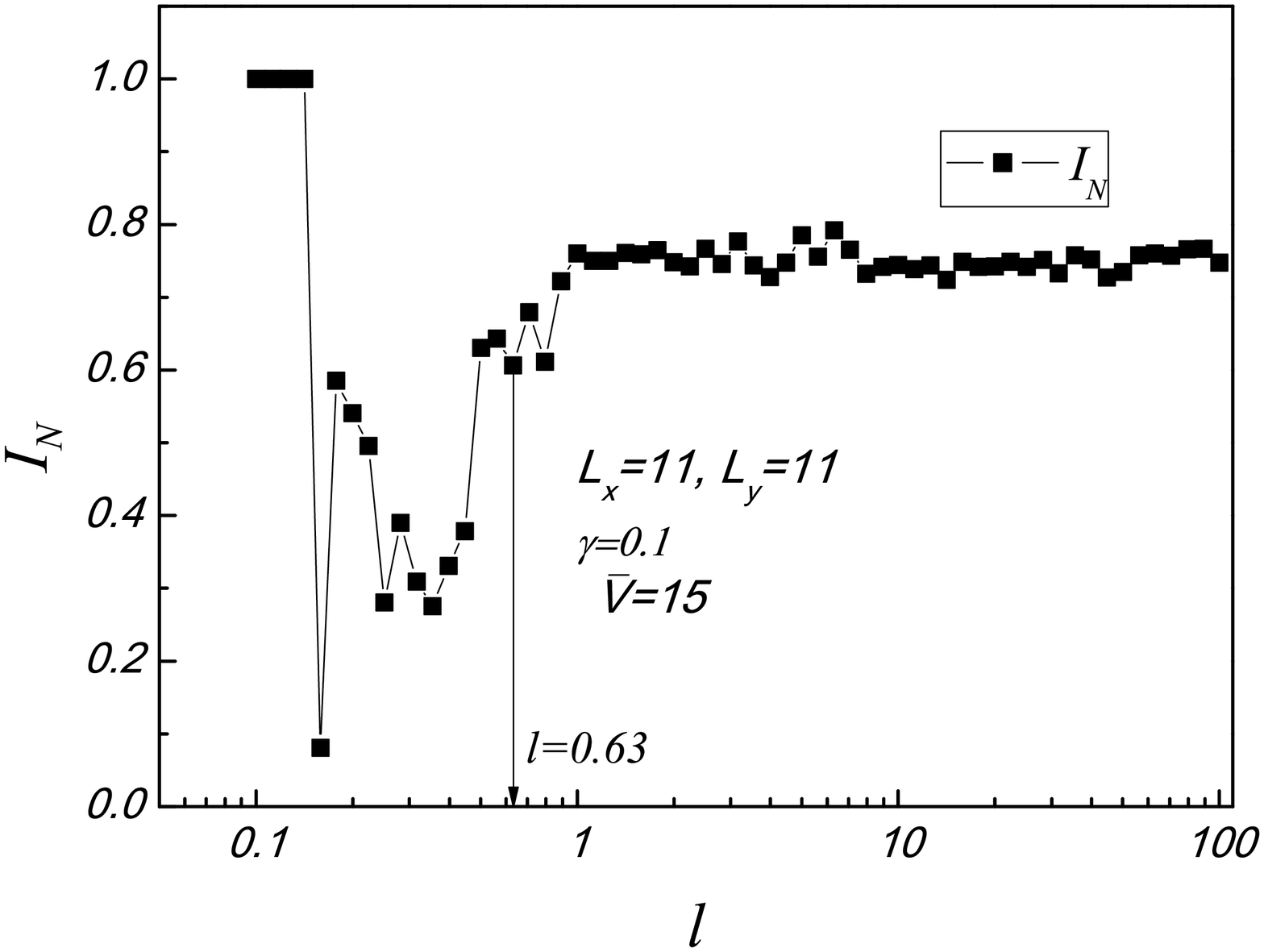}} \subfigure[The corresponding image segmentation
result at the extremum of $V$/$I_N$ in panel
(a)/(b)]{\includegraphics[width= 3in]{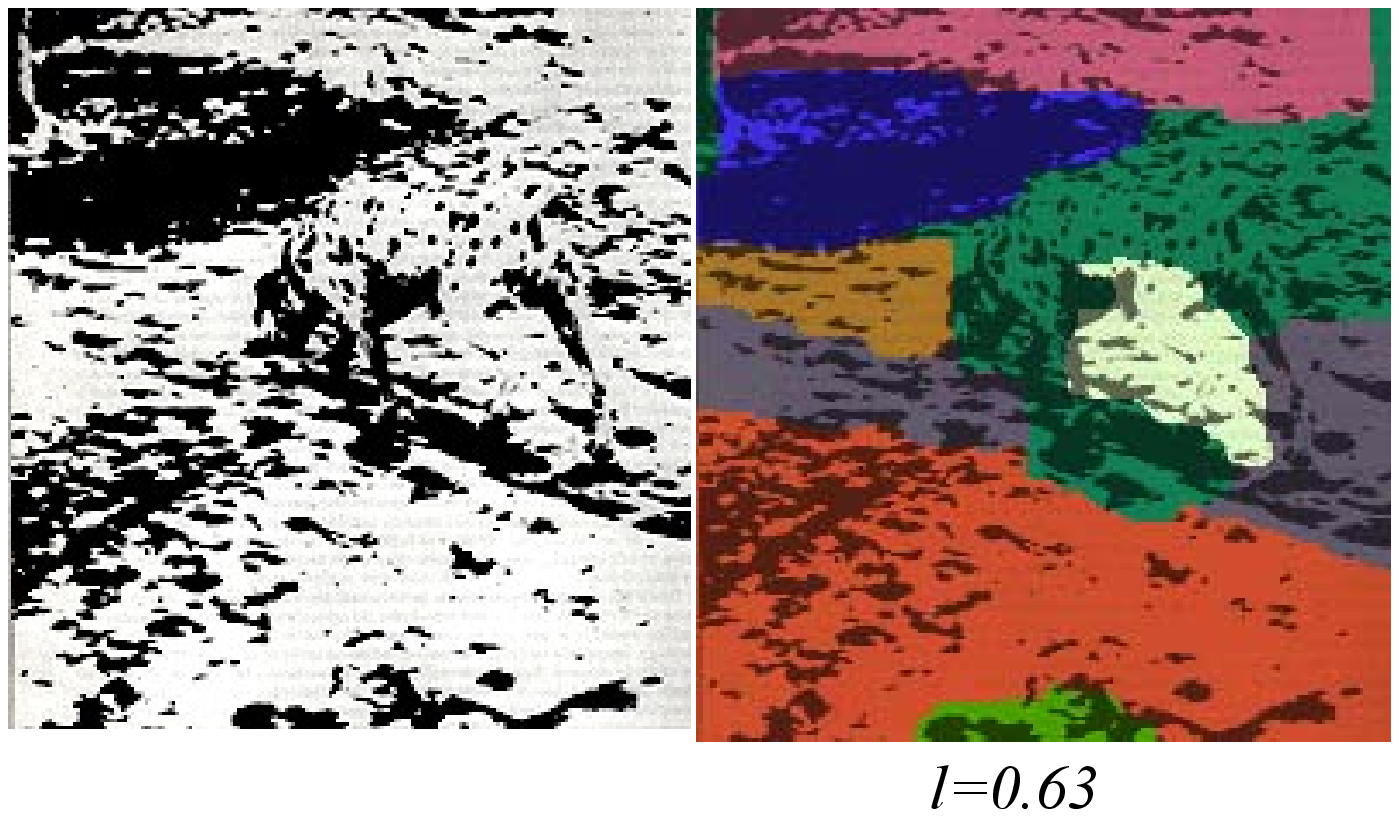}}
\end{center}
\end{figure}
\begin{figure}[]
\begin{center}
\subfigure[The variation of information $V$ as a function of length
$\ell$ at $\gamma=0.05$]{\includegraphics[width= 2.5in]{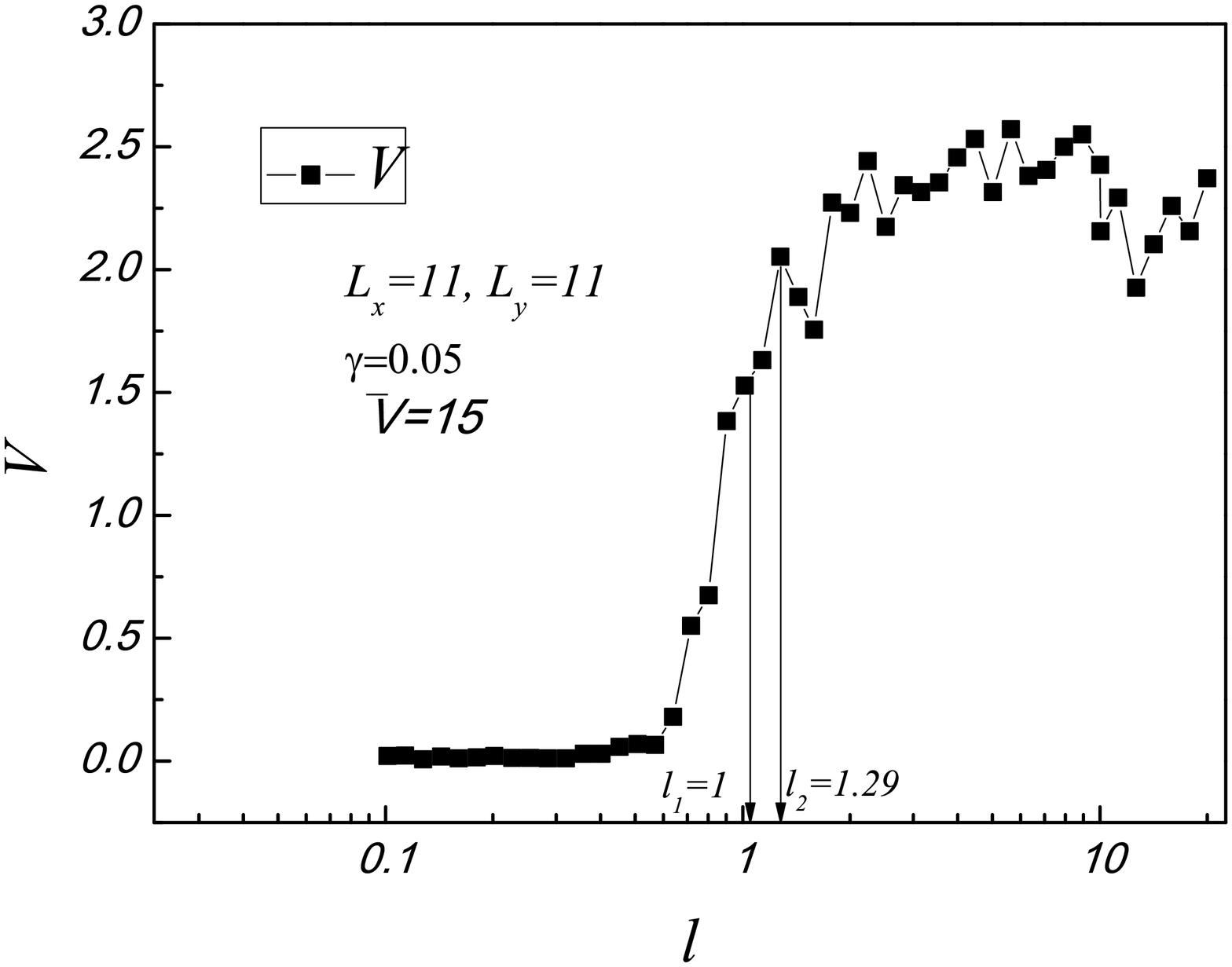}}
\subfigure[The normalized mutual information $I_N$ as a function of
length $\ell$ at $\gamma=0.05$]{\includegraphics[width=
2.5in]{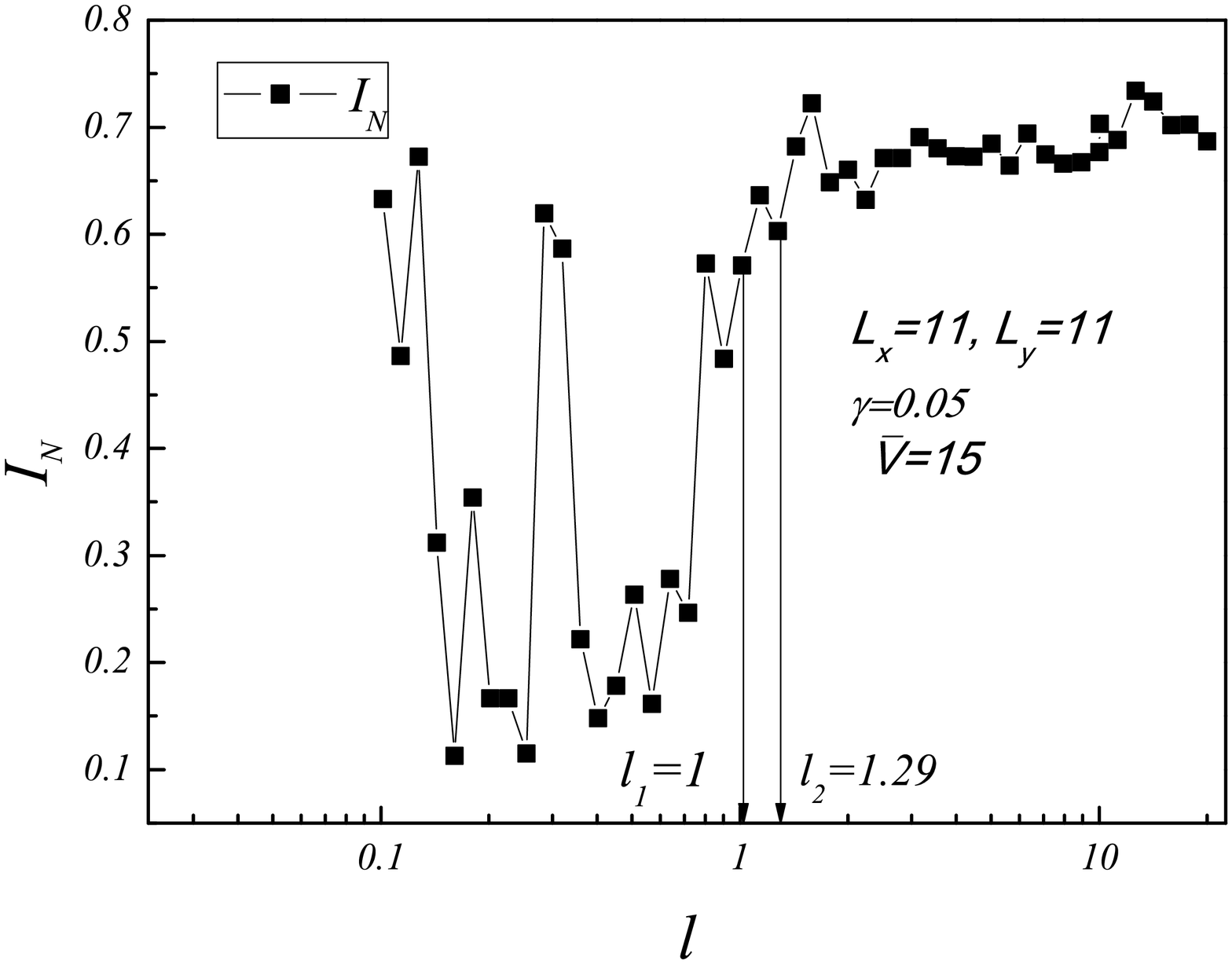}} \subfigure[The corresponding image segmentation
results at the extremum, and at a point close to the peak of
$V$/$I_N$ in panel (d)/(e).]{\includegraphics[width=
3in]{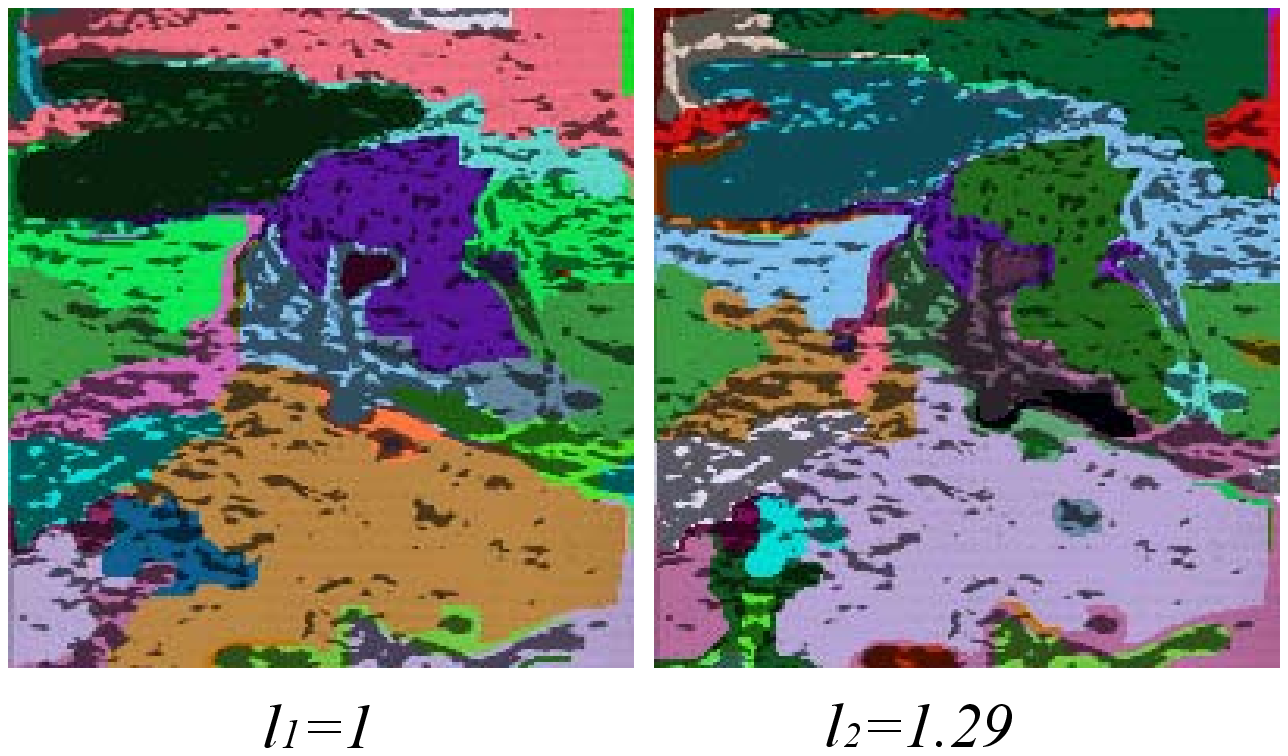}}\caption{[Color Online.] Results of our
algorithm as a function of the length scale $\ell$ in
\eqnref{eqn:vmn} for the dalmatian dog image.
Plots of the variation of information and the normalized mutual
information ($V$, $I_N$) as a function of the length scale $\ell$
appear in panels (a, b)(at resolution of $\gamma=0.1$) and (d, e)
($\gamma=0.05$ ). Panel (c) shows the original image. As seen in
panels (a,b), a coincident local maximum of $V$ and local minimum of
$I_N$ appears (for $\gamma =0.1$) at $\ell =0.63$. Similarly, panel
(f) shows the images corresponding to the peak of $V$ (coincident
with a local minimum of $I_N)$ in panel (d) (and (e)) at
$\ell_2=1.29$ (and $\gamma=0.05$). We examine the results for
$\ell_1=1$ in panel (f). We
are able to detect the body and the back two legs of the dog, even
though with some ``bleeding'' in panel (f). In (c), we are detecting
well except for the inclusion of some ``shade'' noise under the
body. }\label{fig:dalmatian}
\end{center}
\end{figure}

``Camouflage'' refers to a method of hiding. It allows for an
otherwise visible organism or object to remain unnoticed by blending
with its environment. The leopard in the first row of
\figref{fig:camouflage} is color camouflaged. With our algorithm, we
are able to detect most parts of the leopard except the head. The
lizard in the second row uses not only the color camouflage but also
the style camouflage, both the lizard and the ground are composed of
grey spots. We can detect the lizard. The zebra in the bottom row uses the camouflage--
both the background and the zebra have
black-and-white stripes. Our result is very accurate, even though
the algorithm treats the middle portion of the zebra (the
position of the ``hole'') as the background by mistake. This is because,  in this
region, the stripes within the zebra are very hard to
distinguish from the stripes in the background, they are both
regular and vertical.

We applied the ``multiresolution'' algorithm to the zebra image in the
last row of \figref{fig:camouflage} as shown in
\figref{fig:zebra-hard}. The number of communities is $q=3$, the
resolution parameter $\gamma=1$ and the threshold $\bar{V}$
was varied from $\bar{V}=-600$ to $\bar{V}=-1800$. In the low $|\bar{V}|$ area,
some regions inside the zebra tend to merge into the background (the
image with the threshold $\bar{V}=-760$). As the background
threshold $|\bar{V}|$ increases in magnitude, the boundary of the zebra becomes
sharper (the shown segmentation corresponds to a threshold of
$\bar{V}=-1040$). For yet larger values of $|\bar{V}|$, the results
are noisy (the image with the threshold $\bar{V}=-1200$). Thus, in
the range $760\leq|\bar{V}|\leq 1200$, we obtain the clear
detection seen in the last row of \figref{fig:camouflage}.

\begin{figure}[t]
\begin{center}
\subfigure[A 3d plot of the normalized mutual information $I_N$ as a
function of $\log(\ell)$ and $\log(\gamma)$.
]{\includegraphics[width= 2in]{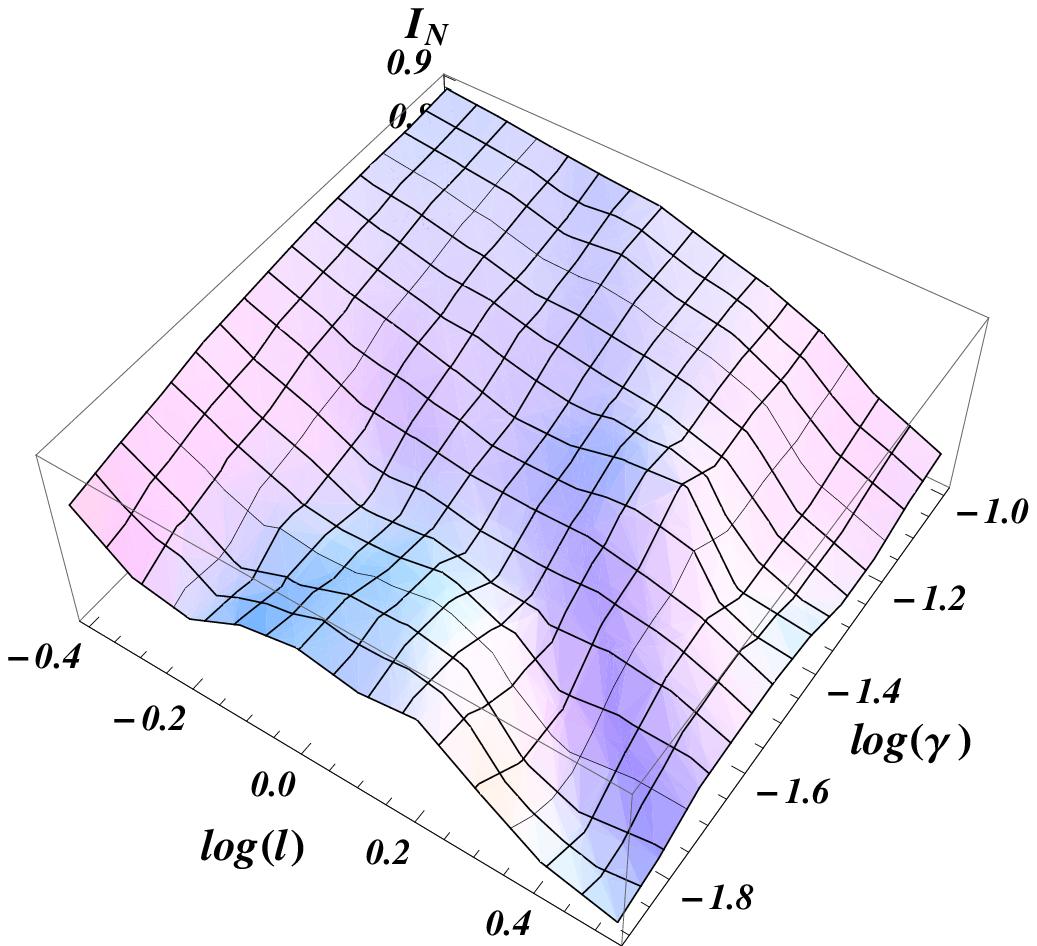}} \subfigure[The 3d
plot of the variation of information $V$ as the function of
$\log(\ell)$ and $\log(\gamma)$]{\includegraphics[width=
2in]{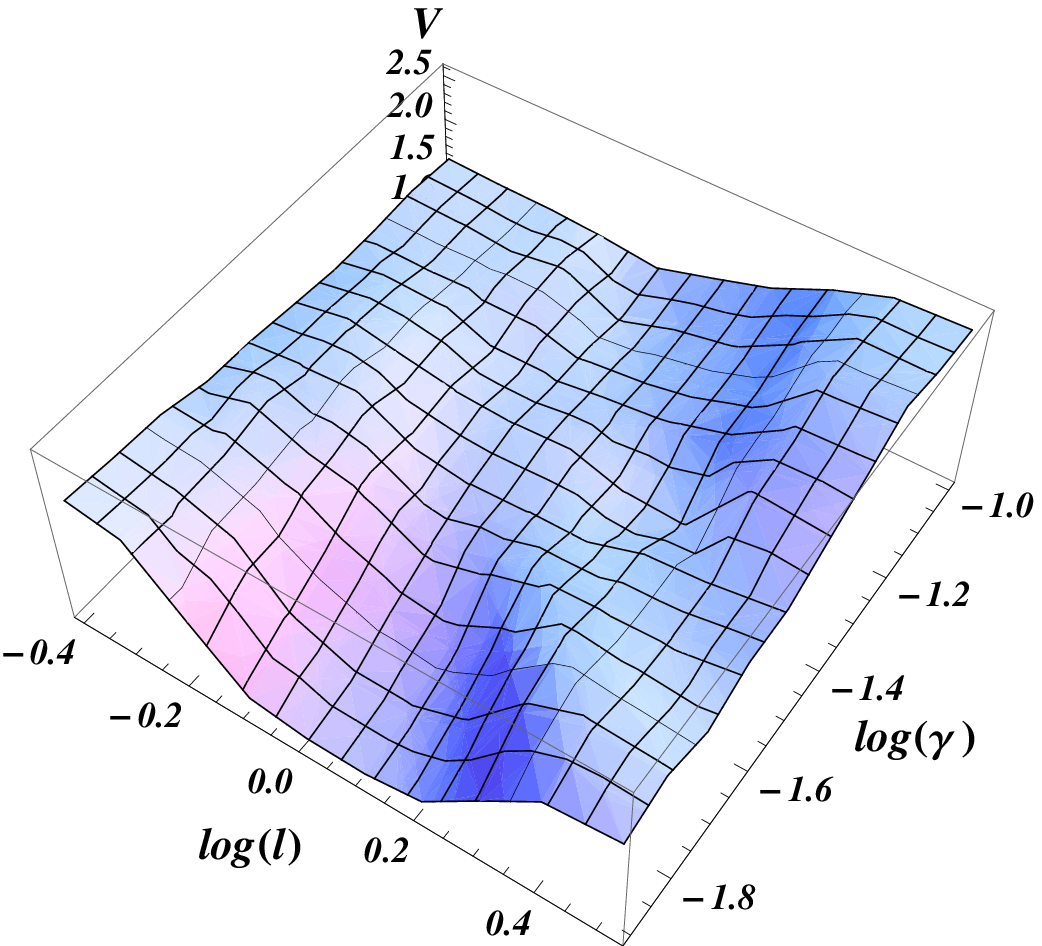}}
\end{center}
\end{figure}
\begin{figure}[t]
\begin{center}
\subfigure[Plots of the susceptibility $\chi$ as the function of
$\log(\ell)$ and $\log(\gamma)$]{\includegraphics[width=
2in]{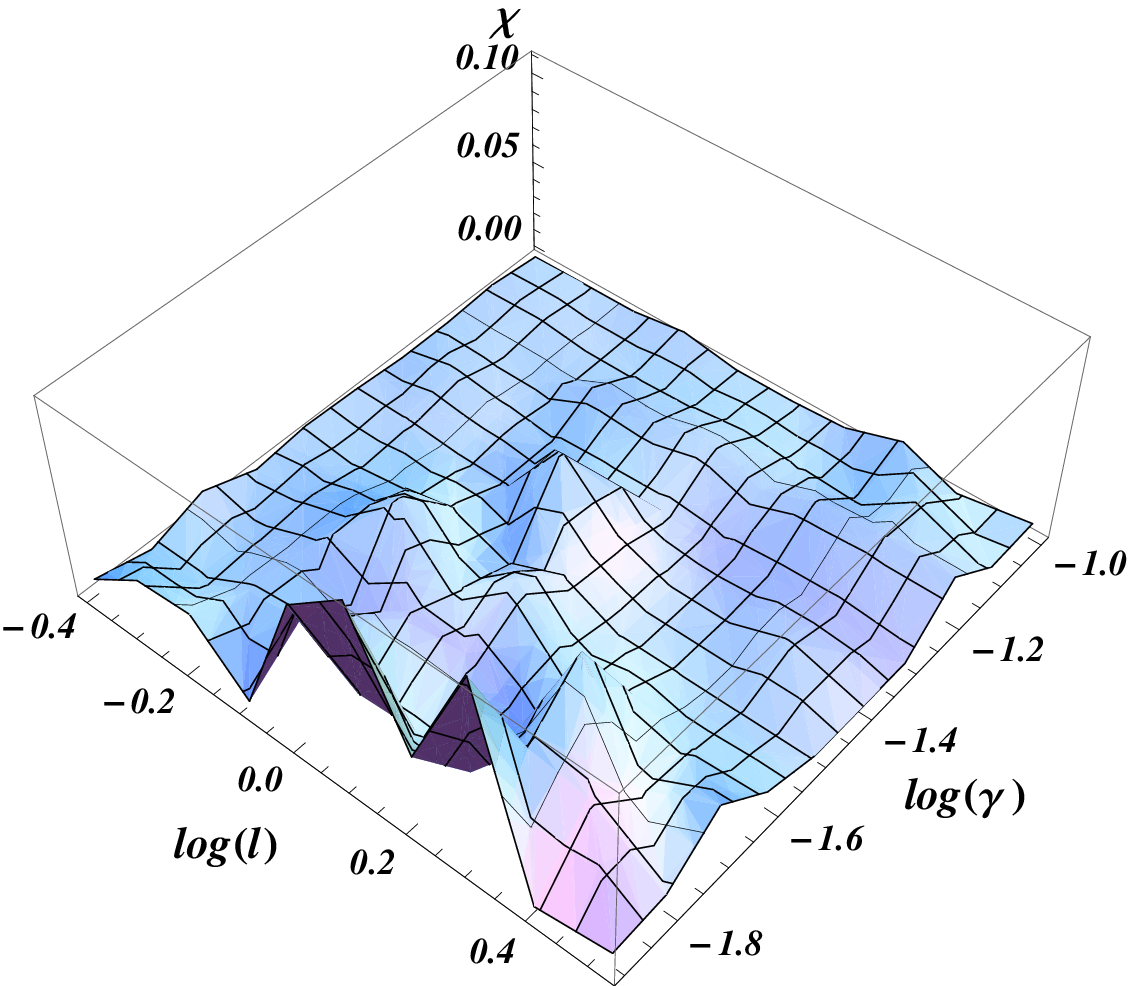}} \subfigure[The Shannon entropy $H$ as a
function of $\log(\ell)$ and
$\log(\gamma)$.]{\includegraphics[width=
2in]{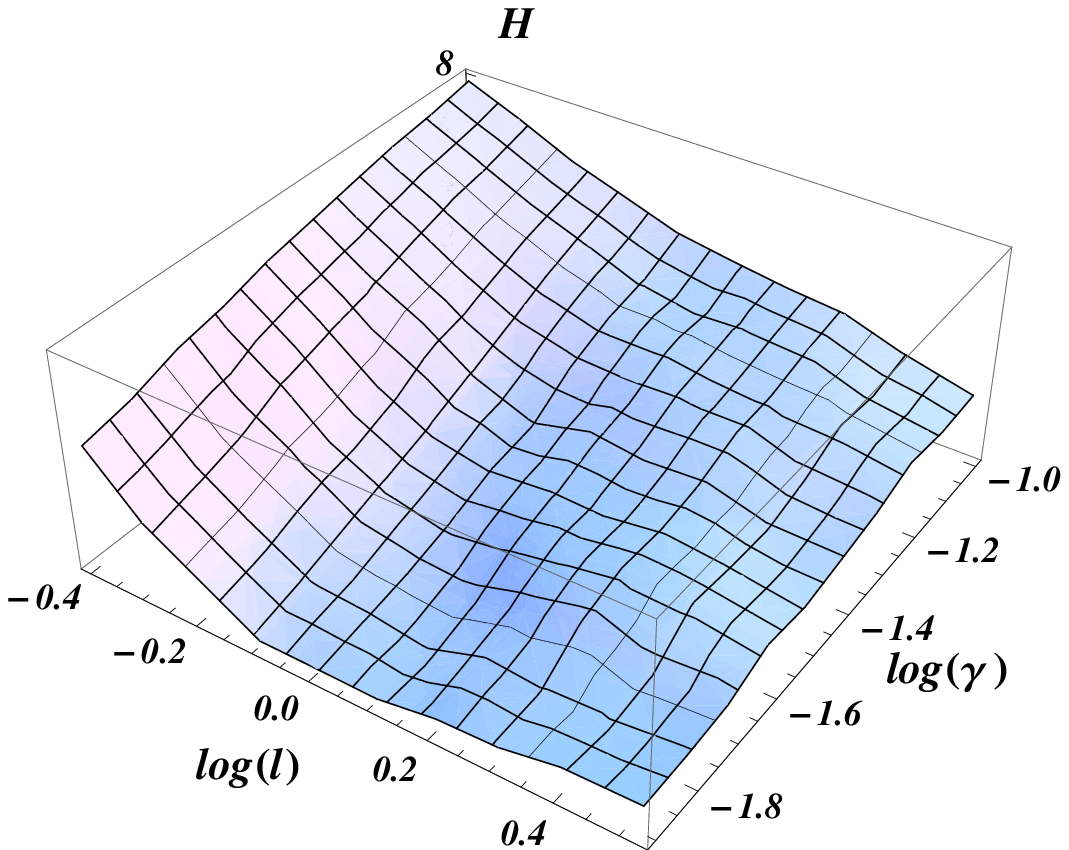}} \subfigure[The energy $E$ as a function
of $\log(\ell)$ and $\log(\gamma)$.]{\includegraphics[width=
2in]{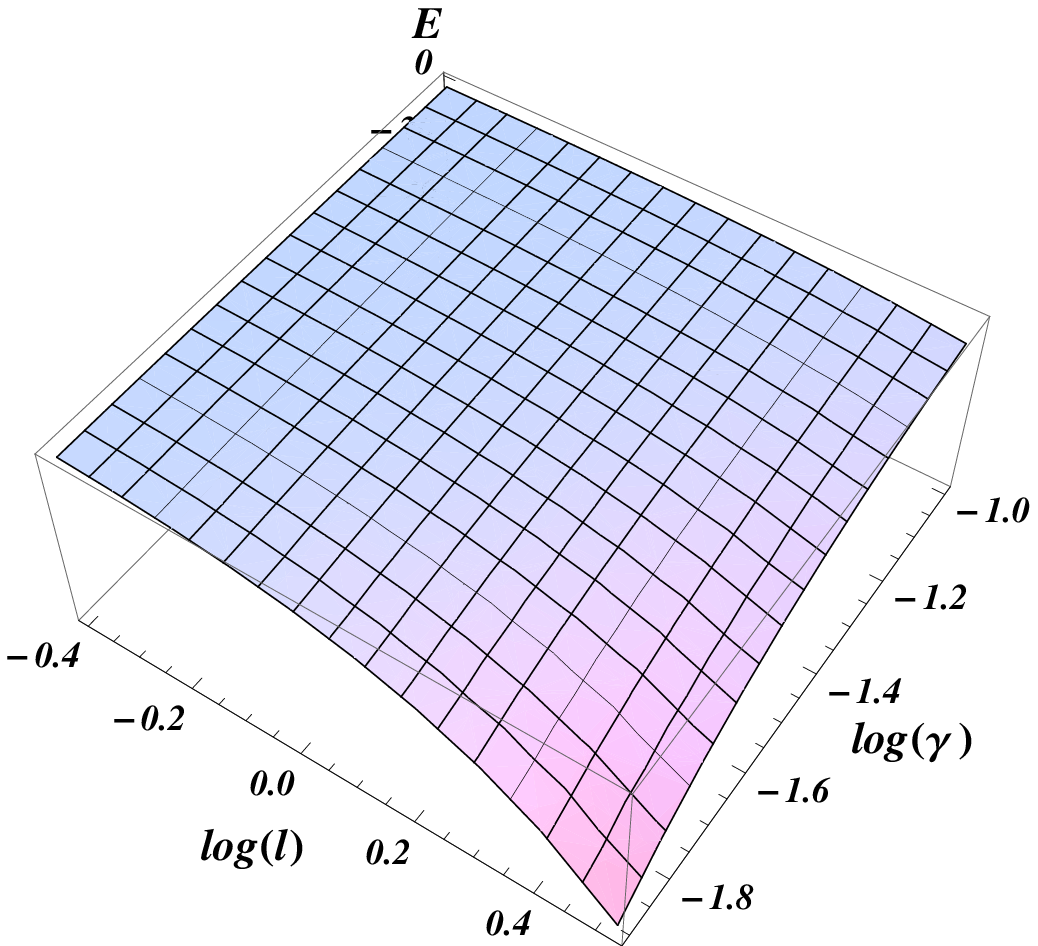}} \caption{Plots of $I_N$, $V$, $\chi$,
$H$ and energy $E$ as the function of $\log(\ell)$ and
$\log(\gamma)$ for the ``dalmatian dog'' image in
\figref{fig:dalmatian}.}\label{fig:3dharddog}
\end{center}
\end{figure}

\subsubsection{Dalmatian dog}\label{sec:dalmatian}

The camouflaged dalmatian dog in panel (c) of \figref{fig:dalmatian}
(and \figref{fig:hardzebradog}) is a particularly challenging image.
We invoke the method detailed in \secref{sec:average} to assign edge
weights. We then apply the multiresolution algorithm to ascertain
the length scale $\ell$ in \eqnref{eqn:vmn}. The inter-replica
averages of the variation of information $V$ and the normalized
mutual information $I_N$ are, respectively, shown in panels (a,b) and panels (d,e) of
\figref{fig:dalmatian}. These information theory overlaps indicate
that, as a function of $\ell$, there are, broadly, two different
regimes separated by a transition at $\ell \sim 1$.  We determine
the value of $\ell$ at the local information theory extremum
that is proximate to this transition and determine the edge weights set by
this value of $\ell$. (See Eq. (\ref{eqn:vmn}).) In Section
\ref{sec:phase-dog}, we will illustrate how we may determine
an optimal value of $\ell$.

We segment the original image of the dalmatian dog via our
community detection algorithm as shown in panels (c) and (f) in
\figref{fig:dalmatian}. The result in panel (c) corresponds to a resolution of $\gamma=0.1$. The image
on the right in panel (c) is the superimposed image of our result and the
original image (on the left) at the particular length $\ell=0.63$.
The ``green'' color corresponds to the dalmatian dog. The method is able to
detect almost all the parts of the dog except the inclusion of
``shade'' noise under the body. The results in panel (f) correspond to a
resolution $\gamma=0.05$. The image on the left in panel (f) is the
superimposed image of our immediate running result and the original
one at the length $\ell_1=1$, which is close to the maximum of $V$ (and the
local minimum of $I_N$). On the right, we provide the result
for  $\ell_2=1.29$ (a value of $\ell$ corresponding to a maximum of
$V$ and a minimum of $I_N$).
The ``purple'' color in the segmented image corresponds to the dalmatian dog. We are
able to detect the body and the two legs in the back, even though
with some ``bleeding''. As we will discuss in the next subsection,
it is possible to relate the contending solutions found in Fig.(\ref{fig:dalmatian})
for different values of $\gamma$ and $\ell$ to the character of the phase diagram.

\subsection{Phase Diagram}\label{sec:phasediagram}

As previously alluded to in \secref{sec:complexity}, we investigated numerically
the phase diagram and the character of the
transitions of the community detection problem for general graphs in \cite{dandan}.
From this, we were
able to distinguish between the ``easy'', ``hard'' and ``unsolvable''
phases as well as additional transitions within contending solutions
within these phases (e.g., our discussion in Section \ref{mra_ex}).
Strictly speaking, of course, different phases appear
only in the thermodynamic limit of a large number of nodes (i.e., $N \to \infty$).
Nevertheless, for large enough systems ($N \gg 1$), different phases are, essentially, manifest.
As we will now illustrate, the analysis of the phase
diagram enables the determination of the optimal parameters
for the image segmentation problem. To make this connection
lucid, we will, in this section, detail the phase diagrams
of several of the images that we analyzed thus far.

\subsubsection{Phase diagram of the Potts model corresponding to the dalmatian dog image}
\label{sec:phase-dog}

We will now analyze the thermodynamic and information theory measures
as they pertain to the dalmatian dog image
(\figref{fig:dalmatian}) for a range of parameters.
In a disparate analysis, in subsection \ref{easy-bird}, we will extend this approach also to finite
temperature (i.e., $T>0$) where a heat bath algorithm was employed.
Here, we will content ourselves with the study of the zero temperature
case that we have focused on thus far.

Plots of the normalized mutual information $I_N$, variation of
information $V$, susceptibility $\chi$, entropy $H$, and the energy $E$
are displayed in \figref{fig:3dharddog}. We set the background intensity to $\bar{V}=15$.
The block size is $L_x\times L_y=11\times 11$. We then varied the resolution
$\gamma$ and the spatial scale $\ell$ within a domain
given by $\gamma \in [0.01,0.1]$ and $\ell \in [0.4,4]$.
In \figref{fig:3dharddog}, all logarithms are in the common basis (i.e., $\log_{10}$).

Several local extrema are manifest in \figref{fig:3dharddog}. In the context of the data to be presented
below, the quantity $Q$ of \eqnref{QZ} can be $I_N$,
$V$, $\chi$, $H$ or $E$, and $z$ may be $\gamma$ or $\ell$.
Examining the squares of the gradients of these quantities, as depicted in
\figref{fig:gradientharddog}, aids the identification of more sharply defined extrema
and broad regions of the parameter space that correspond to different phases.

In \figref{fig:gradientharddog}, we compute the squares
of the gradients of $I_N$, $V$, $\chi$, $H$ and $E$ in panels (a) through
(e). Panel (f) shows the sum of the squares of the gradients of
$I_N$, $V$ and $\chi$. A red dot denotes parameters for a ``good''
image segmentation with the parameter pair being
$(\gamma,\ell)=(0.05,1)$ (or $(\log(\gamma),\log(\ell))=(-1.3,0)$
corresponding to the left hand segmentation in panel (f) of \figref{fig:dalmatian}). Clearly, the
red dot is located at the local minimum in each panel. This
establishes the correspondence between the optimal parameters and
the general structure of the information theoretic and thermodynamic
quantities.

As evinced in \figref{fig:gradientharddog}, there is a local single
minimum which is surrounded by several peaks in the 3D plots of the
squares of the gradients of $I_N$, $V$ (panel(a),(b)) and their sum
(panel (f)). For the dalmatian dog
image (\figref{fig:dalmatian}, setting $Q$ in \eqnref{QZ} to be the square of the gradients
efficiently locates optimal parameters. Note that the other contending solutions in
Fig. (\ref{fig:dalmatian}) relate naturally to the one at $\gamma = 0.05$ and $\ell = 1$.
The $\ell = 1.29$ (i.e., $\log(\ell) = 0.11$) solution on the right hand
side of panel (f) appears in the same ``basin'' as that
of the $\ell =1$ solution. Indeed, both segmentations
of panel (f) of Fig. (\ref{fig:dalmatian})
share similar features. By contrast, the $\gamma=0.1$ and $\ell =0.63$
(i.e., ($\log(\gamma) = -1, \log(\ell) = -0.2$))
segmentation result of panel (c) in Fig.(\ref{fig:dalmatian}) relates
to a different region.

\begin{figure}[t]
\begin{center}
\subfigure[The square of the gradient of $I_N$ (panel (a) of
\figref{fig:3dharddog}) as a function of $\log(\ell)$ and
$\log(\gamma)$. ]{\includegraphics[width=
2in]{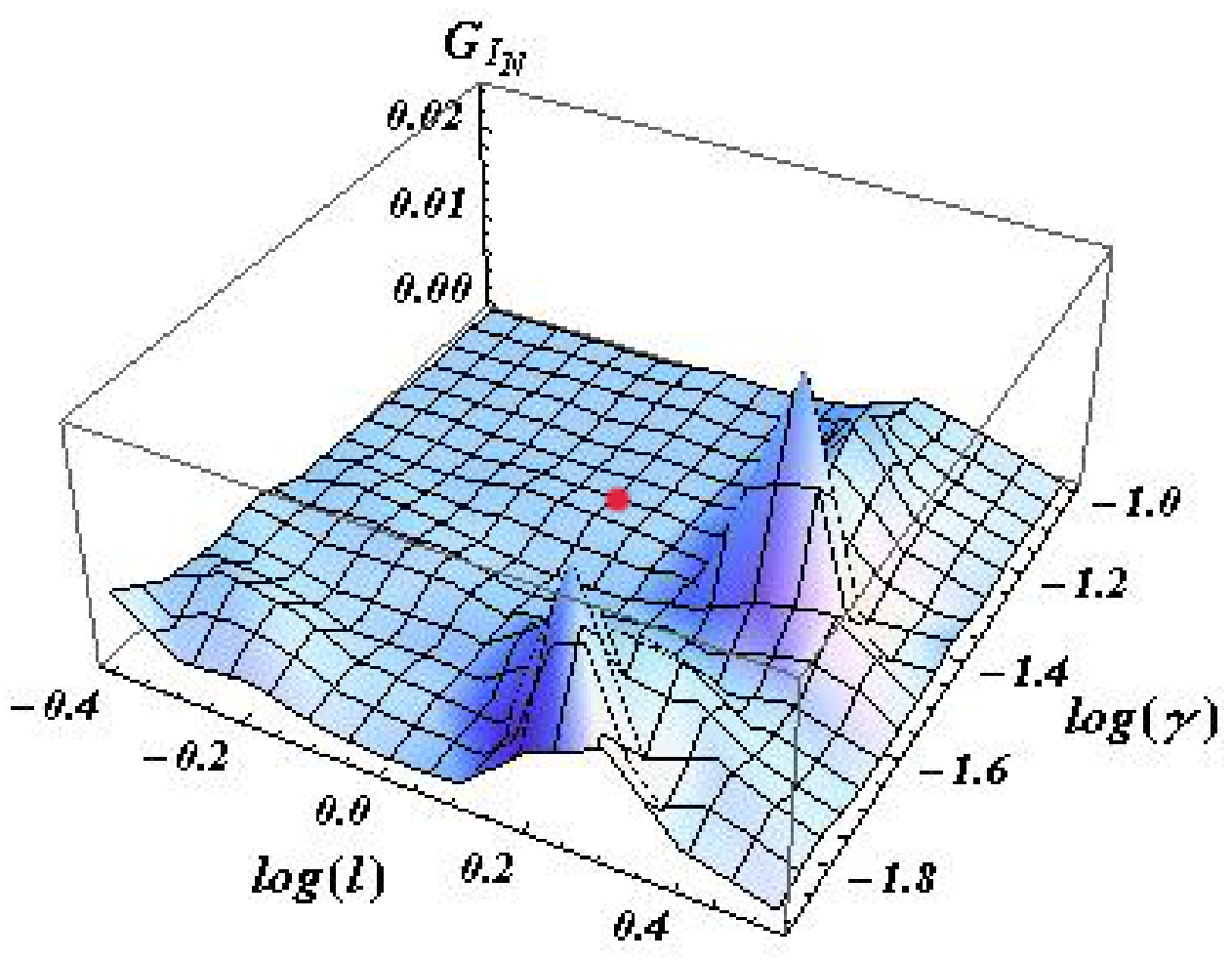}} \subfigure[The square of the
gradient of $V$ (panel (b) of \figref{fig:3dharddog}) as a function
of $\log(\ell)$ and $\log(\gamma)$. ]{\includegraphics[width=
2in]{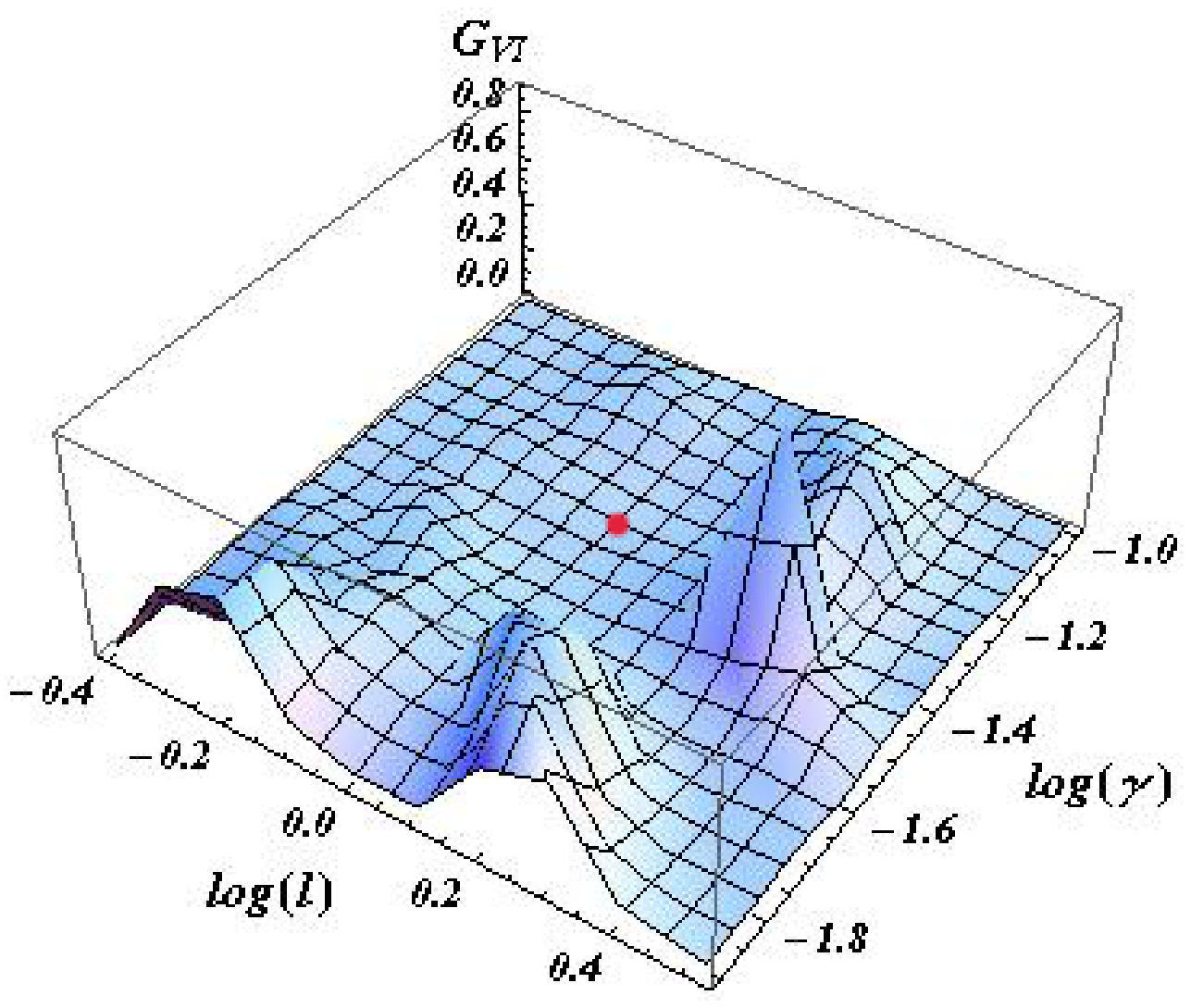}} \subfigure[The square of the
gradient of $\chi$ (panel(c) of \figref{fig:3dharddog}) as a
function of $\log(\ell)$ and $\log(\gamma)$.
]{\includegraphics[width= 2in]{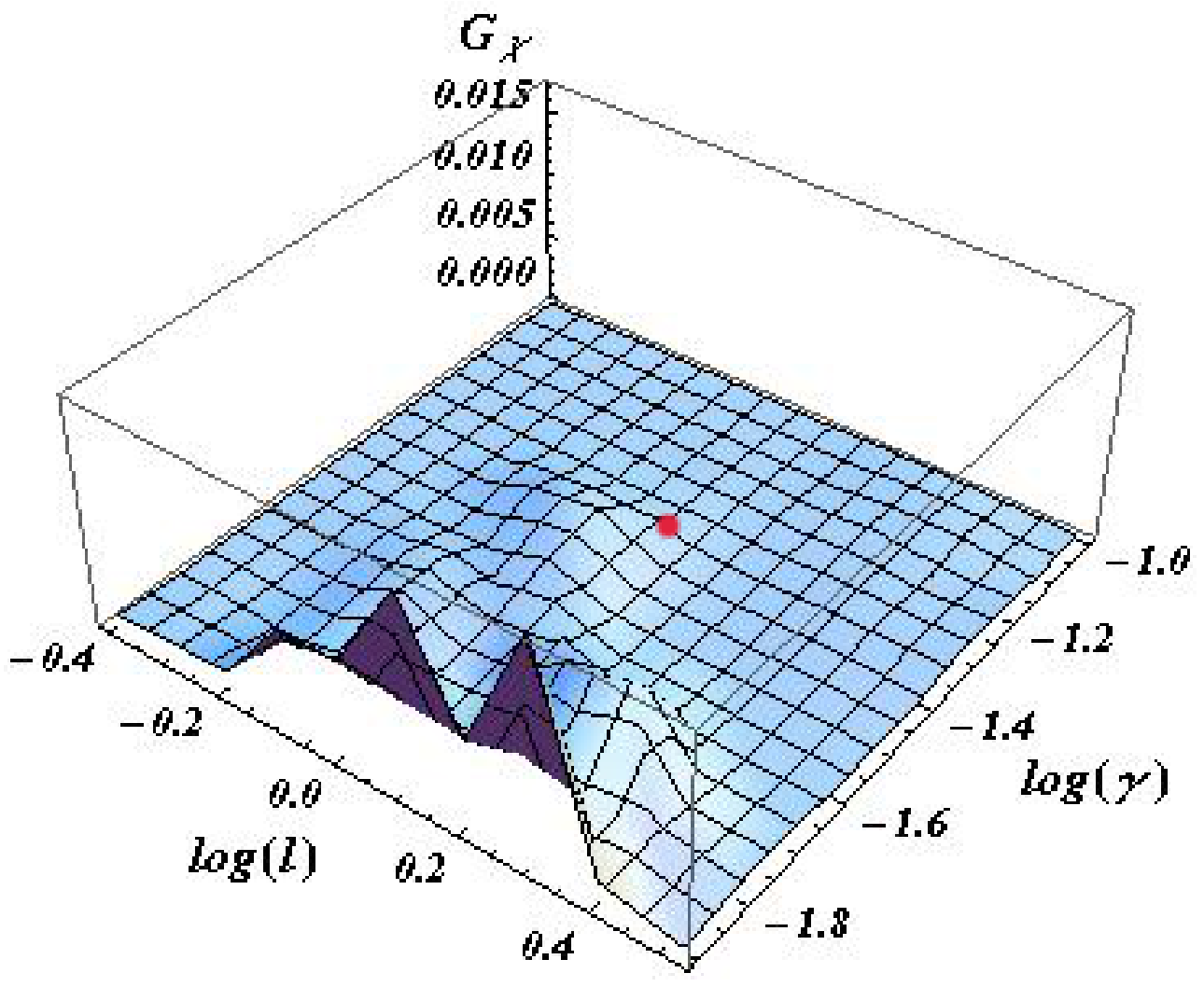}}
\end{center}
\end{figure}
\begin{figure}[t]
\begin{center}
\subfigure[ The square of the gradient of $H$ (panel (d) of
\figref{fig:3dharddog}) as the function of $\log(\ell)$ and
$\log(\gamma)$. ]{\includegraphics[width=
2in]{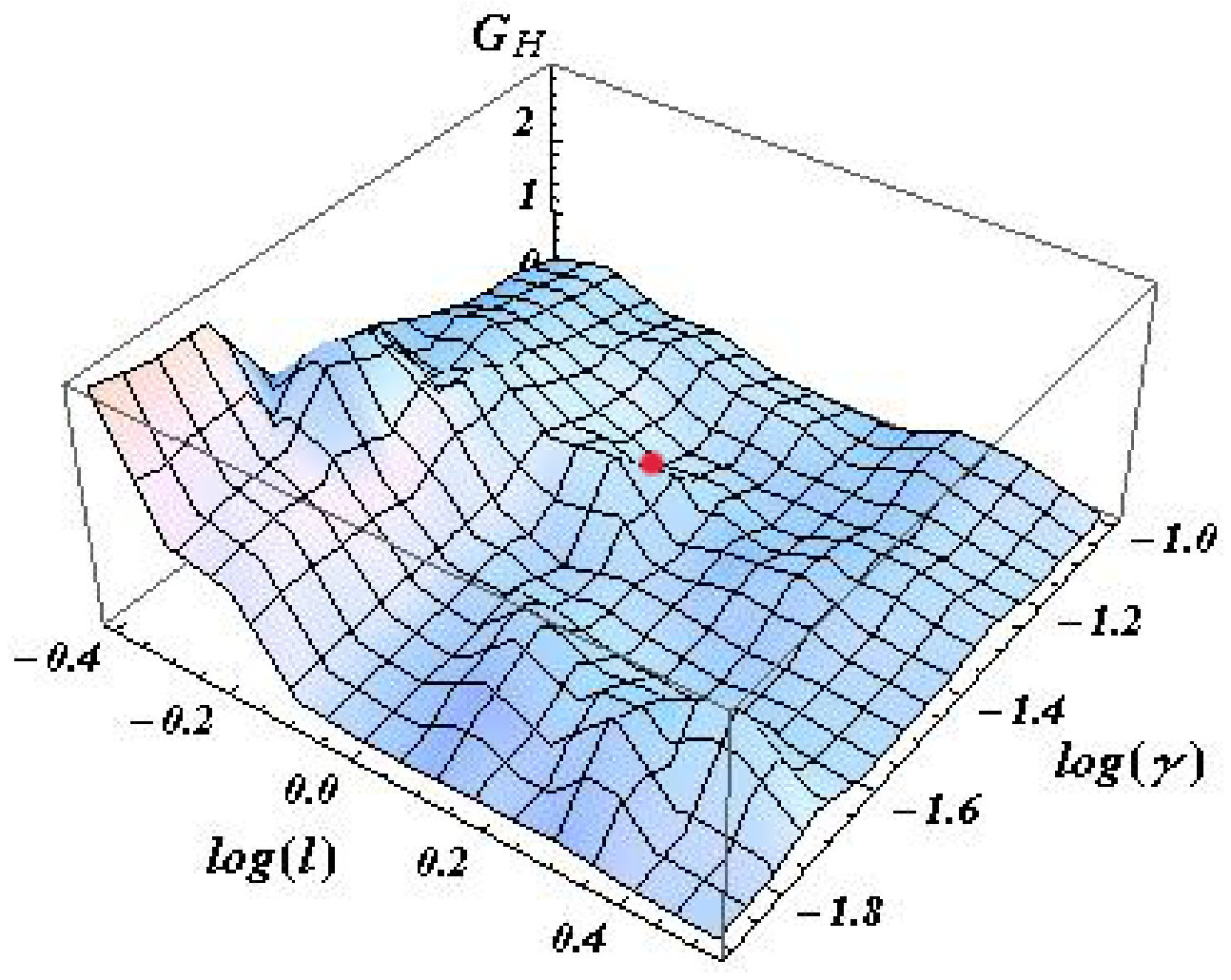}} \subfigure[The square of the
gradient of $E$ (panel (e) of \figref{fig:3dharddog}) as the
function of $\log(\ell)$ and $\log(\gamma)$.
]{\includegraphics[width= 2in]{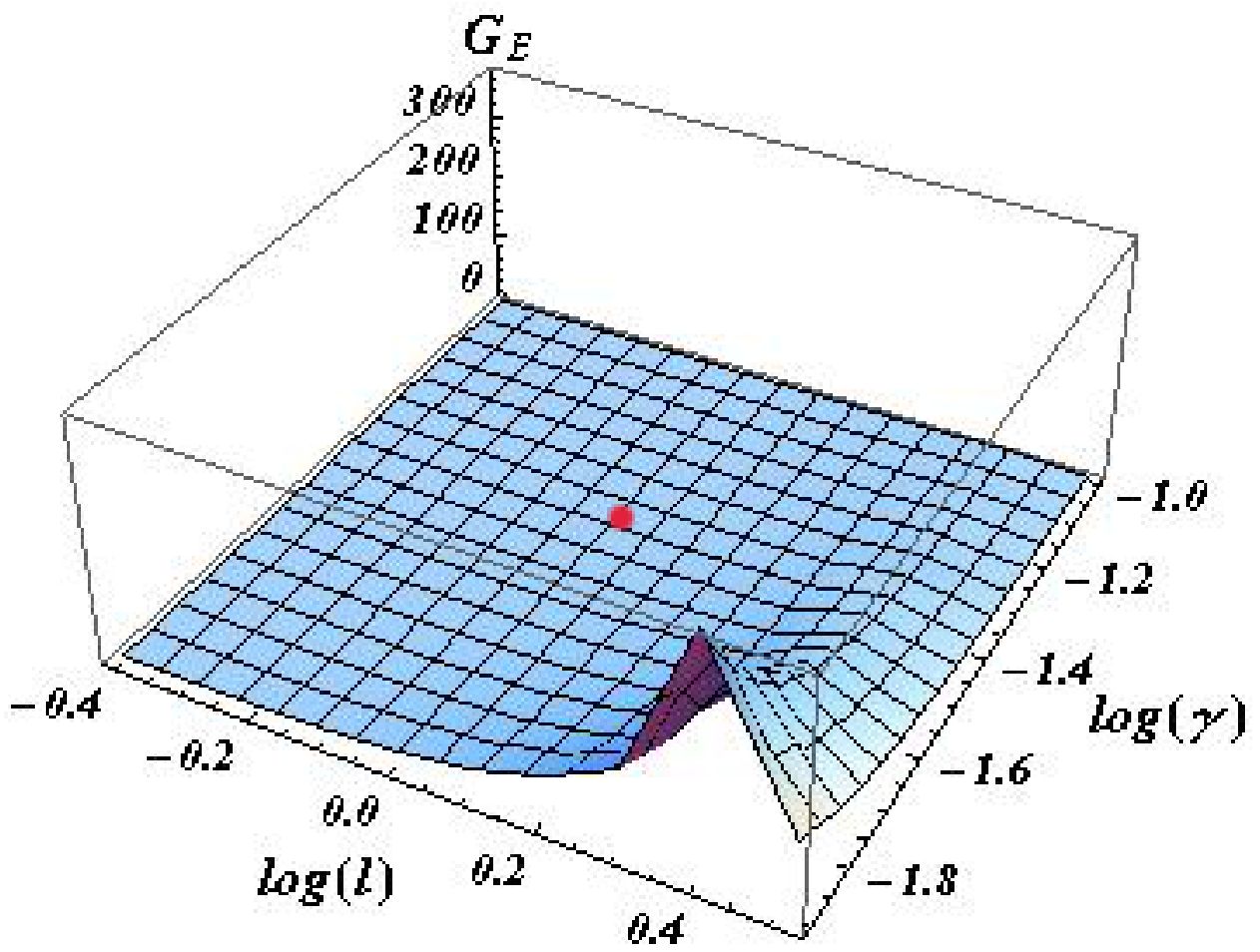}}
\subfigure[The sum of the squares of the gradients of $I_N$, $V$ and
$\chi$ (panel (a),(b) and (c) of \figref{fig:3dharddog}) as the
function of $\log(\ell)$ and
$\log(\gamma)$.]{\includegraphics[width=
2in]{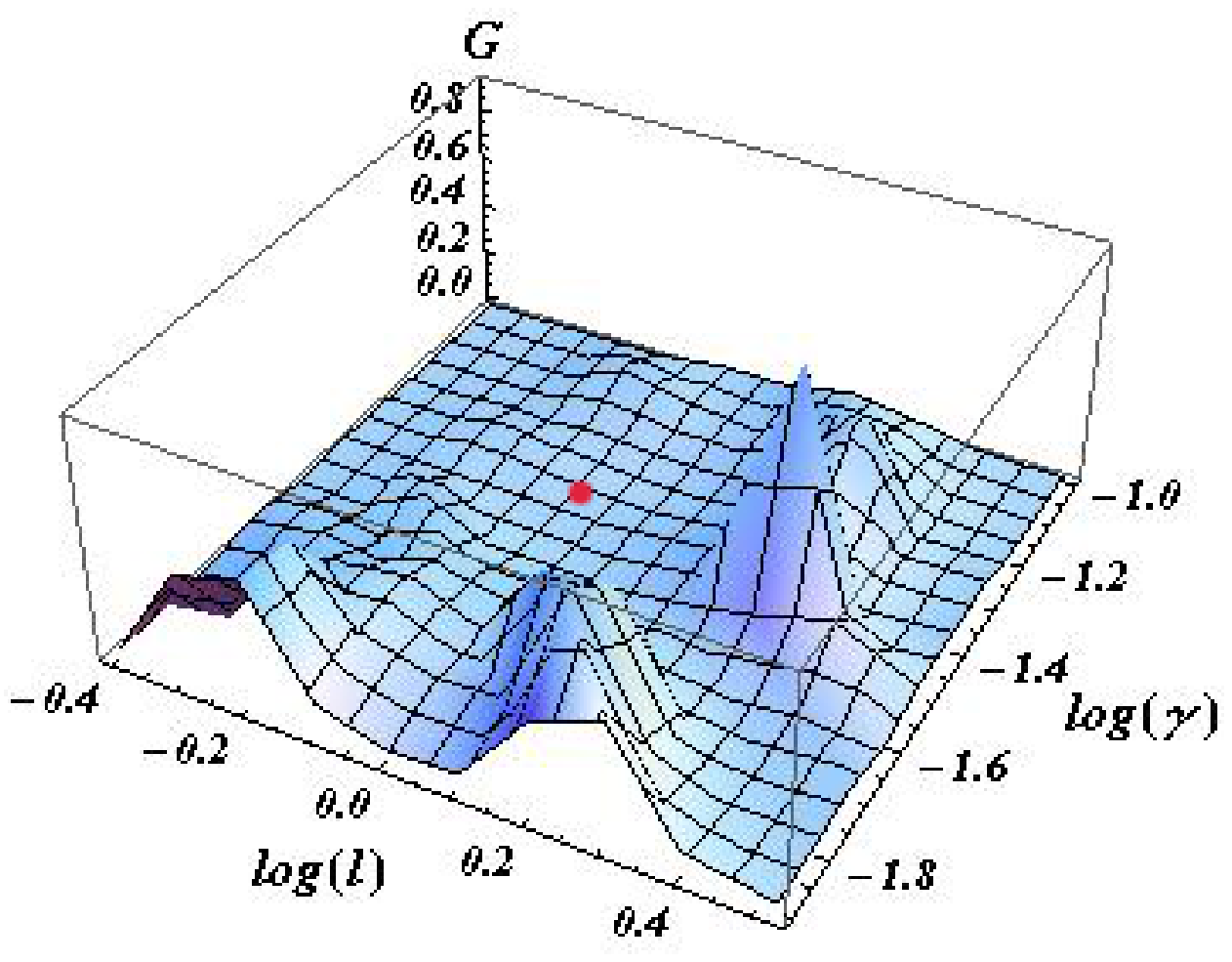}} \caption{Information theory and
thermodynamic measures relating to the dalmatian dog image of Fig.
\ref{fig:dalmatian}. The squares of the gradient of $I_N$, $V$,
$\chi$, $H$, $E$ (panel (a)-(e)) and the sum of the squares of the
gradients of $I_N$, $V$ and $\chi$ (panel (f)) as the function of
$\log(\ell)$ and $\log(\gamma)$. The red dot in each panel denotes
the location of the parameters
($(\log(\ell)$,$\log(\gamma))=(0,-1.3)$ (i.e., $(\ell, \gamma = (1,
0.05)$) of the results in \figref{fig:dalmatian}. This good
segmentation found for these parameters correlates with a local
minimum within each panel. }\label{fig:gradientharddog}
\end{center}
\end{figure}

\subsubsection{A finite temperature phase diagram}
\label{easy-bird}

\begin{figure}[t]
\begin{center}
\subfigure[The normalized mutual information $I_N$ as a function of
the resolution $\log(\gamma)$ and temperature $T$.
]{\includegraphics[width= 2in]{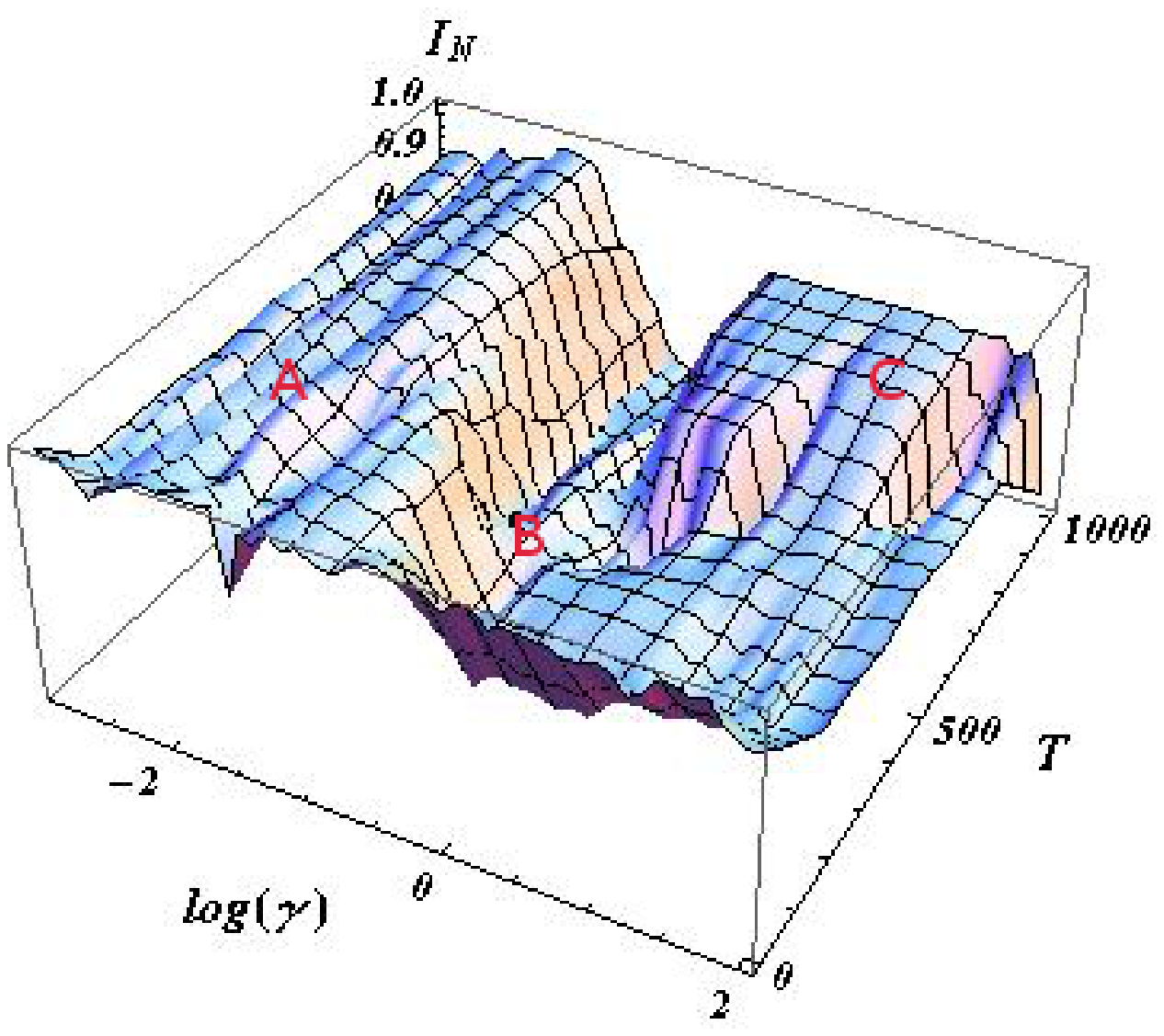}} \subfigure[The
variation of information $V$ as the function of the resolution
$\log(\gamma)$ and temperature $T$. ]{\includegraphics[width=
2in]{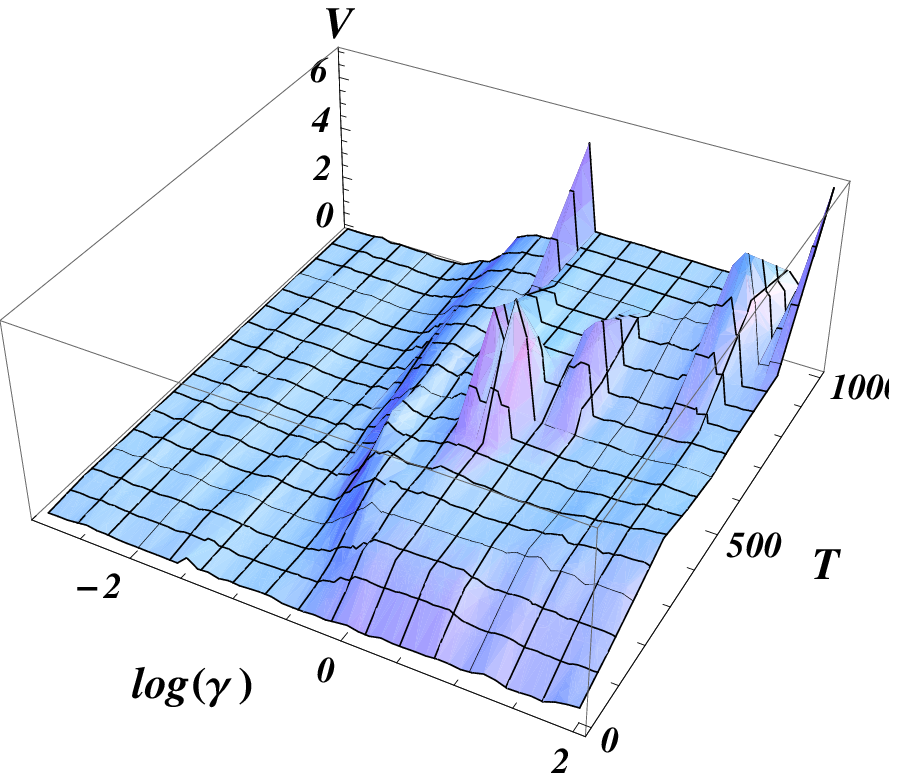}} \subfigure[The susceptibility $\chi$ as the
function of the resolution $\log(\gamma)$ and temperature $T$.
]{\includegraphics[width= 2in]{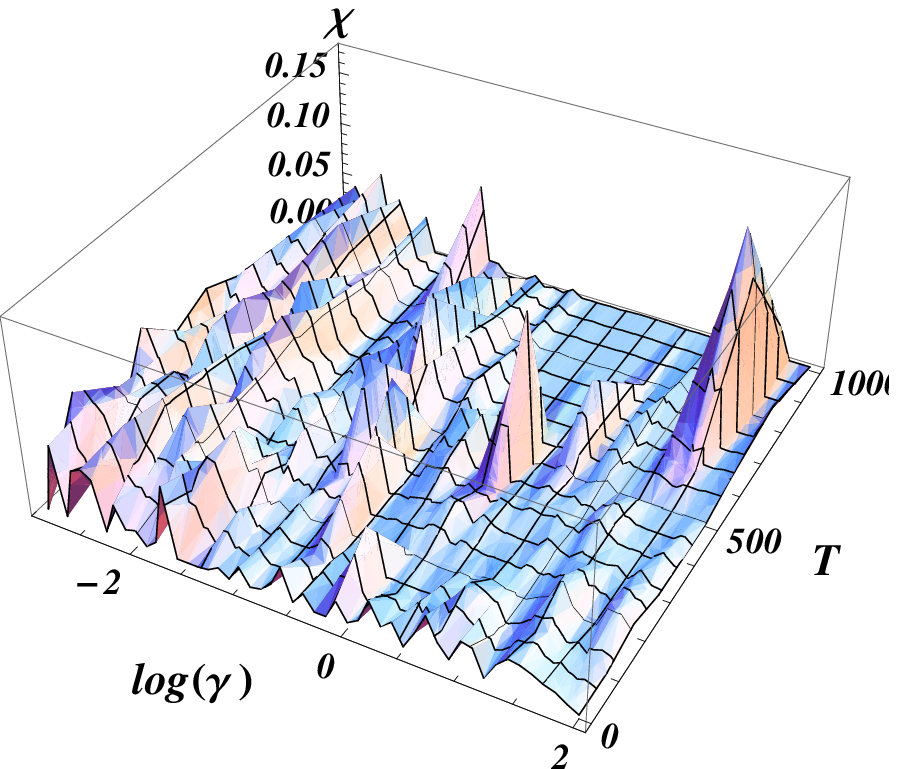}}
\end{center}
\end{figure}
\begin{figure}[t]
\begin{center}
\subfigure[The energy $E$ as a function of the resolution
$\log(\gamma)$ and temperature $T$.]{\includegraphics[width=
2in]{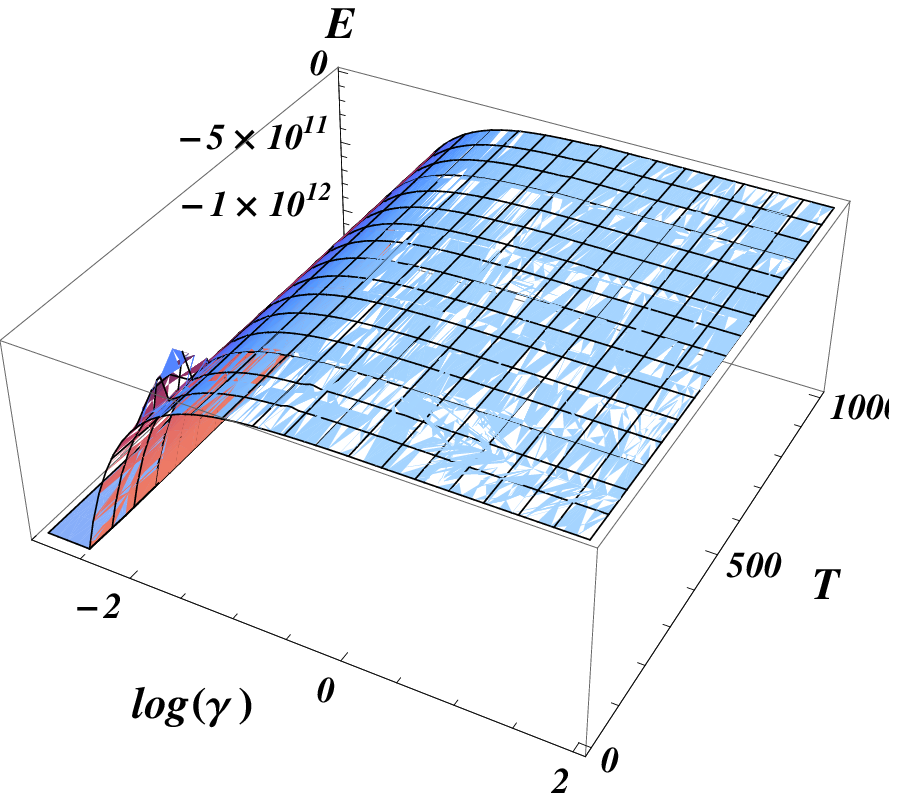}} \subfigure[The Shannon entropy $H$ as the
function of the resolution $\log(\gamma)$ and temperature
$T$.]{\includegraphics[width= 2in]{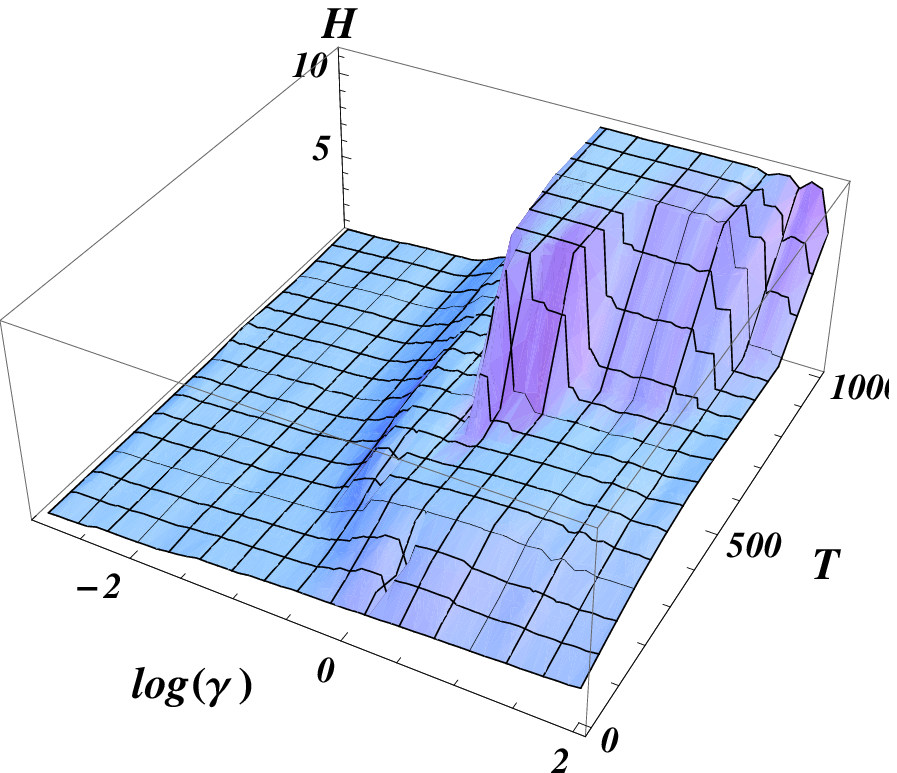}} \caption{The
normalized mutual information $I_N$, variation of information $V$,
susceptibility $\chi$, energy $E$ and Shannon entropy $H$ as the
function of the resolution $\log(\gamma)$ and temperature $T$ for
the ``bird'' image in \figref{fig:birdABC}. In panel (a), we mark
(i) the ``easy'' phase (where $I_N$ is almost $1$) as ``A'', (ii)
the ``hard'' phase (where $I_N$ decreases) by ``B'', and (iii)
denote the ``unsolvable'' phase (where $I_N$ forms a plateau whose
value is less than $1$) by ``C''. The physical character of the
``easy'', ``hard'', and ``unsolvable'' phases is further evinced by
the corresponding image segmentation results in
\figref{fig:birdABC}. We can determine the signatures of the three
phases in all panels apart from panel(c)-the 3d plot of the
susceptibility $\chi$. }\label{fig:3dbird}
\end{center}
\end{figure}

\myfig{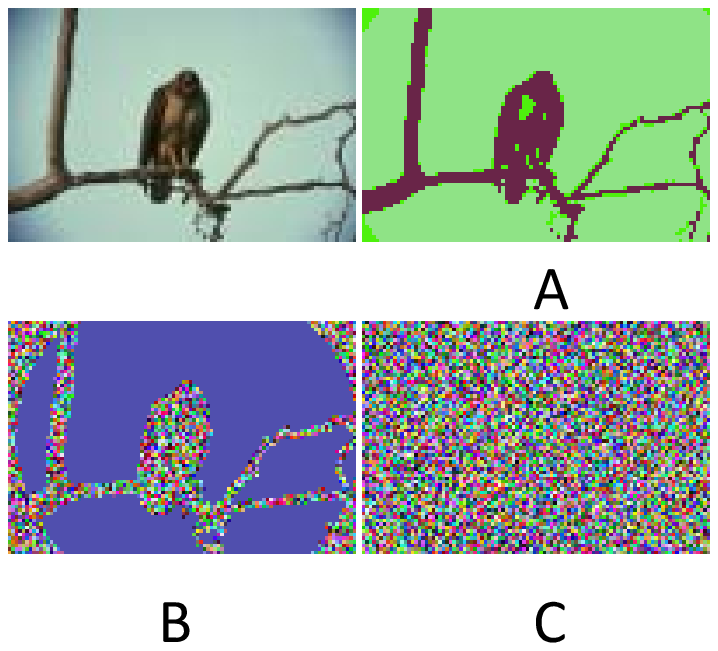}{[Color Online.] The image segmentation results of
the ``bird'' image. The original image is on the upper left. The
segmentations denoted by ``A'', ``B'' and ``C'' correspond to
results with different parameter pairs ($\log(\gamma)$, $T$) that
are marked in panel (a) of \figref{fig:3dbird}. Both results ``A''
and ``B'' are able to distinguish the ``bird'' from the
``background''. However, in panel (b), the ``bird'' is composed of
numerous of small clusters. The segmentation ``C''  does not detect
the ``bird''. The results shown here at points A, B, C correlate
with the corresponding ``easy-hard-unsolvable'' phases in the phase
diagram in \figref{fig:3dbird}.
 }{fig:birdABC}{0.8\linewidth}{}

\figref{fig:3dbird} depicts the finite temperature ($T>0$) phase diagram of the image
of the bird of \figref{fig:birdABC}. We will find that for this easy image,
the phase boundaries between the easy, hard, and unsolvable phases
of the image are relatively sharply defined.

In the context of the data to be presented, we fixed the background
intensity $\bar{V}=15$, set the block size to be $L_x\times
L_y=1\times 1$ and took the spatial scale $\ell \to \infty$. The
varying parameters are the resolution $\gamma$ and temperature $T$.
Instead of applying our community detection algorithm at zero
temperature, we will incorporate the finite temperature
\cite{dandan} in this section. The ranges of the $\gamma$ and $T$ values
are $[0.001,100]$ and $[0,1000]$ respectively. In the panels of
\figref{fig:3dbird}, we show the normalized mutual information
$I_N$, variation of information $V$, susceptibility $\chi$, energy
$E$ and Shannon entropy $H$ as the function of the temperature $T$ and the
logarithm of the resolution $\log(\gamma)$.

We can clearly distinguish the``easy'', ``hard'' and ``unsolvable'' phases from
the 3D plots of $I_N$ (panel (a)), $V$ (panel (b)) and $H$ (panel
(e)). The label``A'' in panel (a) marks the ``easy'' phase, where
$\gamma\in[0.001,0.3]$ for $T\in[0,500]$ and $\gamma\in[0.001,0.01]$
for $T\in[500,1000]$. The ``easy'' phase becomes narrower as
temperature increases. The corresponding image segmentation result
shown in \figref{fig:birdABC} validates the label of the ``easy'' phase. The ``A''
image in \figref{fig:birdABC} is obtained by running our community
detection algorithm with the parameter pairs located in the area
labeled by ``A'' in \figref{fig:3dbird}. The image segmentation denoted by ``A'' can
perfectly detect the bird and the background. The bird is essentially composed
of two clusters and the background forms one contiguous cluster.
This reflects the true composition of the original image on the
upper left. Thus, the bird image can be perfectly segmented in an unsupervised way
when choosing parameters to be in the ``A'' region (corresponding to the computationally
``easy'' phase ).

The region surrounding point ``B'' in panel (b) in
\figref{fig:3dbird} denotes the ``hard'' phase, where $\gamma$ is in
the range of $[0.3,100]$ and $T$ in the range of $[0,500]$. Within
the ``hard'' phase, as the corresponding image labeled by ``B'' in
\figref{fig:birdABC} illustrates, the bird is composed of numerous
small clusters with the background still forming one cluster.  In
this phase, the image segmentation becomes harder and some more
complicated objects cannot be detected.

The label ``C'' in panel (c) in \figref{fig:3dbird} denotes the ``unsolvable''
phase, where the range for $\gamma$ and $T$ is about $[0.1,100]$ and
$[500,1000]$ respectively. The corresponding image in
\figref{fig:birdABC} labeled by ``C'' is composed of numerous small clusters for which it
is virtually impossible to distinguish the bird from the background. In this phase, the normalized mutual information $I_N$ is far less than $1$ (indicating, as expected, the low quality of
segmentations).

Other 3D plots in \figref{fig:3dbird} generally show similar phase
transitions. Especially, the 3D entropy plot (panel (e))
vividly depicts accurate three phases and their clear boundaries.

\section{Conclusions}\label{sec:conclusion}

In summary, we applied a multi-scale replica inference based community
detection algorithm to address unsupervised image segmentation. The
resolution parameters can be adjusted to reveal the targets in
different levels of details determined by extrema and transitions.
In the images with uniform targets, we distributed edge weights
based on the color difference. For images with non-uniform targets,
we applied a Fourier transformation within blocks and assigned the edge
weights based on an overlap. Our image segmentation results were
shown to be, at least, as accurate as some of the best to date (see, e.g., Table.
\ref{table}) for images with both uniform and non-uniform targets.
The images analyzed in this work cover a wide range of categories:
animals, trees, flowers, cars, brain MRI images, etc. Our
algorithm is specially suited for the detection of camouflage
images. We illustrated the existence of the analogs of three
computational phases (``easy-hard-unsolvable'')
found in the satisfiability ($k-$SAT) problem \cite{mezard1,mezard2}
in the image segmentation problem as it was formulated
in our work. When the system exhibits a hierarchal or
general multi-scale structure, transitions further appear between
different contending solutions. With the aid of the structure of the
general phase diagram, optimal parameters
for the image segmentation analysis may be discerned. This general
approach of relating the thermodynamic phase diagram to
parameters to be used in an image segmentation analysis
is not limited to the particular Potts model formulation for unsupervised image
segmentation that was introduced in this work. In an upcoming work, we will illustrate how
supervised image segmentation with edge weights that
are inferred from a Bayesian analysis with prior probabilities
for various known patterns (or training sets),
can be addressed along similar lines \cite{prep}}.
We conclude with a speculation. It may well be that,
in real biological neural networks,
parameters are adjusted such that
the system is solvable for a generic expected
input and critically poised next to the boundaries
between different  contending solutions \cite{caudill}.
 \\

  {\bf{Software.}}\\
 The software package for the ``multi-resolution community detection" algorithm \cite{peter1}
 that was used in this work is available at http://www.physics.wustl.edu/zohar/communitydetection/.
\\

{\bf Acknowledgments}  \\
 We wish to thank S. Chakrabarty, R. Darst,
 P. Johnson, B. Leonard, A. Middleton, D. Reichman,
 V. Tran, and L. Zdeborova
 for discussions and ongoing work and are especially grateful
 to Xiaobai Sun (Duke) for a very careful reading
 of the manuscript, comments, and encouragement.

\appendix

\section*{Appendix A: Improved F-value by removing small high precision features}
\label{app-a}

As seen in \secref{sec:berkeley}, our results in the first three
images except the last one are better than the corresponding ones by
the best algorithm in the Berkeley Image Segmentation Benchmark. One
possible reason to cause the worse result in the last image is that
our algorithm is too accurate. For example, the top image in
\figref{fig:appendix}, our result could detect the small white
spray, which becomes the dots in the background. These small dots
will form small circles in the boundary image shown in the right
column, which are unexpected from the groundtruth, thus will reduce
the value of precision and $F$. (In this case, $F=0.56$.)

Merging these high precision small dots with the background as, e.g.,
fleshed out in the second row in \figref{fig:appendix}, leads to results
that are equivalent to or better than those determined by the algorithm of
global probability of boundary (gPb). A summary is presented in
Table. \ref{table2}.

\myfig{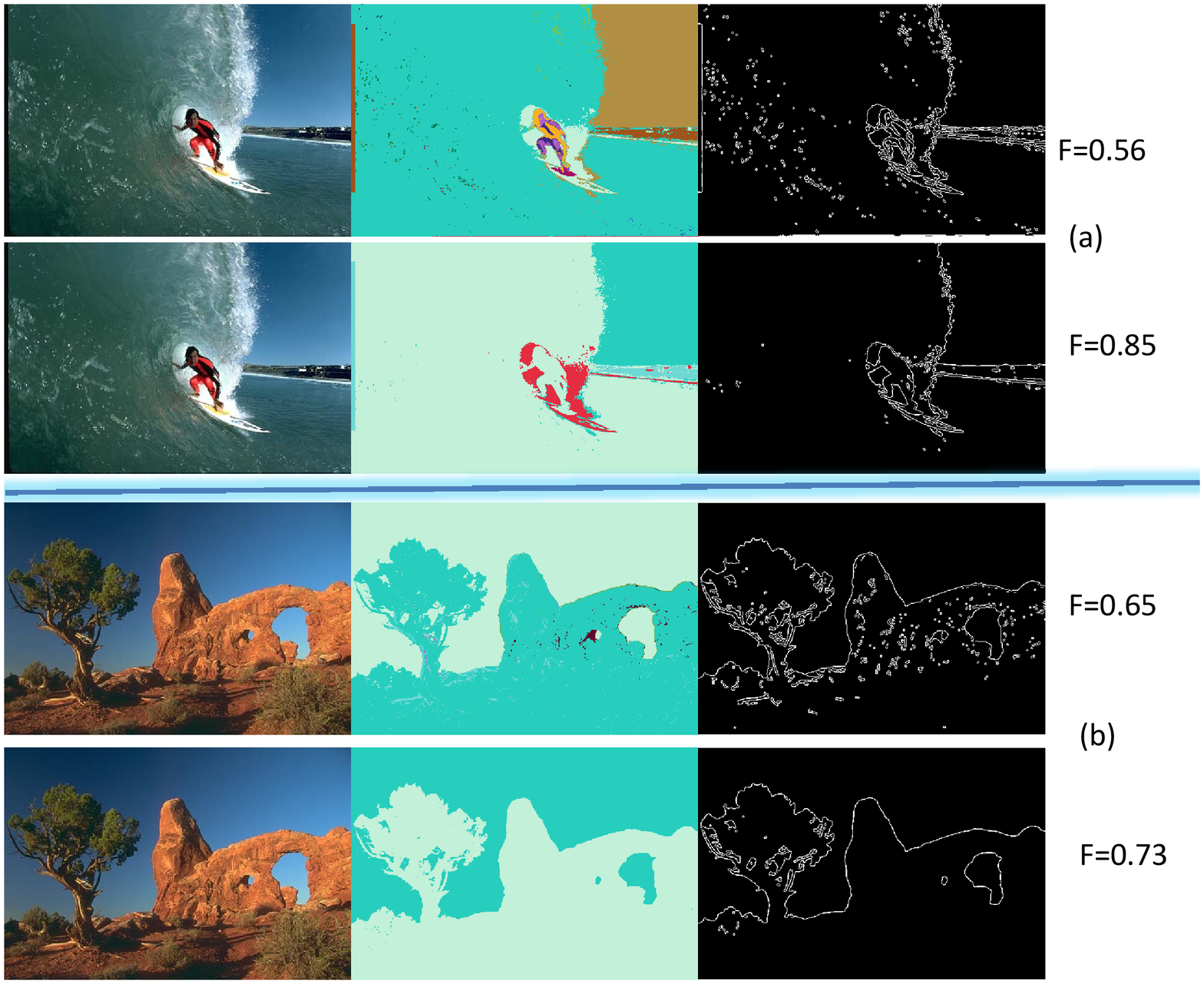}{[Color Online.] The image segmentation
results by our algorithm. The original images in the left most
column are downloaded from Berkeley Image segmentation benchmark.
The central image in the first row/ the third row is the result of
our algorithm at $\gamma=0.01$ and $\bar{V}=20$. The right image in
the first/the third row is the boundary detection result of the
corresponding central image by the software Mathematica. There are
many dots/circles which denote the white spray in original image in
the first row. The small dots/circles in the third row denote the
shadow in the original image. We merge these small dots in the first
and third row into the background and the results shown in the
second and fourth row are more smooth and close to the groundtruth.
This is confirmed by the larger F value shown in Table.
\ref{table2}.
 }{fig:appendix}{1\linewidth}{}

\section*{Appendix B: The image segmentation corresponding to the mutual information ($I_N$) peak }
\label{app-b}

 \begin{figure}[t]
\begin{center}
\subfigure[The curve of $I_N$ as a function of negative threshold
$-\bar{V}$ for the zebra image in panel (b).
]{\includegraphics[width= 3in]{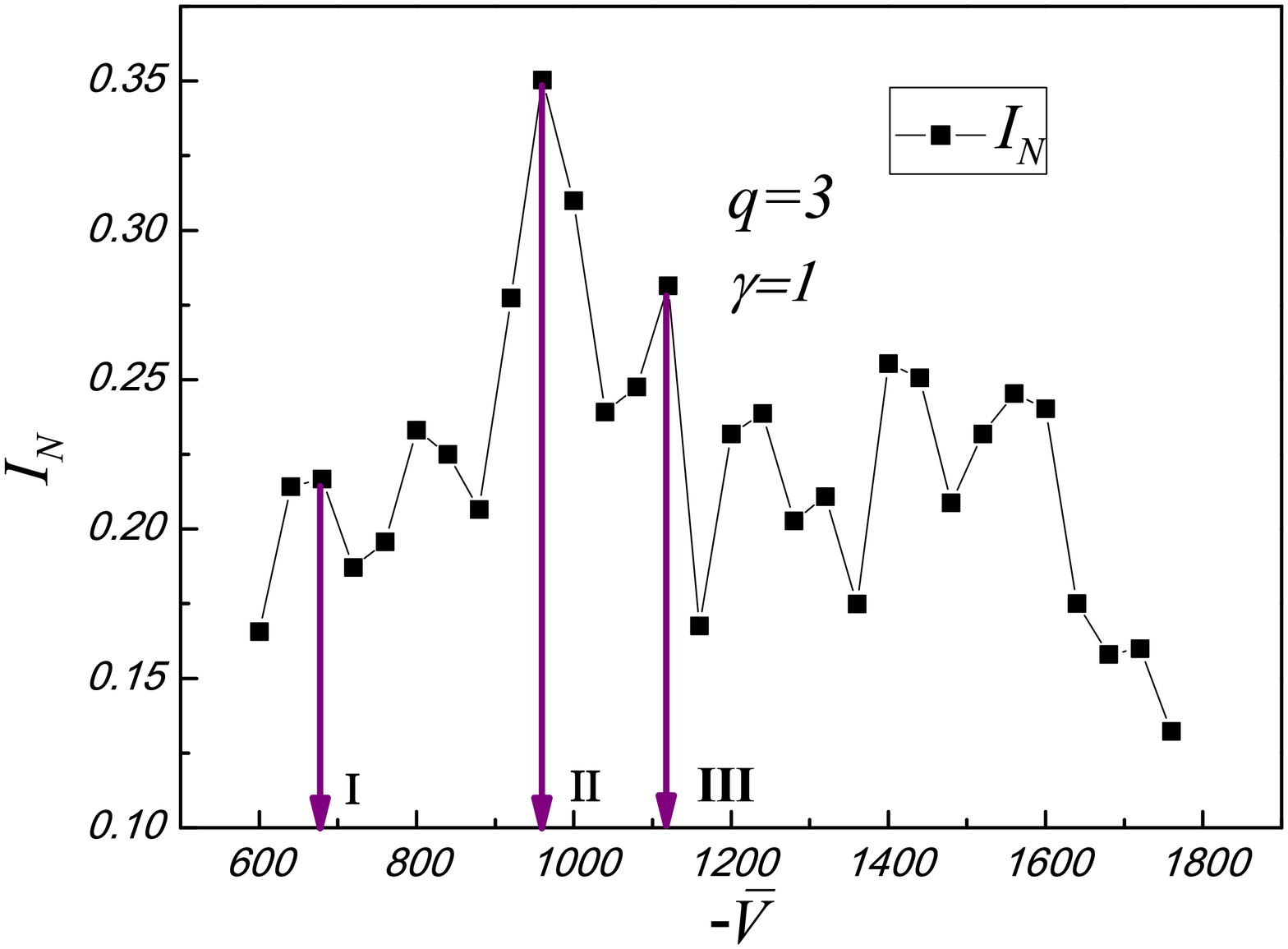}} \subfigure[The
weighted result of the zebra images at the corresponding thresholds:
$\bar{V}_1=-680$, $\bar{V}_2=-960$, and $\bar{V}_3=-1100$.
]{\includegraphics[width= 3in]{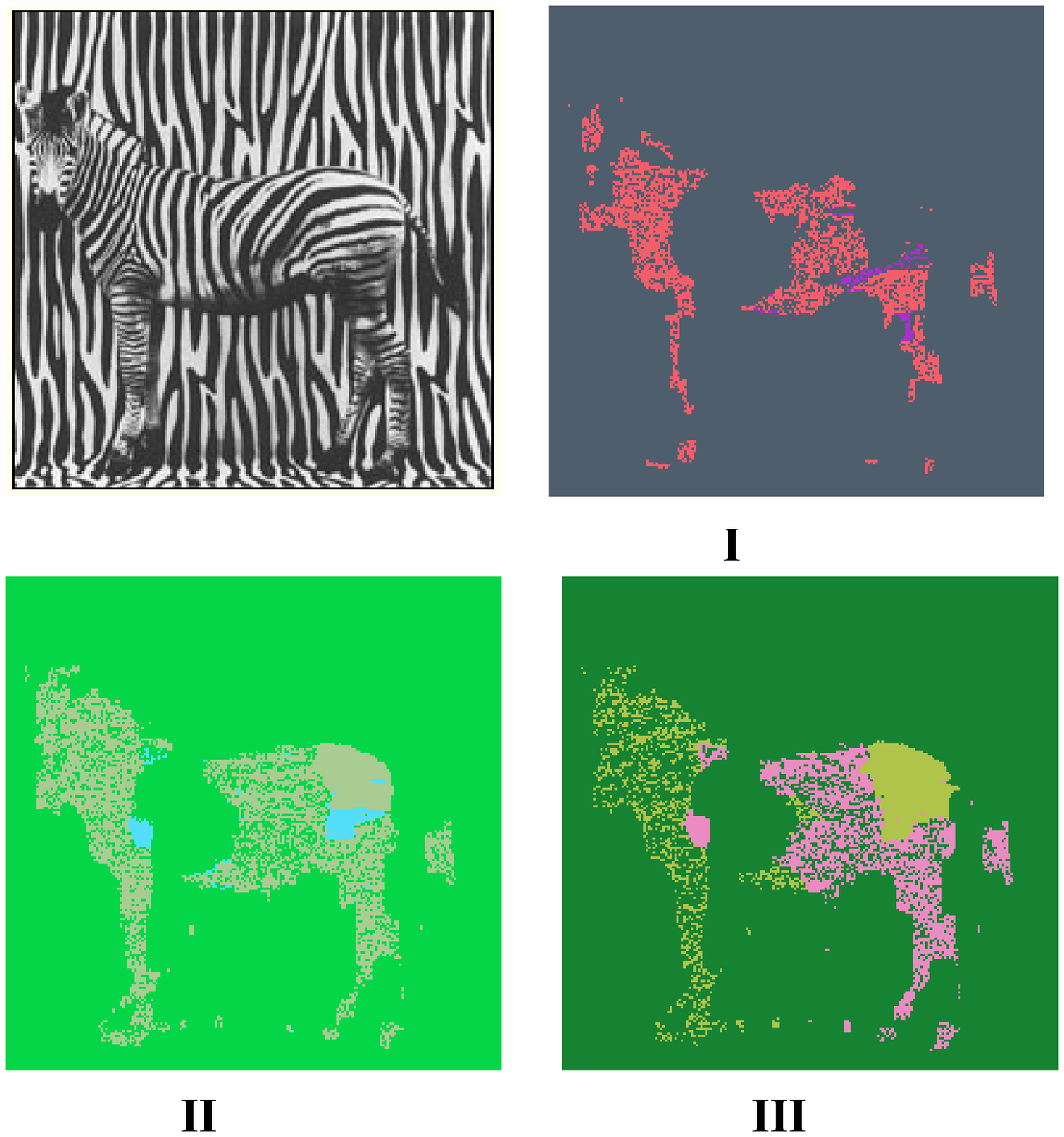}} \caption{[Color
Online] The ``multiresolution'' result of zebra with fixed community
number $q=3$ and resolution $\gamma=1$. In panel(a), we plot the
normalized mutual information $I_N$ as a function of negative
threshold $\bar{V}$. The peaks in $I_N$ also correspond to the
changes of structures. We choose three peaks and run the algorithm
at these three particular thresholds, and the result images are
shown in panel (b). As $|\bar{V}|$ increases, less regions in the
zebra merge to the background, and the boundary becomes more
clear.}\label{fig:zebra-hard2}
\end{center}
\end{figure}

As emphasized throughout this work, we focus on inter-replica
information theory overlap extrema. In some of the earlier examples,
we discussed the results pertaining to variation of information
maxima (often correlating with normalized mutual information
minima).  We now briefly discuss sample results for the normalized
mutual information maxima. We provide one such example in
\figref{fig:zebra-hard2}. Herein, we plot $I_N$ as a function of
$\gamma$ and provide the corresponding segmented images at the peaks
of $I_N$. As shown before, 
in panels I-III of \figref{fig:zebra-hard}, we provide the image
segmentation that correspond to the values of $\gamma$ for which the
variation of information $V$ exhibits a local maximum. In
\figref{fig:zebra-hard2}, we do the same for the normalized mutual
information $I_N$.

\section*{Appendix C: The image segmentation with negative and positive Fourier weight}\label{sec:appendixC}

\begin{table}
\begin{tabular}{|c|c|c|c|}
  \hline
   & F-Our algorithm & F-Our algorithm without noise & F-gPb \\ \hline
  a & 0.56 & 0.85 & 0.82 \\ \hline
  b & 0.65 & 0.73 & 0.74 \\ \hline
  \hline
\end{tabular}
\caption{The F-measure of the images shown in \figref{fig:appendix}.
We provide the comparison with the results by algorithm Global
Probability of Boundary (gPb). Note that after removing the small
dots/noise in both images, the value of F increase significantly.
After this merger, our results become equivalent to (or even better
than) the best results to date.}\label{table2}
\end{table}

In this brief appendix, we wish to compare results obtained with the
weights given by those of Eq. (\ref{VJ}) to those obtained when
$V_{ij}$ is set to be of the same magnitude as in Eq.(\ref{VJ}) but
of opposite sign (referred to below as its ``negative
counterpart''). In the latter case, a large weight $V_{ij}$
corresponds to a large overlap between patterns in blocks. Thus,
minimizing the Hamiltonian will tend to fragment a nearly uniform
background (for which the overlap between different blocks within is
large) and will tend to group together regions that change. The
results of the application of Eq. (\ref{VJ}) and that of its
negative counterpart are shown side by side in
\figref{fig:pos-neg-eq11} II and III. In both cases, the zebra is
successfully detected from the similar stripe-shaped background, as
long as using the right parameters. In (II), the parameters used are
as follows: the background $\bar{V}=-1200$, the resolution parameter is $\gamma=1$
and the block size is $l_x\times l_y=11\times 11$. In (III), we use
a positive background $\bar{V}=900$ but with a negative \eqnref{VJ},
resolution $\gamma=1$ and block size $l_x\times l_y=7\times7$. The
difference shown in \figref{fig:pos-neg-eq11} between result (II)
and (III) due to different fourier weights is that: In (II), the
background forms a large  cluster and the zebra is composed of lots of
small clusters. In (III), the zebra forms a large single community while the
background is composed of many small communities. For the images in
\figref{fig:fourier}, we substitute in Eq. (\ref{eq:newpotts}),
the weights of \eqnref{VJ} along with a negative background
$\bar{V}$.

\myfig{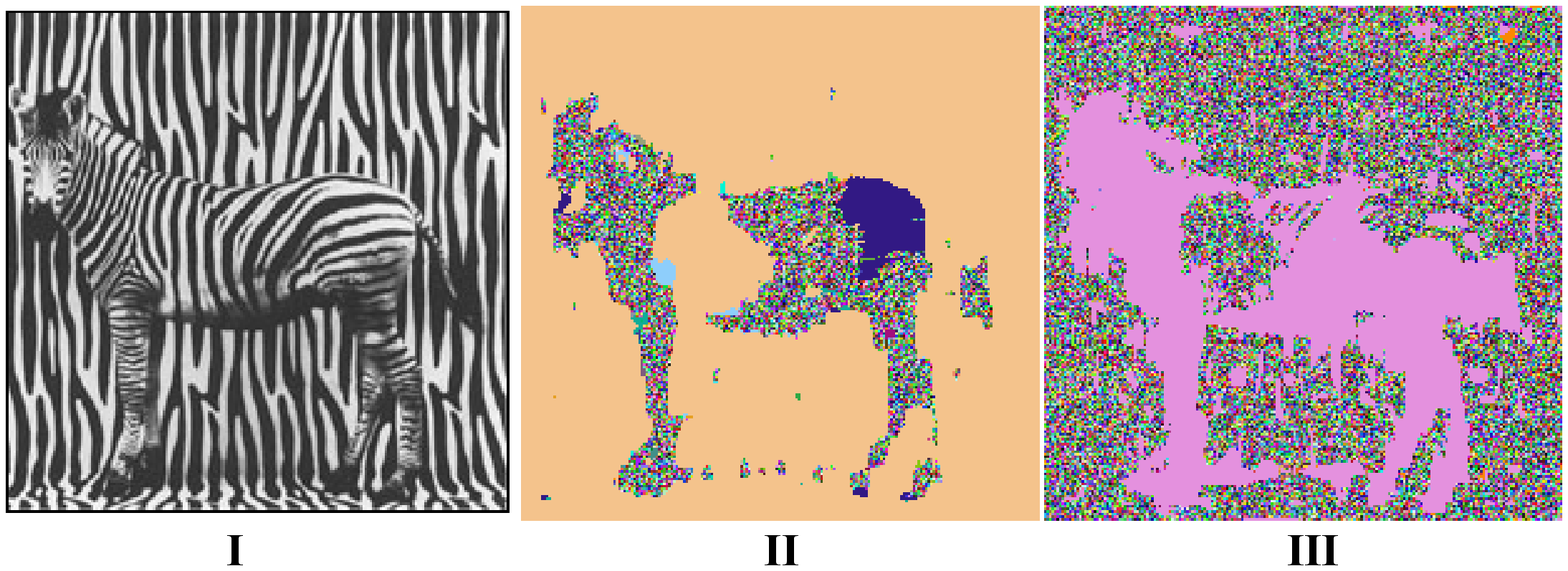}{[Color Online.] The image segmentation
results (II and III) of the original camouflaged zebra in (I). In
panel (II), we used the Fourier based edge weights of \eqnref{VJ}
and with a negative background $\bar{V}=-1200$ (Other parameters are
$\gamma=1$, block size $l_x\times l_y=11\times 11$). (III) The
resulting segmentation when the sign on the right hand side of
\eqnref{VJ} is flipped. Here, we applied a positive background
$\bar{V}=900$ (Other parameters are $\gamma=1$, block size
$l_x\times l_y=7\times7$). Both of the results shown here (i.e., II
and III) are able to detect the zebra.
 }{fig:pos-neg-eq11}{1\linewidth}{}


\begin{thebibliography}{99}
\bibitem{explain_solute} Rather explicitly,
there are four different cases to consider: (i) It is desirable to
have nodes $i$ and $j$ in the same community if $A_{ij} =1$ (i.e.,
if nodes $i$ and $j$ are linked). Thus, a negative energy
$(-A_{ij}/2)$ is introduced for having nodes $i$ and $j$ in the same
community ($\delta(\sigma_{i}, \sigma_{j}) =1$). (ii) It is
undesirable to have nodes $i$ and $j$ in different communities if
they linked $(A_{ij} =1$). Thus, a positive energy contribution
$A_{ij}$ is added in such a case. This is tantamount to augmenting
the energy by $(1- \delta(\sigma_{i}, \sigma_{j})) A_{ij}/2$. (iii)
If nodes $i$ and $j$ are not linked ($A_{ij}=0$) then an energy
penalty should appear when putting them within the same community.
This is captured by the term $\gamma J_{ij} \delta(\sigma_{i},
\sigma_{j})/2$ with $J_{ij} = (1-A_{ij})$. (iv) If nodes $i$ and $j$
are not linked then it is energetically favorable to have them in
different communities. This leads to a negative contribution $(-
\gamma J_{ij} \delta(\sigma_{i}, \sigma_{j})/2)$ in the energy.
Putting all of these pieces together and summing over all node pairs
($ij$) leads, up to an innocuous additive constant, to the
Hamiltonian of Eq. (\ref{eq:ourpotts}). All energy contributions are
``absolute'' and set by the existence of a link (or its absence)
between the two members of each node pair without global (``null
model'' type) comparisons to other nodes in the system. An identical
analysis in the case of more general weighted graphs leads to the
Hamiltonian of Eq.(\ref{eq:newpotts}) that will be employed in the
current work for weighted images.
\end{thebibliography}
\end{document}